\documentclass[review]{elsarticle}

\usepackage[colorlinks,citecolor=blue,linktoc=all,linkcolor=cyan]{hyperref}
\usepackage{graphicx}

\usepackage{multirow}
\usepackage[T1]{fontenc}
\usepackage{dsfont}               
\usepackage{mathrsfs}             
\usepackage{slashed}              
\usepackage{amsmath}
\usepackage{amssymb}
\usepackage{amsbsy}
\usepackage{amsfonts}
\usepackage{bm}
\usepackage{epstopdf}
\usepackage{chemformula}
\usepackage{physics}
\usepackage{graphics}
\usepackage{float}
\usepackage{threeparttable}

\numberwithin{equation}{section}
\numberwithin{table}{section}
\numberwithin{figure}{section}

\journal{Progress in Particle and Nuclear Physics}

\usepackage{titlesec}
\usepackage{sectsty}
\titleformat{\section}{\normalfont\Large\bfseries}{\thesection}{1em}{}
\titleformat{\subsection}{\normalfont\large\bfseries}{\thesubsection}{1em}{}
\titleformat{\subsubsection}{\normalfont\normalsize\bfseries}{\thesubsubsection}{1em}{}

\begin{document}

\begin{frontmatter}

\title{Relativistic atomic structure calculations in support of spectroscopy}

\author[VSI,CU]{L. F. Pašteka}
\author[TAU]{E. Eliav}
\author[VSI]{M. L. Reitsma}
\author[VSI]{A. Borschevsky\corref{corresp}}
\cortext[corresp]{A. Borschevsky}
\ead{a.borschevsky@rug.nl}

\address[VSI]{Van Swinderen Institute for Particle Physics and Gravity, University of Groningen, Nijenborgh 4, 9747 Groningen, The Netherlands}
\address[CU]{Department of Physical and Theoretical Chemistry, Faculty of Natural Sciences, Comenius University, Ilkovi\v{c}ova 6, 84215 Bratislava, Slovakia}
\address[TAU]{School of Chemistry, Tel Aviv University, 6997801 Tel Aviv, Israel}

\begin{abstract}
  Theory can provide important support at all the stages of spectroscopic experiments, from planning the measurements to the interpretation of the results. Such support is particularly valuable for the challenging experiments on heavy, unstable, and superheavy elements and for precision measurements aimed at testing the Standard Model of particle physics. To be reliable and useful in experimental context, theoretical predictions should be based on high-accuracy calculations. For heavy elements, such calculations must treat both relativistic effects and electron correlation on the highest possible level.  Relativistic coupled cluster is considered one of the most powerful methods for accurate calculations on heavy many-electron atoms and molecules. This approach is highly accurate and versatile and can be used to obtain energies and a variety of atomic and molecular properties. Furthermore, its robust and transparent formulation allows for systematic improvement of the accuracy of the calculated results and for assigning uncertainties on theoretical values. The Fock-space coupled cluster (FSCC) variant of this method is particularly useful in the context of spectroscopic measurements as it provides access to atomic spectra and properties of the excited states.
 In this review, we present in detail the relativistic coupled cluster approach and its FSCC variant. We provide a description of the computational procedure used for accurate calculations and for assigning uncertainties. Outstanding recent examples of application to atomic properties, focusing on the experimental context are presented. Finally, we provide a brief  discussion of the  perspectives for future developments and applications of the CC approach.

\end{abstract}


\end{frontmatter}

\newpage

\thispagestyle{empty}
\tableofcontents

\newpage
\section{Introduction}\label{intro}

Spectroscopic and precision experiments on heavy, unstable, and artificial elements provide unique opportunities to investigate the electronic structure in the lower part of the periodic table, probe the nuclear structure of heavy nuclei and unstable isotopes, test the Standard Model (SM) of particle physics, and search for the tiny signatures of physics beyond the Standard Model (BSM). 

Since the 1940s, two naturally occurring elements without stable isotopes, Tc~\cite{perrier1947technetium} and Pm~\cite{marinsky1947chemical}, and  24 artificial elements (Am, $Z=95$, to Og, $Z=118$) have been discovered; elements with atomic numbers above 103 are referred to as transactinides or superheavy elements.  Their production was carried out using a variety of experimental techniques and facilities~\cite{Arm00, NagHir11,Trler2013,DulBloHes22}, such as nuclear reactors, particle accelerators, and even in hydrogen bomb test explosions, as happened in the case of Es and Fm~\cite{GhiThoHig55}. The latest elements added to the periodic table in 2016 are    113 (Nihonium), 115 (Moscovium), 117 (Tennessine), and 118 (Oganesson)~\cite{KarBarShe16a, KarBarShe16b}. Following the discovery of new elements, their spectroscopic and chemical properties are studied in challenging experiments~\cite{NagHir11,Trler2013,Schadel:14}. Such studies can shed light on the lower part of the periodic table, revealing trends which may differ from those shown by their lighter homologs. So far, the heaviest element where spectroscopy was carried out was No ($Z=102$). Sophisticated one-atom-at-a-time experiments were performed to measure the energy levels, the ionization potential, and the hyperfine structure of this unstable and short lived atom~\cite{LaaLauBac16,ChhAckBac18,RaeAckBac18}. The ionization potential or Lr ($Z=103$) was measured using the surface-ionization technique~\cite{SatAsaBor15}, along with those of lighter Fm, Md, and No~\cite{SatAsaBor18}. Spectroscopic measurements on Lr$^+$ and ions of even heavier elements are planned to be performed using a novel technique of ion-mobility-assisted laser spectroscopy, designed to overcome the challenge of decreasing production yields in the region of superheavy elements~\cite{LaaBucVie20,LaaBlo22}. These and other experiments on the heaviest elements push the limits of our knowledge of the behavior and electronic structure of these systems and probe the extreme effects of relativity. 

Spectroscopic measurements can also provide access to nuclear properties, such as nuclear spins, moments, and charge radii, and can be used to test the predictive power of nuclear models. Information on nuclear properties is contained in the shifts and splittings of electronic energy levels and can be extracted in a nuclear model-independent way from the hyperfine structure parameters and isotope shifts measured by laser spectroscopy. A number of recent reviews provide a comprehensive overview of this active research field~\cite{CamMooPea16, BloLaaRae21, YanWanWil23}. An important example of such studies is the spectroscopic investigation of the hyperfine structure of the Th$^{2+}$ ion, which allowed the extraction of the fundamental nuclear properties of the $^{229m}$Th isomer, namely, its magnetic dipole and electric quadrupole moments, as well as its nuclear charge radius~\cite{ThiOkhGlo18}. This isomer is unique in that its energy is only about 7.8~eV above the $^{229}$Th ground state and thus opens the possibility for the construction of a nuclear clock benefiting from high stability and low sensitivity to external perturbations~\cite{CamRadKuz12}. Recently, the frequency of this transition was measured for the first time with spectroscopic accuracy  using VUV frequency combs \cite{zhang2024frequency}. Another outstanding example of spectroscopic studies of nuclei are the challenging measurements of the hyperfine structure in nobelium that yielded the nuclear moments of $^{253}$No and the charge radii of the $^{252-254}$No isotopes and allowed to test the accuracy of nuclear DFT calculations~\cite{RaeAckBac18}.  

Finally, experiments on heavy elements are also extremely promising in the field of the search for signatures of BSM physics. The Standard Model of particle physics is very successful in describing the known physical phenomena and also has strong predictive power~\cite{TanHagHik18}. 
Nonetheless, some observations are inconsistent with the SM, such as the dominance of matter over antimatter~\cite{DinKuz03}. 
This and other inconsistencies prompt the development of new extensions of the SM and motivate the search for new particles and physical phenomena, conducted both in high-energy accelerators and in incredibly precise table-top experiments with atoms and molecules that search for signals of violation of fundamental symmetries beyond the SM predictions.
Here, we will focus on the latter. Table-top precision measurements take advantage of the many accessible energy levels in atoms and even more so in molecules, due to the latter's additional vibrational and rotational motions~\cite{SafBudDeM18,BerDelGed20}.
Atoms and molecules can be used to probe a wide variety of physical phenomena and benefit from strong enhancement effects that amplify the otherwise tiny signals and bring them into the measurable range. Heavy atoms and molecules are particularly advantageous for such experiments as sensitivity to violations of fundamental symmetries tend to scale very rapidly with the proton number, typically as $Z^2$ to $Z^5$, depending on the sought effect~\cite{GauMarIsa19,GauBer24}. 
Furthermore, certain effects experience additional dramatic enhancement due to nuclear deformations. For example, octupole deformed nuclei have an enhanced sensitivity to charge-parity (CP) violating hadronic physics via a nuclear Schiff moment (NSM), in addition to molecular enhancements~\cite{KudPetSkr13,ArrAthAu23}.
Thus, alongside experiments utilizing stable atoms and molecules, such as Hg~\cite{GraCheLin16},  YbF~\cite{HudKarSma11}, ThO~\cite{AndAngDeM18}, HfF$^+$~\cite{RouCalWri23}, TlF~\cite{GraTimKas21}, and BaF~\cite{AggBetBor18}, new experiments are planned, based on radioactive species~\cite{ArrAthAu23}, such as RaF~\cite{GarciaRuiz:20,UdrBriGar21}, which are expected to reach unprecedented sensitivity and discovery potential.  

All the experiments described above have one feature in common -- they are inherently challenging. Artificial elements are usually short lived and are produced in minute quantities, requiring one-atom-at-a-time experimental techniques. Furthermore, their production and use require large experimental facilities and concerted efforts of many research groups. At the same time, experiments that aim to test the Standard Model need to reach unprecedented sensitivity in order to detect the vanishingly small signatures of new physics. Precision measurements with radioactive atoms and molecules, while benefiting from electronic and nuclear structure enhancements, also combine the challenges of the two types of experiments. 

Success of these ambitious experiments thus requires dedicated facilities and specially developed experimental techniques~\cite{CamMooPea16, BloLaaRae21, YanWanWil23,SafBudDeM18,ArrAthAu23}. However, another crucial factor is the availability of strong and reliable theoretical support. Electronic structure theory can provide invaluable contributions in all stages of the experiment, from conception and planning to the interpretation of the results. For example, predictions of transition energies and strengths prior to spectroscopic measurements can help focus on the relevant range and avoid broad wavelength scans, which is crucial when dealing with unstable short-lived elements. The importance of such predictions was seen in the first successful spectroscopy of No, where the scanned range for the measurement of an allowed electronic transition~\cite{LaaLauBac16} was based on prior theoretical predictions~\cite{Fri05,IndSanBou07, BorEliVil07,LiuHutZou07,DzuSafSaf14}, obtained using a variety of computational approaches. 
Theoretical parameters are also often necessary to extract the properties of interest from the measured energy shifts. For example,  electronic  parameters can be used to extract nuclear moments from the hyperfine structure of the measured transition lines. Thus, we see theory and experiment working hand in hand in many recent high-impact spectroscopic studies~\cite{SatAsaBor15, RaeAckBac18,LeiKarGuo20,BarAndRai21,KosYanJia21}.

Atomic and molecular theory has a special place of importance in the field of precision measurements and experiments aimed at testing the Standard Model of particle physics and at searching for BSM physics. Under certain conditions, the electronic structure can act as an amplifier and greatly enhance the signal due to the tiny effects of new physics~\cite{SafBudDeM18}. We can tune these ``amplifiers'' by carefully designing the molecule and by selecting the optimal transitions. High accuracy calculations can be used to propose promising candidates for measurements. Such candidates should benefit both from enhanced sensitivity to signatures of new physics and from practical experimental advantages. The landscape of possible atomic, ionic, and molecular systems and transitions is vast and thus reliable computational methods are needed to investigate many possible systems and select the most promising of them for further experimental study. Some examples of such proposed systems can be found in Ref.~\cite{SafBudDeM18}, and in a number of later publications~\cite{FleDem21,HaoNavNor20,ZulGauGie22,AbeTsuKan23,MaiSkrPen22}.

Another, no less crucial goal of atomic and molecular theory is to provide accurate and reliable
predictions of the properties needed for the planning and executing of precision measurements, thus supporting these challenging endeavors. Examples are atomic or molecular polarizabilities, hyperfine structure parameters, theoretical investigations of possible laser cooling or trapping schemes, or investigations of expected systematic effects~\cite{IsaHoeBer10, HaoPasVis19, BekBorHar19,  DenHaaMoo22, MitPraSri21, OleSkrZai22}.

The interpretation of precision measurements also hinges on reliable theoretical input. Knowledge of atomic or molecular coupling factors is necessary to extract the properties of interest, i.e., the symmetry-violating parameters such as the electron electric dipole moment or the Schiff moment, from the measured energy shifts and splittings. These coupling factors are unique to the given atom or molecule and to the specific electronic state and cannot be measured in principle, necessitating their accurate predictions~\cite{SafBudDeM18}. 
Understanding the results of such experiments and placing them in the context of the Standard Model and its extensions also often requires theoretical input. Perhaps the most prominent case is the measurement of the parity violating 6s--7s transition amplitude in cesium~\cite{WooBenCho97}, the interpretation of which in the context of testing the Standard Model and its extensions changed qualitatively a number of times due to the improving accuracy of the atomic calculations on which the interpretations were based (Ref.~\cite{DzuBerFla12} and references within).

Furthermore, one can go beyond the experimental context and perform purely theoretical investigations of systems and properties where no experiment has yet been possible. For example, such theoretical studies are currently the only way of obtaining any information on the spectroscopic properties of the heaviest transactinides. These properties can not be derived via extrapolations from the lighter homologs, due to the dramatic effect of relativity, which changes the trends in the groups in the Periodic Table~\cite{Per15,EliKalBor18, EliFriKal15}. Even more ambitious are the various attempts to predict the structure of the Periodic Table beyond element 118, that is, to gain basic information about elements that have not yet been produced~\cite{Pyy11, Indelicato:11, EliBorKal19,Savelyev2023}.

Theoretical investigations, whether in experimental context or as part of purely theoretical studies, should be based on accurate and reliable calculations. Dealing with heavy many-electron systems requires state-of-the-at treatment of both relativistic effects and of electron correlation (the description of the instantaneous interaction between the electrons). To be useful for planning and interpreting experiments, the theoretical predictions should have quantifiable and reliable uncertainties. This poses additional requirements on the choice of computational approach, namely, we need a method that is robust, that can be systematically improved, and that allows for analysis of the size of different contributions to the calculated values and the associated uncertainties. 

The combination of relativistic methodology with the accurate treatment of electron correlation requires significant computational resources and becomes intractable even for medium-sized molecules. However, for dealing with single atoms, very sophisticated computational approaches can be used while allowing for realistic calculations regarding both computational time and resources. Currently, a number of approaches are available that are suitable for accurate treatment of heavy many-electron atoms. These are the multiconfigurational Dirac--Fock approach (MCDF)~\cite{Fri02,Grant2007,VerRynJon13,JonGaiBie13}, the relativistic configuration interaction method (CI)~\cite{DzuFlaKoz96,FleOlsMar01, FleOlsVis03}, which can be also augmented by many-body perturbation theory (MBPT)~\cite{DzuFlaKoz96,AMBiT}, yielding the CI+MBPT approach, and the relativistic coupled cluster method (RCC)~\cite{VisLeeDya96,SasPatNay15,EliBorKal14,Eliav2024}. The combination of the CI approach with CC (usually referred to as CI+all-order) is also used for accurate atomic calculations~\cite{SafJoh08,SafKozJoh09,PorKozSaf16}.  In this review, we will focus our attention on the RCC approach and its applications. This method lends itself to systematic improvement of the treatment of electron correlation, relativity, and basis set quality, allowing for achieving extremely high accuracy in the treatment of heavy many-electron systems. At the same time, the robust and transparent nature of this approach makes it possible to devise reliable schemes for evaluating uncertainties on the calculated energies and properties by carrying out extensive investigations of the effect of various computational parameters. \\

This review is organized as follows: Section~\ref{sec:comp} provides an overview of the state-of-the-art methods used in high-accuracy calculations on heavy many-electron systems, focusing on the relativistic coupled cluster approach. Some essential practical aspects of using relativistic coupled cluster methodology are described in Section~\ref{sec:pract}. That includes analysis of the factors that influence the accuracy (Section~\ref{sub:accur}) and the computational procedure for error estimation(Section~\ref{sub:error}).  
The next section describes recent outstanding applications of the relativistic coupled cluster approach to heavy and superheavy (SHE) elements and highly charged ions (Section~\ref{sec:appl}), presenting calculations of fundamental properties such as ionization potentials, spectra, and hyperfine structure parameters. This part of the review showcases the power and the versatility of this method. 
The last part includes perspectives for future developments and applications (Section~\ref{sec:out}), as well as a summary and conclusions (Section~\ref{sec:sum}).

\section{Relativistic computational approaches}\label{sec:comp}

\begin{figure}[h!]
\begin{center}
\includegraphics[trim={0 5cm 0 0},clip,scale=0.5]{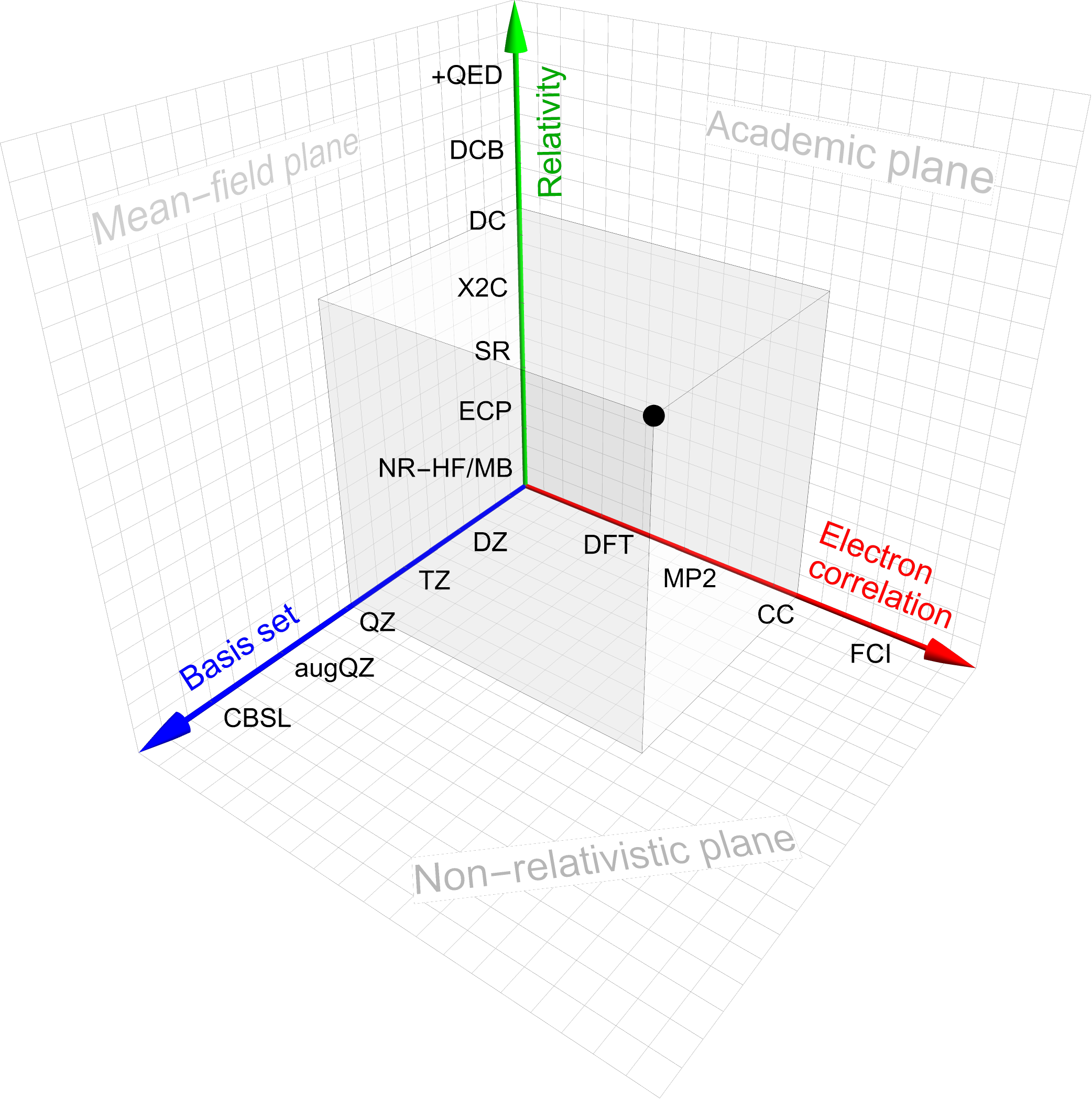}
\caption{Illustration of the three major and (mostly) independent directions of hierarchical improvements in electronic structure calculations. See the text for an explanation of the different labels. 
The black point represents the typical starting point at the DC-(FS)CCSD/QZ level of theory used in most relevant applications.
}
\label{3axes}
\end{center}
\end{figure}

The quality of any \textit{ab initio}, wave function based, calculation on a many-electron system can be described by a point in a three-dimensional space, presented in Figure~\ref{3axes}. The three axes represent the three major computational parameters that determine the accuracy of the employed approach: i) the treatment of relativity, ii) the computational approach used for describing the electron correlation, and iii) the choice of the basis set. 
For the origin of this description, we choose the nonrelativistic (NR) mean-field Hartree--Fock (HF) in a minimal basis (MB), the simplest quantum-mechanical method that can conceivably still be labeled \textit{ab initio}. This can be systematically improved along each of the aforementioned axes.

The first axis we describe here is the treatment of relativistic effects or the choice of the Hamiltonian. At the origin of this axis, the nonrelativistic Schrödinger equation is used, rendering any calculation in the bottom plane inherently nonrelativistic. Progressing along this axis, various approximate approaches for treating relativistic effects are introduced. Initially, staying in the one-component domain, the effective core potentials (ECPs)~\cite{TitMos99,Sch03,DolCao12} and the scalar-relativistic (SR) methodology can be employed for accounting for the effects of relativity; these are particularly effective for closed-shell systems, where spin-orbit effects are not very pronounced. Advancing further along the axis, we encounter the variety of the two-component approaches~\cite{Dyall:07a, Saue:11}, which allow the user to account for the spin-orbit effects. Among those, the relatively recently developed exact two-component relativistic (X2C) Hamiltonian stands out in accuracy and efficiency~\cite{Ilias:07,Knecht:22}. This approach reproduces exactly the positive-energy spectrum of the parent four-component Hamiltonian at a fraction of the computational costs. An extensive review of the two-component approaches can be found in Ref.~\citenum{Liu10}. Further along the axis is the full four-component Dirac--Coulomb (DC) Hamiltonian, followed by the Dirac--Coulomb--Breit (DCB) Hamiltonian, and finally, the quantum electrodynamics (QED) corrections (beyond the DCB description), with the lowest order corresponding to the self energy and vacuum polarization~\cite{Pyy78}. 
It should be noted that the ordering of the methods is illustrative and not strictly exact. For example, while two-component methods generally provide higher quality treatment of relativistic effects compared to their one-component counterparts, the ordering of the various one- and two-component approaches is approximate and does not reflect the complexity of the available methodology.
Additionally, two-component ECPs are also available and in cases where the error introduced by the frozen-core approximation is smaller than that of the neglect of spin-orbit effects, these can easily effectively surpass in performance the theoretically more rigorous all-electron scalar-relativistic Hamiltonians.

The second axis in Figure~\ref{3axes} represents the treatment of electron correlation. At the origin we find the Hartree--Fock (HF) approach, which neglects correlation altogether and models the many-body wave function as a determinant product of single electron orbitals. 
Such calculations are low-cost, but suffer from a correspondingly low accuracy. On the other hand, at the far end of this axis, we find the full configuration interaction (FCI) method, which allows to account for the entirety of electron correlation for a given basis set. Such calculations are unrealistic for all but the smallest systems, combined with very modest basis sets. In between the two extremes reside the plethora of approximate approaches for the treatment of electron correlation, varying in completeness, accuracy, and the corresponding computational costs. Some examples are many-body perturbation theory (MBPT, and in particular its frequently used second order M\o{}ller–Plesset, MP2, variant), multiconfigurational self-consistent field theory (MCSCF/MCDF in the nonrelativistic/relativistic formulation), different levels of approximation of the configuration interaction (CI) approach (e.g. CI with single and double excitations, CISD), and analogously the different orders of the coupled cluster (CC) approach. For a comprehensive overview of electronic structure theory, we refer the reader to Ref.~\citenum{Helgaker2000}. Below, we selectively focus on methods most relevant to the topic of this review.
It is important to stress that the ordering of the methods in Figure~\ref{3axes} is approximate and given for illustrative purposes.
There is a separate systematic hierarchy of improvements embedded within each of the classes of methods, be it the order of M\o{}ller–Plesset perturbation theory or an excitation truncation rank in CI or CC. 
Besides the methods explicitly labeled in the plot, their multireference variants are often essential to model the wave function properly.
Furthermore, not only selection of the electron correlation itself, but also of the orbital space that it is applied to plays an enormous role in the result and brings with it a set of computational parameters subject to systematic improvements.

The third axis in Figure~\ref{3axes} represents the basis set size. Basis sets are sets of functions used to model the electronic wave functions. In the case of many atomic codes, either numerical modeling or B-spline basis sets are usually used~\cite{BacCorDec01}, corresponding to effectively complete basis sets. Molecular codes, however, due to the reduced symmetry and wider scope, rely on analytical basis sets, usually taken as sets of Gaussian functions~\cite{NagJen17,Per21}. These basis sets are optimized for a given element and even for specific types of calculations (for example, relativistic vs. nonrelativistic or generic vs. property specific), and as we increase the basis set size, we gain in the flexibility of the modeled wave function and thus in accuracy, at the cost of the increased computational effort. On the axis in Figure~\ref{3axes}, the basis sets are arranged according to their cardinality (i.e., the number of basis functions provided for the description of each occupied orbital) and culminate in a complete basis set limit (CBSL). Cardinality labels are typically given in multiples (double = D, triple = T, quadruple = Q etc.) of the minimal basis valence set (denoted $\zeta$, z, or Z). While the CBSL can not be reached in realistic calculations, various schemes exist for extrapolation of the results to the CBSL (Section~\ref{sec:basis}). Besides cardinality, the basis sets can be augmented further by specific types of functions, such as, for example, diffuse (low exponent) basis functions needed for the description of the valence electrons, or, conversely, tight (high exponent) basis functions, used to improve the description of the nuclear region. This is simply collectively reflected in the ``augQZ'' point in Figure~\ref{3axes}. 

It is now important to note that the selected computational method, i.e., the combination of the Hamiltonian, the treatment of electron correlation, and the choice of the basis set, should be balanced. That is, for calculations on a heavy system, if high accuracy is desired, both a four-component relativistic framework and a state-of-the-art method for the treatment of electron correlation are required. Furthermore, these should be accompanied by a large basis set to take full advantage of the capabilities of these sophisticated methods, leading to high computational costs. Thus, selecting a computational method is always a compromise between the size of the system, the required level of accuracy, and the available computational resources. 
In general, the full DC or DCB Hamiltonian, along with the QED corrections, are only used in cases where extremely high accuracy is needed and where such a calculation is feasible, i.e., single atoms or ions or small molecules.

Progressing along each axis in  axis Figure~\ref{3axes} entails a significant increase in computational costs along with the improvement in the accuracy of the calculation. 
Computational scaling in terms of basis set size, level of correlation and relativistic treatments are different in nature. It is fairly common to express the cost of correlation methods in terms of the leading (rate-determining) polynomial scaling as 
$\mathcal{O}(N^{x})$
with $N$ representing the number of basis functions or correlated orbitals and $x$ an exponent inherent to the chosen correlation method.
Additionally, the shift from a nonrelativistic to a relativistic framework is associated with a multiplicative prefactor. 
The general scaling for a relativistic electronic structure method can be thus expressed as~\cite{Fleig:12}
\begin{equation}
    S= R\,g\,N^x,
\end{equation}
where all parameters have a direct connection to the respective dimensions of our computational space (Figure~\ref{3axes}) -- prefactor $R$ relates to the level of relativistic treatment, the general prefactor $g$ and the exponent $x$ are specific to the correlation method, and $N$ relates to the basis set size (as well as the system size and active space size).
The mean-field Hartree--Fock method scales are $N^3$ and the corresponding $R$ shifting from NR to four-component DC Hamiltonian is typically 6--10~\cite{Saue:03,Dyall:07a}, depending on the details of implementation and system size. Nevertheless, the real bottleneck lies in the post-HF electron correlation methods.
For coupled cluster, which is the focus of this paper, $x$ grows linearly with the excitation level (see Section~\ref{subsub:CC}) and an approximate formula for the  relativistic prefactor $R$ ($x$-dependent)
has been derived by Fleig \textit{et al.}~\cite{Sorensen:11,Fleig:12}
\begin{equation}
    R(x)=4\sqrt{\pi\left(\frac{x}2{-1}\right)},\ \ x\geq4.
\end{equation}
For a more detailed discussion of computational scaling of relativistic correlation methods and the specific steps involved in the calculations (integral evaluation, integral transformation, cluster iterations, etc), see Refs.~\cite{Visscher1995,Saue:03,Dyall:07a,Sorensen:11,Fleig:12}.
A concrete example of computational costs involved in calculations at different levels of theory is given in Section~\ref{sub:Au}.

\begin{figure}[h!]
    \centering \includegraphics[scale=0.3]{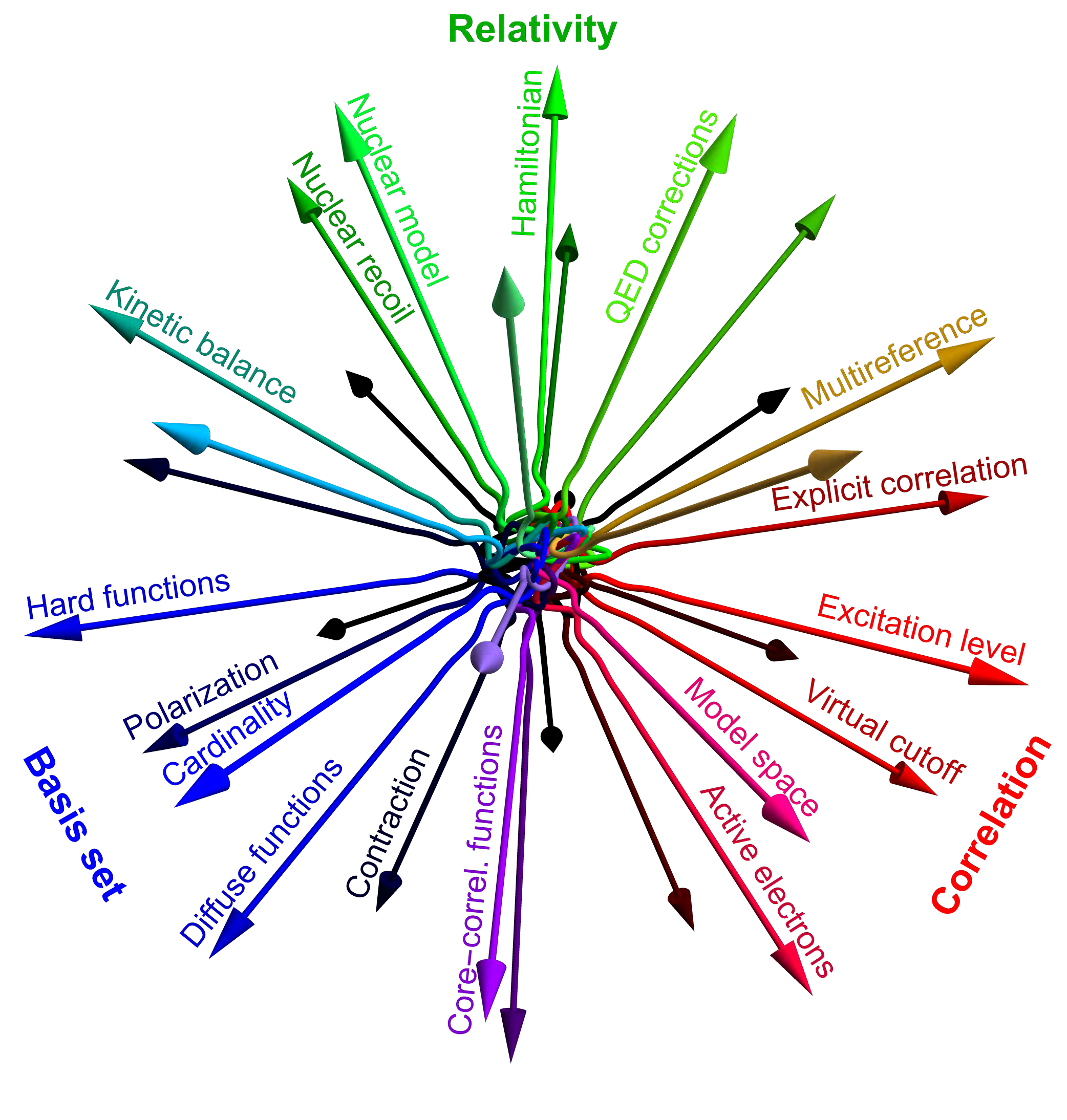}
    \caption{\textit{(New figure) Visual representation of the hidden multidimensionality of the computational parameter space and the increased coupling between these parameters near the origin (corresponding to low level treatment).
    In this representation, the different parameters are nevertheless clustered into three main groups corresponding to the overarching umbrella terms of relativity, electron correlation, and basis sets.}}
    \label{hedgehog}
\end{figure}

The concept of the three-dimensional computational space depicted in Figure~\ref{3axes} is well established in the relativistic quantum chemistry community~\cite{TARCZAY2001,Fossgaard2003,Fleig2006,Saue:11,Fleig:12,Liu2022} and is itself an extension of the broadly adopted two-dimensional version occupying the nonrelativistic plane in Figure~\ref{3axes} dating back to John Pople~\cite{Pople1965}.
It is a very useful and powerful representation of the computational space and it will also serve as the main scaffold around which this review is structured.
Nevertheless, it should be noted that the computational reality is more complex.
While the three axes are independent in Figure~\ref{3axes}, in practice, there is an interplay between the influence of relativistic effects, electron correlation and basis set size. 
This is especially true at the lowest levels of theoretical treatment based on limited correlation and small basis sets.
However, the individual effects gradually detangle as the level theory increases along each axis.
Furthermore, as outlined by the preceding paragraphs, each major axis encapsulates a whole family of different computational parameters. When expanded, each of these could be given their own separate axis in the resulting multidimensional computational space. 
Thus, as a complement to the idealized version of the three-dimensional plot,
we offer a more nuanced representation in Figure~\ref{hedgehog} taking into account more of the intricacies of the computational problem at hand.
Systematic improvements along each axis in this plot can be accompanied by vastly different levels of increase in computational cost and resulting accuracy. Hence, also the balancing act between cost and accuracy described in the preceding paragraph is correspondingly multidimensional in nature.

In the following, we will give an overview of the available high-accuracy computational methods, loosely following the structure of the three axes in Figure~\ref{3axes} and focusing our attention on the relativistic coupled cluster approach.

\subsection{Four-component methodology}

\label{sec:methodology}

The building blocks of the four-component methods are described in this
Section. We start with the many-body QED Hamiltonian and its no-virtual-pair
approximation. The later subsections describe how this formalism is applied to atomic and molecular systems.

\subsubsection{QED origins of relativistic quantum chemistry}

\label{sec:QED_origins}

Relativistic effects play a significant role in the structure, spectroscopy, and chemical behavior of heavy atoms and molecules. Even in lighter elements, these effects manifest as fine and hyperfine structures in electronic states. Since the 1970s, advances in computational power have enabled the development of relativistic many-body methods (see some recent books~\cite{Hess:book,
peterS:book1, peterS:book2, KaldorWilson:book, Hirao:04,
Dyall:07a,Grant2007,Reiher:14,Barysz:10}). 
Traditional quantum chemistry techniques — such as configuration interaction (CI), many-body perturbation theory (MBPT), multi-configuration self-consistent field (MCSCF), coupled cluster (CC) methods, and density matrix renormalization group (DMRG) theory — have been extended to incorporate relativistic effects. These methods are now implemented in user-friendly software for high-precision calculations. The most popular publicly available programs combining relativity and electron correlation on the highest sophisticated levels
are RELCI~\cite{FRITZSCHE2002103}, MCHF~\cite{fischer2007mchf}, MCDFGME~\cite{MCDFGME}, GRASP~\cite{JonGaiBie13,GRASP2018}, AMBIT~\cite{AMBiT} and CI-MBPT~\cite{CI+MBPT} for atoms, and MOLFDIR~\cite{MOLFDIR}, UTCHEM~\cite{utchem}, BERTHA~\cite{BERTHA}, DIRAC~\cite{Saue:20}, RAQET~\cite{hayami2018raqet}, BAGEL~\cite{BAGEL}, and EXP-T~\cite{Oleynichenko:EXPT:20} for
atoms and molecules.

The relatively delayed development of relativistic many-body methodologies stems from the mathematical challenges of integrating relativity with quantum many-body theory in a consistent computational framework. A fully covariant and consistent quantum description requires treating matter (electrons and positrons) and radiation (photons) on equal footing, as established in quantum electrodynamics (QED). Developed in the late 1940s by pioneers like Feynman, Dyson, Schwinger, and Tomonaga
~\cite{Schwinger:49,Schwinger:49a, Feynman:49,Feynman:49a, Tomonaga:48, Dyson:49,Dyson:49a}, QED is one of the most successful and precise physical theories, accurately describing microscopic phenomena down to energy scales of
$10^{-13}$ cm. It provides a natural foundation for accurate relativistic quantum chemistry.

Despite its success, QED's impact on quantum chemistry has been limited due to its technical complexities. Advanced methodologies for handling QED effects in many electron systems exist but have yet to be fully applied to heavy atoms and molecules~\cite{Lindgren:11,Shabaev:02,Andreev:08}. Currently, only leading QED effects including vacuum polarization and self-energy are approximated using effective model potentials in actual heavy elements calculations~\cite{pyykko2003search,flambaum2005radiative,Shabaev:13}.

More elaborate effective relativistic Hamiltonians, properly derived from QED and  describing the
structure and interactions of the quanta of fermionic fields (e.g., electrons
and positrons) with quanta of electromagnetic fields (e.g., photons) in
forms suitable for developing a relativistic many-body procedure, with the
aim of applying this procedure later in atomic and molecular electronic structure
calculations, can be found in Refs.~\cite{Lindgren:11,Shabaev:93,Liu:14,Liu2013,Nonn:QED:2024}.

  The total number of particles is not conserved in QED, and
electron-positron pair creation processes are included in calculated systems.
The number of photons is also variable, depending on the particular fermionic
interaction process. This is why the generalized Fock-space with variable numbers of electrons/positrons and photons is considered to be
the only appropriate mathematical framework for the development of QED-consistent many-body
approaches~\cite{Kutzelnigg:12,Liu2013}. 

Probably the most consistent and potentially
powerful  many-body QED method, called the covariant evolution operator (CEO) method,
with a structure resembling that of stationary multi-particle approaches used in quantum chemistry and atomic physics, has been
developed in the Fock-space constraints by Lindgren and coworkers~\cite{Lindgren:11,Lindgren:04}.
It offers the possibility of being merged with quantum chemical machinery
based on the Bloch equation to provide a unified tool suitable for application
to general quasidegenerate atomic and molecular configurations.
The CEO\ method has a particularly mathematically suitable form when formulated in the generalized
Fock space with variable numbers of fermions and so-called uncontracted
virtual photons. It is, therefore, considered a natural framework for
implementing Fock-space many-body quantum chemical approaches, capable of
describing systems with a variable number of particles. In particular, the
relativistic Fock-space coupled cluster (FSCC) approach, which is an
all-order, size extensive, multi-root, multireference method (for a recent
review, see~\cite{ELIAV202479}), is an ideal candidate for merging with CEO. The
FSCC method and its recent applications in the effective and
intermediate Hamiltonian formulations are described in Sections~\ref{sec:fscc}--\ref{sec:pract}. Recently, a more advanced, so-called double Fock-space coupled
cluster method (DFSCC), based on CEO-QED, has been presented
~\cite{Eliav:10,Eliav:10a}. This approach includes the treatment
of both electronic and photonic degrees of freedom on equal quantum footing.
DFSCC yields a possible avenue for covariant high-precision treatment of heavy relativistic
multielectronic systems. Another promising option, which
can be applied within the algebraic approximation, is to use (at least at the
mean-field HF level) the variational QED procedure~\cite{Saue:03,salman2023calculating}, where the
explicitly filled Dirac negative energy sea is included in the system core
states, thus defining the HF state formally as having a large but finite charge
and mass. The renormalization procedure is then explicitly included in the SCF iterations. A modification of this procedure, described in Ref.~\cite{Eliav:10}, is based on using negative energy states of the free
electron as well as single retarded photon exchange effective potential. This
will incorporate the leading radiation effects (Lamb shift) in the
renormalized HF energy and wave function self-consistently so that the
appropriate reducible multiphotonic part of the vacuum polarization and the
self-energy will be included in the direct and exchange SCF terms,
respectively. Both the 
above-mentioned 
advanced relativistic many-body
QED-based approaches (namely, DFSCC and variational QED) are still under development and will not be discussed further here.

\subsubsection{Relativistic framework and the no-virtual-pair
approximation}

\label{sec:NVPA}

Heavy atomic and molecular systems are both of great interest for research based on modern, highly precise spectroscopy and for the searches for the effects of new physics beyond the Standard Model. Thus, computational treatment of atoms and molecules will be regarded here from a common theoretical viewpoint. However, we will then concentrate on the practical atomic applications, referring the readers to some recent reviews describing outstanding recent molecular investigations within the relativistic coupled cluster approach~\cite{ELIAV202479,LiuChe21,Pototschnig:2021,saue2015relativistic}.    
Atoms and molecules are described in quantum chemistry as systems with a finite number of
particles interacting via instantaneous, energy-independent two-body
potentials. This picture ignores partially or fully some fundamental
QED phenomena, such as the existence of the negative energy states continuum,
radiative effects and retardation of the interparticle interactions, which
are important for a fully covariant description. Fortunately, many of these
QED corrections are numerically small for real atomic and molecular systems,
explaining the success of low-order approaches directly derived from QED by constructing relativistic many-body Hamiltonians via
summation of one-electron Dirac Hamiltonians and interparticle
two-body instantaneous
potentials. A rigorous form of such four-component stationary Hamiltonians is
the Dirac--Coulomb (DC) Hamiltonian, which uses the nonrelativistic Coulomb
form of interparticle interactions,
\begin{equation}
{\displaystyle\hat{H}=\sum_{i}\hat{h}_{D;V}(i)+\frac{1}{2}\sum_{i\neq j}%
\hat{g}^{{\text{Coulomb}}}(i,j)+\hat{V}_{nn};\text{ \ }\hat{V}_{nn}=\frac
{1}{2}\sum_{A\neq B}\frac{Z_{A}Z_{B}e^{2}}{\left\vert \mathbf{R}%
_{A}-\mathbf{R}_{B}\right\vert }}, \label{eq:DCH}%
\end{equation}
where $\hat{h}_{D;V}(i)$ are the one-electron four-component Dirac operators in
the molecular field $V$, which usually includes finite-size nuclear electric potential (see Refs.~\cite{Visscher:97,Andrae2000} for different popular nuclear models) and can also incorporate effectively leading QED effects in the form of model potentials (see~\cite{pyykko2003search,flambaum2005radiative,Shabaev:13}) and external fields  
\begin{equation}
\quad\hat{h}_{D;V}=\beta mc^{2}+c\left(  {\boldsymbol{\alpha}}\cdot
\mathbf{p}\right)  -eV. \label{eq:Dirac_Ham}%
\end{equation}
Here the $4\times4$ vector ${\boldsymbol{\alpha}}$ and scalar $\beta$
quantities are related to the Dirac matrices,
\begin{equation}
\ \ {\mbox{\boldmath$ \alpha  $\unboldmath}}=\left(
\begin{array}
[c]{cc}%
0_{2} & {\mbox{\boldmath$ \sigma  $\unboldmath}}\\
{\mbox{\boldmath$ \sigma  $\unboldmath}} & 0_{2}%
\end{array}
\right)  ,\quad\beta=\left(
\begin{array}
[c]{cc}%
I_{2} & 0_{2}\\
0_{2} & -I_{2}%
\end{array}
\right),
\end{equation}
with $\boldsymbol{\sigma}$ the Pauli spin matrices. $\hat{V}_{nn}$ is the
classical internuclear potential. The inter-electronic potential $\hat
{g}^{{\text{Coulomb}}}(i,j)$ is the Coulomb term%
\begin{equation}
\hat{g}^{{\text{Coulomb}}}(i,j)=e^{2}\frac{I_{4}\cdot I_{4}}{r_{ij}},
\label{eq:Coulomb term}%
\end{equation}
where the $4\times4$ identity matrices $I_{4}$ have been inserted to stress
that while the Coulomb term looks like the nonrelativistic
electron-electron interaction, its physical content is different. Upon
reduction to the nonrelativistic form through a Foldy--Wouthuysen
transformation~\cite{Chraplyvy:53a,Chraplyvy:53b,Barker:55,Itoh:65,Sjovoll:97}, one finds that the relativistic operator contains, for instance
spin-own orbit interaction, in addition to the instantaneous Coulomb
interaction. The first scheme based on the DC Hamiltonian was developed in
1935 by Swirles~\cite{Swirles:35}, who generalized the SCF approach of her
scientific advisor Hartree to the relativistic realm. However, practical relativistic quantum chemical calculations did not appear until the 1970s.

The transformation of the interelectronic interaction to covariant form may
be made by adding the missing effects of retardation and magnetic interaction
to the nonrelativistic limit represented by the instantaneous Coulomb
interaction. 
The lowest-order relativistic corrections to the
Coulomb electrostatic interaction between the electrons were considered for
the first time in the Feynman (Lorentz)\ gauge by Gaunt in 1929~\cite{Gaunt:29},
when a magnetic interaction of order $\alpha^{2}$ was added to the DC
Hamiltonian (\ref{eq:DCH}). This magnetostatic term is called the Gaunt
interaction and has the form
\begin{equation}
\hat{g}^{{\text{Gaunt}}\text{ }}(1,2)=-e^{2}\frac{c{\mbox{\boldmath$ \alpha
$\unboldmath}}_{1}\cdot c{\mbox{\boldmath$ \alpha  $\unboldmath}}_{2}}%
{c^{2}r_{12}}. \label{eq:Gaunt term}%
\end{equation}
The Gaunt interaction is instantaneous, similar to the Coulomb term. One can
furthermore show by reduction to the nonrelativistic form that the Gaunt term
carries all spin-other-orbit interactions~\cite{Saue:PhD:96}. It was later shown
by Breit~\cite{Breit:29,Breit:30,Breit:32} that the retardation of the Coulomb interaction
gives rise to effects of the same order, $\alpha^{2}$. This leads, together
with the magnetic interaction to the so-called Breit interaction
~\cite{Breit:29,Breit:30,Breit:32}
\begin{equation}
\hat{g}^{{\text{Breit}}\text{ }}(1,2)=-\frac{e^{2}}{2c^{2}r_{12}}\left[
\left(  c{\mbox{\boldmath$ \alpha  $\unboldmath}}_{1}\cdot
c{\mbox{\boldmath$ \alpha  $\unboldmath}}_{2}\right)  +\frac{1}{r_{12}^{2}%
}\left(  c{\mbox{\boldmath$ \alpha
$\unboldmath}}_{1}\cdot\mathbf{r}_{12}\right)  \left(  c{\mbox{\boldmath$
\alpha  $\unboldmath}}_{2}\cdot\mathbf{r}_{12}\right)  \right]  .
\label{eq:Breit1}%
\end{equation}
The Breit term is written here in a slightly unusual form~\cite{Saue:03}%
, using explicitly the relativistic velocity operator
$c{\mbox{\boldmath$ \alpha  $\unboldmath}}$. This term can be derived as the
low-frequency limit of the single virtual photon exchange interaction in the
Coulomb gauge as described by QED. The final form of the Dirac--Coulomb--Breit
(DCB) Hamiltonian is
\begin{equation}
{\displaystyle\hat{H}_{\mathrm{DCB}}=\sum_{i}\hat{h}_{D;V}(i)+\frac{1}{2}%
\sum_{i\neq j}\{\hat{g}^{{\text{Coulomb}}}(i,j)+\hat{g}^{{\text{Breit}}\text{
}}(i,j)\}+\hat{V}_{nn}}. \label{eq:DCBH}%
\end{equation}

The Breit interaction (\ref{eq:Breit1}) is instantaneous, although it
compensates for the leading effect of the retardation of the Coulomb
interaction. In a full QED treatment, there is an additional time/energy
dependent retardation effect of the Breit interaction of order $\alpha^{3}$.
If energy is conserved, the exchange of a single transverse retarded photon
yields the following form of the frequency-dependent Breit interaction for the
case of ``on-shell'' interactions,
\begin{equation}
\hat{g}_{\text{${\omega}$}}^{{\text{Breit}}\text{ }}(1,2)=-\frac{e^{2}}%
{2c^{2}r_{12}}\left[  \left(  c{\boldsymbol{\alpha}}_{1}\cdot
c{\mbox{\boldmath$ \alpha  $\unboldmath}}_{2}\right)  -\frac{\left(
c{\mbox{\boldmath$
\alpha  $\unboldmath}}_{1}\cdot{\mbox{\boldmath$ \nabla  $\unboldmath}}%
_{1}\right)  \left(  c{\mbox{\boldmath$ \alpha  $\unboldmath}}_{2}%
\cdot{\mbox{\boldmath$ \nabla  $\unboldmath}}_{2}\right)  (\exp(i|\omega
|r_{12})-1)}{\omega^{2}}\right]  . \label{eq:Breit_omega}%
\end{equation}
In the SCF approximation, the photon frequency $\omega$ is defined by the
orbital energy difference. \ In the zero-frequency or energy-independent limit
$\omega\rightarrow0$, the expression (\ref{eq:Breit_omega}) transforms into
(\ref{eq:Breit1}).

In the alternative Feynman gauge, the frequency-dependent Gaunt interaction
has the form
\begin{equation}
\hat{g}_{\omega}^{{\text{Gaunt}}\text{ }}(1,2)=-e^{2}\frac
{c{\mbox{\boldmath$ \alpha
$\unboldmath}}_{1}\cdot c{\mbox{\boldmath$ \alpha  $\unboldmath}}_{2}%
\exp(i|\omega|r_{12})}{c^{2}r_{12}}. \label{eq:fdGaunt}%
\end{equation}
In the zero-frequency (energy-independent) limit, this interaction is
transformed into the instantaneous Gaunt interaction (\ref{eq:Gaunt term}).
The instantaneous Gaunt interaction does not contain any retardation, and
therefore the retardation correction to this interaction is of order
$\alpha^{2}$, an order of magnitude larger than the energy-independent interaction
in the Coulomb gauge, which is the Breit term (\ref{eq:Breit1}). This implies
that when the frequency-independent Gaunt potential (Feynman gauge) is used in
the quantum chemical calculations of heavy element compounds, considerable
errors may be introduced~\cite{Gorceix:88,Lindroth:89,Sucher:88,Lindgren:90}. In
bound-state QED calculations, on the other hand, when the retardation is properly taken
care of, this error is eliminated, and the Feynman gauge is often used due to
its simplicity.

Most of the benchmark relativistic approaches implemented with the DCB Hamiltonian were adapted from the nonrelativistic realm by using special relativistically invariant double
point groups, as well as Kramers (time-reversal) symmetry when applicable. In
the atomic case, the high symmetry allows the separation of radial and angular
degrees of freedom. The angular part can be solved analytically with the help
of Racah algebra~\cite{Grant:61}, whereas the radial equations can be solved
by finite difference methods. In molecular calculations, one has to resort to
the \emph{algebraic approximation}, using finite basis set expansions.
This approximation is often used for atoms, too. The first basis set
calculations led to rather disastrous results (see~\cite{Schwarz:82a} for
references), caused by the fact that the relativistic four-component Hamiltonian
(\ref{eq:DCH}) is not bounded from below. This is due to the existence of the
negative energy continuum, which in the usual quantum chemical practice is
kept unfilled, in contrast to the true QED methodology, where it is filled
according to Dirac's original idea (``Dirac sea''). Thus, the relaxation of
occupied atomic/molecular electronic levels by the negative states orbitals in
a true QED and the DCB Hamiltonian cases is different. The difference between
the ``relaxations'' within the two approaches has the same order of magnitude
as the Lamb shift ($O(\alpha^{3})$), according to Ref.~\cite{Liu:14}. Special care
must be taken for the correct formulation of the SCF and correlation methods based
on such a non-QED Hamiltonian. 

Talman and LaJohn~\cite{Talman:86,LaJohn:92} pointed out that it is feasible to find
the electron-like positive energy solutions of the DCB Hamiltonian by using a variational minimax SCF principle,
where the energy is minimized with respect to rotations in the virtual
positive energy spinor space (the optimization process used in the
nonrelativistic approach) and maximized with respect to rotations in the
negative energy spinor space. 
The positive and negative energy parts of the
spectrum are well separated energetically, and the optimization procedure is
feasible. The coupling of the large and small components of the Dirac equation
leads to a difference in parity of the two components and thus requires
separate basis set expansions for each component. However, the small and large
component basis sets should not be chosen independently. 
To make the
minimax variational procedure stable, one has to impose a special condition,
known as \textquotedblleft kinetic balance\textquotedblright, connecting the
small and large component basis sets, or else face the so-called
\textquotedblleft variational collapse\textquotedblright%
~\cite{Schwarz:82a} or \textquotedblleft basis set
disease\textquotedblright~\cite{Schwarz:82}.

Brown and Ravenhall~\cite{Brown:51} proposed to include the
interelectronic interactions between projection operators in the positive
energy spectrum, $\Lambda^{+\text{ }}$, to avoid the mixing of the
negative and positive continuum determinants, thus overcoming the so-called continuum dissolution or
\textquotedblleft Brown--Ravenhall disease\textquotedblright, where no
bound state is obtained. This strategy has been further explored by Sucher and
others (see~\cite{Kutzelnigg:87} for a review), based on QED theory.
The final projected Dirac--Coulomb--(Breit) Hamiltonian has the form
\begin{equation}
\hat{H}_{\mathrm{DCB}}^{+}  =\sum_{i}\hat{h}_{D;V}(i)+\frac{1}{2} \sum_{i\neq j}\Lambda_{i}^{+\text{ }}\{\hat{g}^{{\text{Coulomb}}}(i,j)+\hat
{g}^{{\text{Breit}}\text{ }}(i,j)\}\Lambda_{j}^{+\text{ }}+\hat{V}_{nn}. \label{eq:PDCB}%
\end{equation}
Here, 
$\Lambda_{i}^{+\text{ }}  =\overset{+}{\underset{n}{\sum}}\psi_{n}^{+}(x_{i})\psi_{n}^{{}}(x_{i})$ is the projection to the positive energy states of the single particle Dirac Hamiltonian $\hat{h}_{D;V}$. 
$H_{\mathrm{DCB}}^{+}$ is correct to second order in the fine-structure
constant $\alpha$, but is not covariant. This Hamiltonian is expected to be
highly accurate for most neutral and weakly-ionized atoms and molecules
~\cite{Lindgren:11,Lindgren:88,Lindgren:89}.

Additional QED terms are required for benchmark calculations of super-heavy elements (SHE) or multiply ionized species of heavy-element compounds; these corrections are considered in several recent reviews~\cite{IndSanBou07,Indelicato2019,Indelicato2021,shabaev2024quantum}. Recent methodological developments and calculations of leading QED effects in heavy systems may be found in Refs.~\cite{Shabaev:13,Pyykk2011,flambaum2005radiative,Tupitsyn:13,Lowe:13,Goidenko:09,Goidenko:03,Roberts:13,Shabaev:13,Schwerdtfeger2015,ginges2016atomic,SunSal22,salman2023calculating,shabaev2024quantum}.

If one constructs the projection operators from the same independent
particle basis used in the expansion of the second quantized form of
$H_{\mathrm{DCB}}^{+}$, which is almost always the case, the effect of the
projection operators is simply to limit the eigenfunctions of $H_{\mathrm{DCB}%
}^{+}$ to configurations with positive energy spinors only. In practice, the
continuum dissolution problem was solved in the algebraic approximation by
simply ignoring the negative energy branch in the correlation step, and
four-component relativistic molecular calculations are routinely carried out
today. When the finite element method is used for atoms, the two problems
listed above are solved by imposing the electron-like boundary conditions at
$r=0$ and $r\rightarrow\infty$ for bound solutions~\cite{Grant:88}. The
approximation based on $H_{\mathrm{DCB}}^{+}$ is called no-virtual pair
approximation (NVPA), since the virtual electron-positron pairs which cause
the Brown--Ravenhall disease are eliminated.

The $H_{\mathrm{DCB}}^{+}$ Hamiltonian is not unique since the distinction
between electron and positron creation and annihilation operators, as well as
the operators $\Lambda^{+}$, depend on the orbital set in which the field
operators are expanded. One possible choice is the eigenorbitals of the
free-particle Dirac equation, giving the \textquotedblleft
free\textquotedblright\ picture. Another choice is the solutions of the Dirac
equation in the molecular field $V$~\eqref{eq:Dirac_Ham}, leading to the Furry
picture~\cite{Furry:51}. A third possibility is to continuously update the
projection operators in each SCF iteration so that they correspond at
convergence to the solutions of the combined molecular and mean-field
potentials of the Hartree--Fock equations. This so-called \textquotedblleft
fuzzy\textquotedblright\ picture, proposed by Mittleman~\cite{Mittleman:81},
corresponds to the standard four-component approach. Some authors do not
distinguish between the \textquotedblleft fuzzy\textquotedblright\ and
 Furry pictures (see, for instance,
~\cite{Labzowsky:QED:96}). The optimal choice of the projection operators is
discussed in~\cite{Visscher:02}.

Properly designed four-component many-electron NVPA methods are currently the
most advanced and precise approaches for molecules and most atoms and are
used for benchmark calculations. The separate basis set expansion of the large
and small components used in NVPA leads to higher computational costs compared
with nonrelativistic or more approximate relativistic methods. Careful
analysis shows that the cost difference between four-component NVPA schemes and
nonrelativistic methods is in the prefactor, not in the scaling
~\cite{Saue:03}. Still, this difference is large enough to encourage the
use of more cost-efficient relativistic approximations, which replace
four-component approaches by two- or one-component schemes. One-component methods
include only the so-called scalar (or kinematic) relativistic effects. 
Two-component approaches incorporate, in addition, electron spin effects.
Exact (or infinite order) two-component (X2C) methods have been developed
recently (see original papers
~\cite{Dyall:97,Dyall:99,Filatov:06,Barysz:01,Wolf:02,Kutzelnigg:05,Ilias:07,Knecht:22} and reviews~\cite{Liu10,Saue:11,Peng:12}). Most X2C methods are iterative and
based on either elimination techniques for the small component or special
unitary transformations that decouple the NVPA\ Hamiltonian. They are capable
of reproducing the energies of four-component NVPA in the iterative numerical
limit if the same projection operators are used in the NVPA Hamiltonian. This
condition is not trivial to satisfy. Four-component methods usually allow the
continuous update of the NVPA Hamiltonian and, therefore, the complete
relaxation of the electronic wave function. In contrast, in most X2C approaches,
this relaxation is absent due to the use of predefined projection operators
before performing an approximate decoupling of the electronic and positronic
degrees of freedom. X2C may lead to excellent (almost exact) approximations of
the four-component NVPA method and allow relativistic calculations at reduced
computational cost. Still, it is incorrect to state that it provides complete
equivalence with the four-component methods. In addition, since the X2C
Hamiltonian and property operators are formulated in matrix algebra, the
real-space representation of the charge and current density in this approach
could be substantially different from the parent four-component case.
Additional care in X2C methods is required for properties with large response
contribution from negative-energy states (e.g., NMR parameters). Recently, novel
X2C approaches were formulated in a way to cast electric and magnetic fields,
as well as electron self-energy and vacuum polarization into a unified form
~\cite{Liu:14}. It should also be noted that the reduction of computational
cost in the X2C approaches relative to four-component NVPA occurs only at the SCF
and integral transformation steps; the correlation step has the same
computational scaling in two- and four-component methods~\cite{Saue:03}.
Since the correlation stage in the vast majority of practical applications is
the cost-determining step of the entire calculation, then using a
four-component formalism due to its formal and program simplicity often
becomes the rational method of choice.

A different approach to computational cost saving retains the four-component
framework and seeks savings by the reduction or elimination of intermediate
quantities (e.g., two-electron integrals) appearing in the calculations,
exploiting the atomic nature of the small component density~\cite{Saue:03}.
Another popular, efficient, and practical approach to a low number of components in wave function is the relativistic effective core potential (RECP) method~\cite{TitMos99,Sch03,DolCao12}. Recently, an updated version of the tiny-core two-component Generalized RECP (GRECP with a special non-local treatment of outer-core atomic region and inclusion of Breit and leading QED effects) has been developed and proved to be extremely precise and efficient for heavy and superheavy species~\cite{petrov2004accounting,mosyagin2020generalized,zaitsevskii2023generalized,oleynichenko2023libgrpp}.

The four-component approach is mandatory if one wishes to go beyond the NVPA
approximation and formulate a strictly covariant many-body theory. Such
formulations, which have a fundamental character and may substantially impact quantum
chemistry science, are not yet available. This is mainly due to the
impossibility of expressing covariant particle-particle interactions in a closed
analytical energy-independent form of the type used in building quantum
mechanical Hamiltonians for stationary atomic and molecular states. As stated
above, deriving a covariant relativistic many-body method requires switching
to the QED framework instead of the quantum mechanics basis. The QED
description includes four-component fermionic quantum fields and explicit
treatment of photons. The evident drawback is that the already high
computational scaling with system size of four-component methods becomes even
worse due to the photonic degrees of freedom. This presently makes many-body-QED
methods applicable to benchmark calculations of small systems only. However, the
rapidly increasing computational resources and algorithm improvements make it feasible to follow this rather uncompromising route.

\subsection{Electron correlation and modeling of the wave function}\label{sub:corr}

Here, we provide a brief overview of electron correlation methods broadly divided into two major groups that are regularly used in high-accuracy atomic applications -- CI-based and CC-based methods, with a strong focus on the latter. We do not review alternative correlation methods such as relativistic many-body perturbation theory (MBPT), random phase approximation (RPA), and density functional theory (DFT), which are reviewed elsewhere, e.g.,~\cite{Saue:03,Lindgren:11,Johnson1989,Dreizler2007,vanWllen2010,Chen2017}. 
\subsubsection{CI-based methods}\label{subsub:CI}

The configuration interaction approach is conceptually perhaps the simplest method of including electron correlation effects in an atomic or molecular calculation. Under this general umbrella term, we include all variational methods based on linear expansion of the wave function. Origins of the method reach as far back as the early 1930s~\cite{Slater1929,Condon1930,Bacher1933,Ufford1933}.
A comprehensive historical account of CI can be found in a review by Shavitt~\cite{Shavitt1998} 
and for a review of relativistic CI methods, see Ref.~\cite{Fischer2016}.

The linear expansion of the CI wavefunction in terms of Slater determinants $|\Phi_i\rangle$ can be expressed simply as
\begin{equation}\label{eq:CI1}
    |\Psi_{\text{CI}}\rangle = \sum_iC_i|\Phi_i\rangle,
\end{equation}
where the coefficients $\boldsymbol{C}$ are determined by a variational minimization of the energy expectation value
\begin{equation}\label{eq:CImini}
E_\text{CI}=\min_{\boldsymbol{C}} \frac{\langle\Psi(\boldsymbol{C})|\hat{H}|\Psi(\boldsymbol{C})\rangle}{\langle\Psi(\boldsymbol{C})|\Psi(\boldsymbol{C})\rangle}.
\end{equation}
Rather than individual Slater determinants, most implementations expand CI in terms of spin-symmetrized configuration state functions (CSFs)~\cite{Fischer2016}, hence the name of the method.
The conceptual simplicity and variational nature of CI are both attractive features leading to its widespread adoption and application in atomic electronic structure calculations.

Recasting equation~\eqref{eq:CI1} into the language of second quantization conveniently reveals an underlying structure grouping different determinants in terms of excitation rank
\begin{equation}\label{eq:CI2}
|\Psi_{\text{CI}}\rangle = C_0 |\Phi_0\rangle
    +\sum_{i,a} C_i^a \hat{a}_a^\dagger \hat{a}_i |\Phi_0\rangle
    +\sum_{ijab} C_{ij}^{ab} \hat{a}_a^\dagger \hat{a}_b^\dagger \hat{a}_j \hat{a}_i |\Phi_0\rangle
    +\sum_{ijkabc} C_{ijk}^{abc} \hat{a}_a^\dagger \hat{a}_b^\dagger \hat{a}_c^\dagger \hat{a}_k \hat{a}_j \hat{a}_i |\Phi_0\rangle
    +\dots
\end{equation}
Here, $|\Phi_0\rangle$ is the reference (typically Hartree--Fock) determinant, indices $i,j$ refer to occupied orbitals, $a,b$ to virtual orbitals, and $\hat{a}_a^\dagger$/$\hat{a}_i$ are the creation/annihilation operators, respectively. Individual terms in the expansion following the reference correspond then to single (S), double (D), triple (T) etc. excitations.

Full CI (FCI) expansion \eqref{eq:CI2}, where all possible configurations are included in the calculation, is formally exact (within the finite basis of the system); however, due to its quasi-exponential scaling~\cite{Helgaker2000}, it is completely impractical for any but the smallest of systems. Generally, the expansion \eqref{eq:CI2} is truncated to a chosen level of excitation from the ground state configuration, most commonly singles and doubles (CISD).
Truncated CI schemes are, however, not size-extensive (energy-separable), leading to errors in energies when comparing systems that differ in the number of particles.
This problem is partially alleviated by introducing one of the popular non-exact \textit{a posteriori} size-extensitivity corrections due to Davidson and others correcting for the dominant missing quadruple excitations~\cite{Davidson1974,Duch1994,Szalay2011} or alternatively using \textit{a priori} corrections to the CI scheme based on the multireference coupled electron-pair approximation reviewed in Ref.~\cite{Szalay2011}.

We also include the multiconfigurational (Dirac)--Hartree--Fock (MC(D)HF) method in this subsection. However, in a strict sense, it is separate from CI, with the distinction being the variational optimization of the expansion coefficients and the individual orbital wave functions in the former case and the expansion coefficients alone in the latter, i.e., compare \eqref{eq:CImini} to
\begin{equation}\label{eq:MCSCF}
E_\text{MCHF}=\min_{\boldsymbol{c},\boldsymbol{C}} \frac{\langle\Psi(\boldsymbol{c},\boldsymbol{C})|\hat{H}|\Psi(\boldsymbol{c},\boldsymbol{C})\rangle}{\langle\Psi(\boldsymbol{c},\boldsymbol{C})|\Psi(\boldsymbol{c},\boldsymbol{C})\rangle},
\end{equation}
with the orbital mixing coefficients $\boldsymbol{c}$.
In practical terms, the minimization with respect to orbital coefficients $\boldsymbol{c}$ and CI expansion coefficients $\boldsymbol{C}$ alternate in the so-called micro- and macro-iterations until self-consistency is achieved.
The selection of individual configurations (determinants) included in the MCHF expansion differs significantly from CI. Rather than truncation based on excitation rank, the configurations are selected based on their ``chemical'' relevance. Two main strategies are in popular use, either a) hand-picked configurations -- this is recommended only in case of a deeper understanding of the intricacies of the modeled system; or b) partitioning the orbital space into active and inactive part and performing full or truncated CI within these spaces, corresponding to complete active space (CAS) and restricted active space (RAS) approaches, respectively. If multiple layers of orbital space partitioning are performed, this leads to the generalized active space (GAS) framework.
The convergence and stability of MCDHF as well as the resulting energy and wavefunction are rather sensitive to the choice of included configurations. Furthermore, the multiparameter minimization \eqref{eq:CImini} is often riddled with numerous local minima. Consequently, MCDHF should not be treated a a black-box method and requires some level of expertise from the user. 
Some of the recent implementations of relativistic MCDHF can be found in popular atomic~\cite{GRASP2018,MCDFGME,JAC} and molecular program packages~\cite{DIRAC23,BAGEL}. This method has been successfully applied to calculations of spectra and properties of various heavy and superheavy elements \cite{Fri02,SewBacDre03, EliFriKal15}. For a technical treatise of the relativistic MCDHF method and its practical algorithmic implementation, we refer the reader to the Ref.~\cite{Jensen:96}

The CI and MCDHF methods are closely tied together, and MCDHF typically constitutes the first step forming a reference for the subsequent relativistic multireference (MR) CI calculations, i.e., $|\Phi_0\rangle$ in \eqref{eq:CI2} is a linear combination of determinants/CSFs. The set of determinants forming the reference in MR-CI constitutes the model space.
In this context, the correlation problem is often divided into so-called static and dynamic parts. The static correlation part covers a limited number of dominant configurations necessary for the qualitatively correct description of the electronic state (CSF). These are typically treated using MCDHF and form the (multi)-reference basis for the subsequent CI calculation. The dynamic correlation part modeled by CI covers the remaining instantaneous interactions resulting from the small but numerous contributions of all remaining configurations.

The truncated CI converges rather slowly with the number of configurations and excitations towards the FCI limit, somewhat limiting its applicability for highly accurate description of electronic structures. One approach to circumvent this problem is partitioning the occupied orbital space into non-overlapping core and valence parts. The smaller valence space can then be treated with the MCDHF and CI methods, allowing for bigger configuration sets, and the missing core-valence correlation contribution is added through MBPT correction based on the Dirac--Hartree--Fock description of the closed-shell core. The resulting CI+MBPT method was originally introduced in~\cite{Dzuba1996}, subsequently developed and implemented by others~\cite{Savukov2002,Dzuba2007,CI+MBPT,AMBiT} and was successfully applied in numerous high-accuracy spectroscopy studies, e.g.~\cite{Porsev1999,Berengut2006,Berengut2010}.

\subsubsection{CC-based methods}\label{subsub:CC}
A general coupled cluster approach is a size-extensive computational scheme involving infinite-order summation of definite classes of 
perturbation terms, accomplished by a specific form of the exponential parameterization of the wave operator
~\cite{Coester:58,Coester:60,Cizek:69,Bartlett:91}. 
The CC approach provides a systematic way to improve upon the
Hartree--Fock method by including higher-order electron correlation
effects through efficient self-consistent iterative solutions of the appropriate coupled cluster equations.

In the simplest single-reference formulation, the coupled cluster wave function is expressed as
\begin{equation}
    |\Psi_{\text{CC}}\rangle = e^{\hat{S}}|\Phi_0\rangle,
\end{equation}
where $|\Phi_0\rangle$ is the reference (usually Hartree--Fock)
Slater determinant, defining the Fermi vacuum of the system, and $\hat{S}$ is the cluster operator
\begin{equation}
    \hat{S} = \hat{S}_1 + \hat{S}_2 + \hat{S}_3 + \cdots + \hat{S}_N
\end{equation}
with $N$ being the number of electrons in the system. The individual cluster operators are defined as
\begin{equation}
    \hat{S}_1 = \sum_{i,a} s_i^a \hat{a}_a^\dagger \hat{a}_i,
\end{equation}
\begin{equation}
    \hat{S}_2 = \frac{1}{4} \sum_{ijab} s_{ij}^{ab} \hat{a}_a^\dagger \hat{a}_b^\dagger \hat{a}_j \hat{a}_i,
\end{equation}
\begin{equation}
    \hat{S}_3 = \frac{1}{36} \sum_{ijkabc} s_{ijk}^{abc} \hat{a}_a^\dagger \hat{a}_b^\dagger \hat{a}_c^\dagger \hat{a}_k \hat{a}_j \hat{a}_i, \dots
\end{equation}
for singles, doubles, triples etc., respectively, where $i,j$ refer to occupied orbitals, $a,b$ to virtual orbitals, $\hat{a}_a^\dagger$/$\hat{a}_i$ correspond to the creation/annihilation operator of an electron on appropriate orbital
and $s_i^a$, $s_{ij}^{ab}$, $s_{ijk}^{abc}$ are the cluster amplitudes.

The correlation energy is determined by projecting the Schrödinger equation onto the reference state
\begin{equation}
    E_{\text{CC}} = \langle\Phi_0|(H e^{\hat{S}})_c|\Phi_0\rangle.
\end{equation}
The subscript $c$ designates the connected part of the corresponding operator expression. For CCSD (coupled cluster with singles and double excitations), this reduces to
\begin{equation}
    E_{\text{CCSD}} = E_{\text{HF}} + \sum_{ia} f_i^a s_i^a + \frac{1}{4}\sum_{ijab} \langle ij||ab\rangle s_{ij}^{ab},
\end{equation}
where $f_i^a$ are Fock matrix elements and $\langle ij||ab\rangle$
are antisymmetrized two-electron integrals.

The equations for the $s_i^a$ amplitudes are obtained by projection onto singly excited determinants
\begin{equation}
    \langle\Phi_i^a|(\hat{H}e^{\hat{S}})_c|\Phi_0\rangle = 0.
\end{equation}
Similarly, for doubles
\begin{equation}
    \langle\Phi_{ij}^{ab}|(\hat{H}e^{\hat{S}})_c|\Phi_0\rangle = 0.
\end{equation}

The computational cost of CC methods increases rapidly with the included
level of excitation
\begin{itemize}
    \item CCSD: $\mathcal{O}(N^6)$
    \item CCSD(T): $\mathcal{O}(N^7)$
    \item CCSDT: $\mathcal{O}(N^8)$
    \item CCSDTQ: $\mathcal{O}(N^{10})$
\end{itemize}
where $N$ is the number of active (correlated) orbitals/spinors. 

The CCSD(T) approximation listed above, which includes perturbative triple excitations and scales as $\mathcal{O}(N^7)$ is often
called the ``gold standard'' of quantum chemistry due to its
excellent balance between accuracy and computational cost. Within this approximation, the energy is given as
\begin{equation}
    E_{\text{CCSD(T)}} = E_{\text{CCSD}} + E_{(T)},
\end{equation}
where $E_{(T)}$ is the perturbative triples correction, introduced in 1989 by Raghavachari et. al. \cite{raghavachari1989fifth}. This energy correction includes all contributions correct to the third-order perturbation theory derived from the approximate triple excitation amplitudes $S_3$ calculated by the contraction $H$ and the $S_2$ converged on the CCSD level of theory.
This approximation provides chemical accuracy ($\sim$1 kcal/mol) for many molecular properties when used with an appropriate basis set (see review \cite{bartlett2007coupled}).

Historically, the first four-component coupled cluster calculations of atomic systems appeared in the early 1980s, followed by molecular applications from the mid-1990s. 
A numerical procedure for solving the
relativistic many-body Dirac--Coulomb equation, based on the pair approximation
of the CC approach has been developed by Lindgren and coworkers
~\cite{Lindgren:80}. A different approach employs discrete basis sets of local
or global functions. Summation over an infinite set of bound states and
integration over the positive energy continuum is replaced by finite
summation over the pseudospectrum. The implementation of the projection
operators is made easy by the clean separation of positive and negative energy
states; it amounts to limiting summations to the positive energy branch of the
one-electron pseudospectrum. A relativistic CC technique based on local
splines (piecewise polynomial fitting) has been developed and implemented to a
number of atoms~\cite{Blundell:89,Blundell:89a,Blundell:90,Blundell:91,Liu:91}. Another kind of local basis set has been
introduced to relativistic CC by Salomonson and \"{O}ster~\cite{Salomonson:89},
who discretized the radial space. This technique is similar in spirit to the
spline method and may be regarded as its limiting case (single-point
representation rather than polynomial fitting). Single reference relativistic
four-component coupled cluster methods for molecules were introduced in the
1990s~\cite{Eliav:SnH4:96,VisLeeDya96}, followed by remarkable methodological
evolution up to the recently introduced coupled cluster approaches of general
excitation rank, namely the generalized active space (GAS)-CC implementation
by S\o rensen et al.~\cite{Sorensen:11,Fleig:07} and the
state-specific CC implementation by Nataraj et al.~\cite{Nataraj:10}.

When the reference determinant is no longer dominant and substantial, \emph{static} (nondynamical) correlation becomes important. This effect is often present for bond dissociation, open‑shell excitations, and in heavy atoms with near degeneracies. In such cases, the SR assumption fails and multireference extensions of teh coupled cluster approach are required.  Two complementary theoretical strategies have been followed since the early 1980s (see review \cite{Mukherjee:89}):

1. \textbf{Hilbert‑space multireference CC (HS–MRCC)} – also called
\emph{state‑universal} multireference CC (SU–MRCC) because it is based on multifunctional reference states spanned by a chosen \emph{model space} $\mathcal{P}$ in which each zero-order configuration function (Slater determinant) serves as a Fermi vacuum for the particular state-specific wave operator's compound included in the state-universal  Jeziorski--Monkhorst (JM) ansatz \cite{Jeziorski:81}. It is particularly suitable for computing global molecular PESs where multiple states of the same symmetry and electron number are quasi-degenerate.

2. \textbf{Fock‑space multireference CC (FS–MRCC)} – sometimes called
\emph{valence‑universal} because a single (usually closed-shell) Slater determinant serves as a Fermi vacuum state to generate all multireference states in different valence electronic Fock space sectors. The separate Fock sectors ($N\pm k$ electrons, different spins) are treated by the same valence-universal wave operator (usually taken in the normal ordered form, as proposed by Lindgren  \cite{Lindgren:86}). This approach is practical for calculating energy differences such as ionization potentials and excitation energies, where multiple states related to changes in the number of electrons or excitation level are targeted simultaneously \cite{Mukherjee:89,Kaldor:91}.  Because only one closed‑shell determinant enters the common reference state (Fermi vacuum), the numerical stability of FS–MRCC is excellent, and its transparent perturbative structure, similar to that of the single-reference CC method, allows uncertainty quantification and straightforward relativistic extensions; these features make it ideally suited for precision spectroscopy. Currently, relativistic CC applications for multi-reference problems are heavily dominated by FS-MRCC. The theoretical background of this approach and its numerous benchmark applications are the primary focus of this review and are presented in detail in the following sections. 

Both strategies originate from the generalized many‑body Bloch equation and share the exponential parametrization while differing in partitioning and projection. We summarize below the essential features of the above approaches and outline the key relativistic developments.

While many methodological efforts have addressed the electronic structure of heavy-element systems, a general four-component (4c) spinor-based, accurate MRCC method of state-universal type that treats all reference functions equally is highly desirable. However, because of many technical difficulties, attempts to develop such an approach are scarce.
Following Jeziorski and Monkhorst~\cite{Jeziorski:81}, the correlated \emph{multi‑root and multireference} wave functions are expressed using the following state-universal ansatz
\begin{equation}
|\Psi_k\rangle = \sum_{\mu \in \mathcal{P}} C_k^\mu e^{S_\mu}|\Phi_\mu\rangle,
\label{eq-jm}
\end{equation}
where $\mathcal{P}$ is the model space, $|\Phi_\mu\rangle\in \mathcal{P}$ are reference determinants, $S_\mu$ are cluster operators specific to each reference, $C_k^\mu$ are CI expansion coefficients.

Inserting Eq.~\eqref{eq-jm} into the Schrödinger equation as a generalized Bloch equation yields the SU–MRCC working equations \cite{Mukherjee:89,li1997reduced}.  The JM exponential parametrization ensures size‑extensivity but introduces highly non‑linear, multi‑root coupled equations that are very complicated and often suffer from \emph{intruder state} divergences, symmetry breaking, and spin contamination problems. Although Hoffmann and Khait \cite{hoffmann1999self} have suggested an intruder-free HS-MRCC method using a unitary MRCC ansatz and a Hermitian effective Hamiltonian scheme, their approach, remaining in the nonrelativistic framework, did not transform the multi-root Hilbert space CC to a popular practical method due to the many remaining technical difficulties.

An effective way to circumvent convergence problems and inaccuracies of the multi-root JM approach are offered by state-specific (SS) or single-root (sr) MRCC theories. These methods target one specific state at a time and are briefly enumerated below.

1. Mukherjee's SSMRCC (MkCC) \cite{mahapatra1998state}: This well-studied HS-MRCC method, developed by Mukherjee and co-workers, is rigorously size-extensive and size-consistent (with localized active orbitals due to its non-invariance). It has shown potential for a variety of problems.

2.	Brillouin-Wigner MRCC (BWMRCC) (see review \cite{hubavc2010brillouin} and reference therein): While BWMRCC methods can be very accurate, they are not fully size-extensive, lacking a crucial trait of standard CC methods. A posteriori size-extensivity corrections have been suggested, but these can reintroduce intruder state problems.

3.	MRexpT Method \cite{hanrath2005exponential}: This is another SSMRCC approach. Like BWMRCC, it can lack rigorous size-extensivity.

4.	Internally Contracted MRCC (icMRCC): These methods, such as those developed by Banerjee and Simons \cite{BanSim81}, and later extended within CC theory (e.g., by Evangelista and Gauss \cite{evangelista2011orbital}, Hanauer and Köhn \cite{hanauer2011pilot} ), are very effective for quasi-degenerate situations. They use a fixed set of contraction coefficients for the reference configurations, simplifying the wave function ansatz and reducing computational cost.

5.	Block Correlated CC (BCCC): Developed by Li \cite{li2004block}, BCCC is another functional SS formalism extensively explored for MR systems. However, BCCC is not size-extensive in terms of scaling the number of active electrons.

Until now, only the first of the enumerated SS-MRCC approaches, namely MkCC, has been adopted to the fully relativistic four-component regime by Ghosh at. al. \cite{ghosh2016relativistic}. The relativistic MkCC method does not require a dominant configuration in the model space function, making it more general. Therefore, it differs from the relativistic state-selective methods of Sørensen--Fleig--Olsen \cite{Sorensen:11,Fleig:07} and Nataraj--Kállay--Visscher \cite{Nataraj:10} mentioned earlier, because the 4c-MkCC allows treatment of all reference functions on an equal footing via a state-specific parameterization of the JM ansatz for a state-universal wave operator. In contrast, the latter methods formally use an SR formalism where the cluster expansion is with respect to one ``formal'' determinant, leading to a lack of invariance concerning its choice and unequal footing for reference functions. As far as we know, the work of Ghosh \textit{et al.} \cite{ghosh2016relativistic} remains the only published application of the relativistic MkCC.

In this review, we highlight the \textbf{relativistic multi-root multireference FSCC method} as the workhorse of modern high-precision heavy element spectroscopy while also recognising the complementary potential role of 4c SSMRCC and related HS-based formalisms in strongly multiconfigurational regimes.
Because FSCC starts from a single closed‑shell vacuum and accesses excited, electron-attached, and electron-detached states through well-defined sectors, it marries naturally with four‑component relativistic Hamiltonians and allows a transparent inclusion of Breit and leading QED corrections (vacuum polarisation, self‑energy) \cite{Eliav:10,Eliav:10a}.
This method yields outstanding agreement with the experiment for many transition energies and other properties of
heavy atoms (see reviews~\cite{Kaldor:98,Eliav:10,Eliav:10a,EliFriKal15,EliKalBor18,EliBorKal19,ELIAV202479}) and recently also of their molecular compounds
~\cite{Eliav:CdH:98,Visscher:01,Infante:06,Real:09,Gomes:10,Tecmer:11,Zaitsevskii:17,tecmer2018electron,HaoPasVis19,Zaitsevskii:TDM:20,Oleynichenko:CCSDT:20,Isaev:21,HaaDoeBoe21,Osika:22,OleSkrZai22,DenHaaMoo22,lu2022intermediate,skrzynski2023benchmark,SkrOleZai23,zaitsevskii2023theoretical}. This success makes the scheme a useful tool for reliable predictions of the
structure and spectrum of heavy and superheavy elements, which are difficult to access
experimentally. The Fock space methodology is the only quantum chemical approach suitable
for treating systems with a variable number of particles. This and other
methodological benefits of the FSCC approach make it an ideal candidate for
merging with QED theory to create an infinite order size-extensive covariant
many-body method.

Similar (but usually more approximate) approaches have been developed and
implemented by different groups in a fully relativistic regime to use
for heavy and superheavy elements. Among these methods are equation-of-motion
coupled cluster (EOM-CC)~\cite{Nandy:14,SheSauVis18, LiuChe21,RanDut21}, ``all-order-CI'' method~\cite{SafKozJoh09}, and linear response CCSD (LR-CCSD)~\cite{Chaudhuri:13}. The numerical instability caused by intruder states in non-zero valence sectors, which poses a significant challenge in the conventional Fock-space multi-reference coupled cluster (FSMRCC) method, is effectively avoided in equation-of-motion CC, all-order CI, and coupled cluster linear response theory approaches. This improved stability arises from two key structural features of these alternative methods: first, their use of linear excitation operators rather than the exponential parameterization employed in FSMRCC, and second, their implementation of a CI-like eigenvalue framework for treating electronic states in higher valence sectors, which provides a more robust mathematical foundation for handling multi-reference character. All these methods have
the same two-step structure. The first step applies the coupled cluster
method to a reference closed shell configuration. The \textquotedblleft
dressed\textquotedblright\ effective Hamiltonian in the appropriate Fock-space
valence sector (e.g., subspace of the Hilbert space with the appropriate number of valence particles/holes) is then generated and diagonalized to yield the relevant
energies and wave functions.

A brief description of the relativistic NVPA\ FSCC method is given below. 
The novel double Fock-space coupled cluster method
~\cite{Eliav:10} based on the CEO approach of Lindgren
~\cite{Lindgren:04} is a further step on the way to a covariant many-body
technique, suitable for benchmark electronic structure calculations.

\subsubsection{Relativistic FSCC method}

\label{sec:fscc}

The NVPA\ Dirac--Coulomb--Breit Hamiltonian $H_{\mathrm{DCB}}^{+}$~\eqref{eq:PDCB} can be
rewritten in second-quantized form~\cite{Ishikawa:89,Ishikawa:90,Ishikawa:90a,Ishikawa:93,Ishikawa:94,Sucher:87} in terms of
normal-ordered products of spinor creation and annihilation operators
$\{a_{r}^{+}a_{s}\}$ and $\{a_{r}^{+}a_{s}^{+}a_{u}a_{t}\}$, corresponding to
the ``fuzzy'' picture,
\begin{equation}
H=H_{\mathrm{DCB}}^{+}-\langle0|H_{\mathrm{DCB}}^{+}|0\rangle=\sum_{rs}%
f_{rs}\{a_{r}^{+}a_{s}\}+\frac{1}{4}\sum_{rstu}\langle rs||tu\rangle
\{a_{r}^{+}a_{s}^{+}a_{u}a_{t}\}. \label{eq:h}%
\end{equation}
Here $f_{rs}$ and $\langle rs||tu\rangle$ are, respectively, the elements of the
one-electron Dirac--Fock--Breit and the antisymmetrized two-electron Coulomb--Breit
interaction matrices over Dirac four-component spinors. The effect of the
projection operators $\Lambda^{+}$ is now taken over by normal ordering,
denoted by the curly brackets in (\ref{eq:h}), which requires annihilation
operators to be moved to the right of creation operators as all
anticommutation relations vanish. The Fermi level is set at the top of the
highest occupied positive energy state, and the negative energy states are ignored.

The development of a general multi-root multireference scheme for treating
electron correlation effects usually starts from consideration of the
Schr\"{o}dinger equation for a number ($d$) of target states,
\begin{equation}
H\Psi^{\alpha}=E^{\alpha}\Psi^{\alpha}\text{ \ , }\alpha=1,...,d\text{} .
\end{equation}
The physical Hamiltonian is divided into two parts,
\begin{equation}
H=H_{0}+V ,
\end{equation}
so that $V$ is a small perturbation to the zero-order Hamiltonian $H_{0}$,
which has known eigenvalues and eigenvectors,
\begin{equation}
H_{0}|\mu\rangle=E_{0}^{\mu}|\mu\rangle.
\end{equation}

In many open-shell heavy compound systems, exact or near-degeneracy occurs when certain energy levels $E_{0}^{\alpha}$ are equal or nearly equal. 
By adopting the NVPA approximation, a natural and
straightforward extension of the nonrelativistic open-shell CC theory emerges.
The multireference valence-universal Fock-space coupled cluster approach is
presented here briefly; a fuller description may be found in
~\cite{Mukherjee:89,Kaldor:91}. FSCC defines and calculates an effective Hamiltonian in
a $d$-dimensional model space $P=\sum|\mu\rangle\left\langle \mu\right\vert ,$
$\mu=1,..,d$, comprising the most strongly interacting zero-order many-electron 
wave functions. All other functions are in the complementary
$Q$-space, so that $P+Q=1$. All $d$ eigenvalues of $H_{\mathrm{eff}}$ coincide
with the relevant eigenvalues of the physical Hamiltonian,
\begin{equation}
H_{\mathrm{eff}}\Psi_{0}^{\alpha}=E^{\alpha}\Psi_{0}^{\alpha}\text{ \ ,
}\alpha=1,...,d\text{} .
\end{equation}
There is no summation over the index $\alpha$, and
\begin{equation}
\Psi_{0}^{\alpha}=C_{\mu}^{a}|\mu\rangle,\text{ \ \ }\alpha=1,...,d \text{.
\ }%
\end{equation}
$\Psi_{0}^{\alpha}$ describes the major part of $\Psi^{\alpha}$ for all
$\alpha=1,...,d$,
\begin{equation}
P\Psi_{{}}^{\alpha}=\Psi_{0}^{\alpha}\ ,\text{ \ \ }\alpha=1,...,d .
\end{equation}
The effective Hamiltonian has the form~\cite{Lindgren:80,Lindgren:86}
\begin{equation}
H_{\mathrm{eff}}=PH\Omega P\;, \ H_{\mathrm{eff}}=H_{0}+V_{\mathrm{eff}}.
\label{eq:heff}%
\end{equation}
$\Omega$ is the normal-ordered wave operator, mapping the eigenfunctions of
the effective Hamiltonian onto the exact ones, $\Omega\Psi_{0}^{\alpha}%
=\Psi^{\alpha}$, $\alpha=1,...,d$. It satisfies intermediate normalization,
\begin{equation}
P\Omega P=P .
\end{equation}
The effective Hamiltonian and wave operators are connected by the generalized
Bloch equation, which for a complete model space $P$ may be written in the
compact linked form~\cite{Lindgren:86}
\begin{equation}
Q[\Omega,H_{0}]P=Q(V\Omega-\Omega H_{_{\mathrm{eff}}})_{\mathrm{linked}}P.
\label{eq:Bloch}%
\end{equation}
$\Omega$ is parameterized exponentially in the coupled cluster method. A
particularly compact form is obtained by using the normal ordered form,
\begin{equation}
\Omega=\{\exp(S)\}. \label{eq:exps}%
\end{equation}

The Fock-space approach starts from a reference state (usually closed shell, but other single-determinant functions may also be used \cite{stanton1992fock, MajChaSud24}),
correlates it, then adds and/or removes electrons one at a time, recorrelating
the whole system at each stage. The sector $(m,n)$ of the Fock space includes
all states obtained from the reference determinant by removing $m$ electrons
from designated occupied orbitals, called valence holes, and adding $n$
electrons in designated virtual orbitals, called valence particles. Till recently, the
practical limit was $m+n\leq2$, although higher sectors have also been tried earlier in the nonrelativistic framework
~\cite{Hughes:92,Hughes:93,Hughes:93a,Hughes:93b,Hughes:95}. In 2020, the above limit has been raised to $m+n\leq3$ within a fully relativistic framework and implemented in the highly efficient open-source code EXP-T~\cite{Oleynichenko:EXPT:20}. Since then, this method has been successfully applied to various heavy systems~\cite{Skripnikov:21,ELIAV202479}. A similar theoretical development is also happening in parallel in the nonrelativistic community~\cite{Musial2024}.

The excitation operator $S$, serving as the variable of
exponential parameterization of the wave operator $\Omega,$ is partitioned
into sector operators
\begin{equation}
S=\sum_{m\geq0}\sum_{n\geq0}S^{(m,n)}. \label{eq:sect}%
\end{equation}
This partitioning allows for partial decoupling of the open-shell CC equations
according to the so-called subsystem embedding condition~\cite{Mukherjee:89}. The
equations for the ($m,n$) sector involve only $S$ elements from sectors
$(k,l)$ with $k\leq m$ and $l\leq n$, so that the very large system of coupled
nonlinear equations is separated into smaller subsystems, which are solved
consecutively: first, the equations for $S^{(0,0)}$ are iterated to
convergence; the $S^{(1,0)}$ (or $S^{(0,1)}$) equations are then solved using
the known $S^{(0,0)}$, and so on. This separation, which does not involve any
approximation, reduces the computational effort significantly. The effective
Hamiltonian (\ref{eq:heff}) is also partitioned by sectors. An important
advantage of the method is the simultaneous calculation of many states.

In the usual way, each sector excitation operator is a sum of virtual
excitations of $l$ electrons,
\begin{equation}
S^{(m,n)}=\sum_{l}S_{l}^{(m,n)}\,, \label{eq:ccsd}%
\end{equation}
with $l$ going, in principle, to the total number of electrons. In practice,
$l$ has to be truncated. The level of truncation reflects the quality of the
approximation, i.e., the extent to which the complementary $Q$ space is taken
into account in the evaluation of the effective Hamiltonian. The series
(\ref{eq:ccsd}) is truncated either at $l=2$ (CCSD) or at $l=2$ (CCSDT) in practical applications. The current highest approximation within FSCC is the CCSDT (coupled cluster with single, double, and triple excitations) scheme (see the implementation details in the EXP-T  program description~\cite{Oleynichenko:EXPT:20,Oleynichenko:CCSDT:20} and the recent review of its applications~\cite{ELIAV202479}). The implementation
involves the fully self-consistent, iterative calculation of all one-,
two-, and three-body virtual excitation amplitudes and sums all diagrams with these
excitations to infinite order. 

The FSCC equations for a particular $(m,n)$
sector of the Fock space are derived by inserting the normal-ordered wave
operator (\ref{eq:exps}) with Fock-space exponential parameterization of the
excitation operator (\ref{eq:ccsd}) into the Bloch equation (\ref{eq:Bloch}).
The final form of the FSCC equation for a complete model space includes only
\emph{connected} terms~\cite{Lindgren:80,Lindgren:86},
\begin{align}
Q[S_{l}^{(m,n)},H_{0}]P  &  =Q\{(V\Omega-\Omega H_{_{\mathrm{eff}}}%
)_{l}^{(m,n)}\}_{\mathrm{conn}}P,\label{eq:FSCC}\\
H_{\mathrm{eff}}^{(m,n)}  &  =P(H\Omega)_{\mathrm{conn}}^{(m,n)}P\; .
\label{eq:heff_fs}%
\end{align}

As negative energy states are excluded from the $Q$ space, the diagrammatic
summations in the CC equations are carried out only within the subspace of the
positive energy branch of the HF spectrum. After converging the FSCC equation
(\ref{eq:FSCC}), the effective Hamiltonian (\ref{eq:heff_fs}) is diagonalized,
yielding directly transition energies. The effective Hamiltonian in the FSCC
approach has a block-diagonal structure with respect to the different Fock-space
sectors. From (\ref{eq:heff_fs}), it follows that two different Fock-space sectors
belonging to a common Hilbert space (with the same number of particles) do not
mix even if they have strongly interacting states. This means that important
nondynamic correlation effects are approximated. The mixed-sector CC presented
briefly below avoids this problem.

The FSCC equation (\ref{eq:FSCC})\ is solved iteratively, usually by the
Jacobi algorithm. As in other CC approaches, denominators of the form
$(E_{0}^{P}-E_{0}^{Q})$ appear, originating in the left-hand side of the
equation. 
The well-known intruder state problem \cite{lowdin1951note} appears when some $Q$
states are close to and strongly interacting with $P$ states, which may lead to
divergence of the CC iterations. 
A generalization of the effective
Hamiltonian, the intermediate Hamiltonian (IH) approach,
has been developed in recent years in many variants for use within the FSCC approach~\cite{Landau:IH:99,Landau:04,Eliav:XIH:05,Zaitsevskii:18a,zaitsevskii2023generalized}. It eliminates intruder state problems caused by energy overlap of $P$ and $Q$ spaces, which spoils the convergence of the CC iterations. Thereby, this development allows the use of much larger model spaces, leading to enhanced applicability and accuracy. An additional advantage of using extended model spaces is that it reduces the need to include high excitations in the formalism. The need for high excitations (triples and
higher) is usually limited to a small group of virtual orbitals. If such orbitals are brought into $P$, all excitations involving them are included in infinite order by diagonalizing the effective Hamiltonian, avoiding the need for the (usually expensive) treatment of their contribution to dynamical correlation. The Intermediate Hamiltonian approach is described in some detail below.

\subsubsection{The intermediate Hamiltonian FSCC method}

\label{sec:ih}

The intermediate Hamiltonian (IH) method has been proposed by Malrieu
~\cite{Malrieu:85} in the framework of degenerate perturbation theory. The $P$
space is partitioned into the main $P_{m}$ subspace, which includes all the
states of interest, and the intermediate $P_{i}$ subspace, serving as a buffer
between $P_{m}$ and the rest of the functional space $Q$. The corresponding
operators satisfy the equations
\begin{equation}
P_{m}+P_{i}=P\ ,\ \ \ \ \ \ \ P+Q=1\ . \label{eq:pq}%
\end{equation}

The rationale for this partitioning is the following: the higher states in $P$ contribute significantly to the states of interest, which evolve from the lower $P$ states but couple strongly with intruders from $Q$ and
spoil the convergence of the iterations; they should, therefore, be treated
differently from the lower states. This goal is achieved by partitioning $P$
and allowing a more approximate treatment of $P_{i}$ states. The intermediate
Hamiltonian $H_\text{I}$ is constructed in $P$ according to the same rules as the
effective Hamiltonian,
\begin{equation}
H_\text{I}=PH\Omega P\;, \label{eq:phrp}%
\end{equation}
but only $|\Psi_{m}\rangle$ states with their largest part in $P_{m}$ are
required to have energies $E_{m}$ closely approximating those of the physical
Hamiltonian,
\begin{equation}
H_\text{I}P|\Psi_{m}\rangle=E_{m}P|\Psi_{m}\rangle\;. \label{eq:ih}%
\end{equation}
The other eigenvalues, which correspond to states $|\Psi_{i}\rangle$ with the
largest components in $P_{i}$, may be more or less accurate. This leads to
some freedom in defining the relevant eigenfunctions and eigenvalues and,
therefore, in evaluating problematic $QSP_{i}$ matrix elements. To restrict this freedom while enhancing the generality and flexibility of the approach compared to the standard effective Hamiltonian method, we also incorporate the following partitioning.
\begin{equation}
Q=Q_{i}+Q_{m} . \label{eq:qq}%
\end{equation}
This additional partitioning narrows the overlap of the $P$ and $Q$ energies,
and only $P_{i}$ and $Q_{i}$ subspaces can now overlap.
The number of problematic amplitudes, now $Q_{i}SP_{i}$,
is thus reduced.

Partitioning the $P$ and $Q$ projectors of the FSCC equation (\ref{eq:FSCC})
into the main and intermediate parts by formulas (\ref{eq:pq},\ref{eq:qq})
yields four coupled CC equations,
\begin{align}
Q_{m}[S,H_{0}]P_{m}  &  =Q_{m}\{V\Omega-\Omega H_{_{\mathrm{eff}}%
}\}_{\mathrm{conn}}P_{m}\label{eq:ihcc}\\
Q_{i}[S,H_{0}]P_{m}  &  =Q_{i}\{V\Omega-\Omega H_{_{\mathrm{eff}}%
}\}_{\mathrm{conn}}P_{m}\label{eq:ihcc2}\\
Q_{m}[S,H_{0}]P_{i}  &  =Q_{m}\{V\Omega-\Omega H_{_{\mathrm{eff}}%
}\}_{\mathrm{conn}}P_{i}\label{eq:ihcc3}\\
Q_{i}[S,H_{0}]P_{i}  &  =Q_{i}\{V\Omega-\Omega H_{_{\mathrm{eff}}%
}\}_{\mathrm{conn}}P_{i}. \label{eq:ihcc4}%
\end{align}
Only the last of these can cause convergence problems. Successful replacement
of this equation by another, based on physical considerations, is the central
point of the IH method. The new equation to be used instead of (\ref{eq:ihcc4}%
) will be called the \emph{IH condition} (IHC). Ideally, it should satisfy the
following demands:

\begin{itemize}
\item  be free of convergence problems;
\item  have minimal impact on the other coupled equations (\ref{eq:ihcc}%
--\ref{eq:ihcc3}).
\end{itemize}
Subject to these demands, we would like the IHC to be as close to
(\ref{eq:ihcc4}) as possible.

Several IH-FSCC methods have been developed and applied recently based on
different IH conditions. The first such approach~\cite{Landau:IH:99}, denoted IH1,
uses the condition
\begin{equation}
Q_{i}\Omega P_{m}H\Omega P_{i}=Q_{i}H\Omega P_{i}, \label{eq:ih1}%
\end{equation}
that is similar to the equation proposed by Malrieu and applied up to the third
order of degenerate perturbation theory~\cite{Malrieu:85}. While Malrieu's scheme
could not go beyond the third order because terms with small denominators appear; the later IH-FSCC variants are all-order and may be used in the framework of any multireference CC formulation \cite{Landau:IH:99}.

The second IH-FSCC scheme (IH2) \cite{Landau:04} is based on the perturbation theory expansion of
the problematic $Q_{i}SP_{i}$ amplitudes. In the lowest order, we take
\begin{equation}
Q_{i}SP_{i}=0 . \label{eq:ih2}%
\end{equation}
This type of IH condition has also been used for developing a new type of
hybrid multireference coupled cluster schemes, including the mixed sector CC
presented below.

Another IH condition leads to the most flexible and useful scheme, the
extrapolated IH (XIH)~\cite{Eliav:XIH:05,Eliav:Alk:05}, which can yield correct solutions both
for $P_{m}$ and $P_{i}$, thus, recovering the whole effective Hamiltonian
spectrum in the extended model space $P$.
The IH condition for the XIH approach has the form
\begin{equation}
Q_{i}[S,H_{0}+P_{i}\Delta P_{i}]P_{i}=Q_{i}\{\beta\Delta S+V\Omega-\Omega
H_{_{\mathrm{eff}}}\}_{\mathrm{conn}}P_{i}. \label{eq:xih}%
\end{equation}
$\Delta$ is an energy shift parameter, correcting small energy denominators
for the problematic intruder states. A compensation term with the
multiplicative parameter $\beta\leq1$ is added on the right-hand
side. For $\beta=1$, the $P_{i}\Delta P_{i}$ term on the left-hand side is
fully compensated, so that (\ref{eq:xih}) is equivalent to (\ref{eq:ihcc4}).
Proper choice of the two parameters makes it possible to reach convergence in
(\ref{eq:xih}) and thus in the non-problematic equations (\ref{eq:ihcc}%
--\ref{eq:ihcc3}). Several calculations with different values of the
parameters allow extrapolation of both $P_{m}$ and $P_{i}$ level energies to
the limit $\Delta\rightarrow0$ or $\beta\rightarrow1$. This extrapolation was
found to be robust, mostly linear for $P_{m}$ states and quadratic for
states in $P_{i}$ \cite{Eliav:XIH:05}. In the extrapolation limit, the IH method transforms into
the effective Hamiltonian approach. The XIH approach is asymptotically size
extensive and, in many cases, size consistent, even for incomplete $P_{m}$,
requiring only that the entire model space $P$ is complete (see the recent development and application of the Incomplete Model Space (IMS) XIH approach in~\cite{Eliav2024,zaitsevskii2023generalized}. A somewhat similar
IH FSCC scheme has been proposed by Mukhopadhyay et al.\ in 1992
~\cite{Mukhopadhyay:92}, but to the best of our knowledge, it has never been implemented. A very useful extension of XIH, which is based on the efficient Padé-extrapolation of specially constructed intermediate Hamiltonians rather than energies, has been proposed recently~\cite{Zaitsevskii:18a}.

The intermediate Hamiltonian approaches presented here may be applied within
any multi-root multireference infinite order method. For example, this method was implemented
within another all-order relativistic multi-root
multireference approach, the Hilbert space or state universal CC, which is the
main alternative to the Fock-space CC~\cite{Eliav:HSCC:09}. The HSCC is based on the
Jeziorsky--Monkhorst parameterization of the wave operator
~\cite{Jeziorski:81},
\begin{equation}
\Omega=\underset{\mu=1}{\overset{d}{\sum}}\Omega_{\mu}=\underset{\mu
=1}{\overset{d}{\sum}}\{\exp(S^{\mu})\}P^{\mu}\text{; \ }P^{\mu}=|\mu
\rangle\left\langle \mu\right\vert . \label{eq:hscc_JMparameter}%
\end{equation}
Here, every determinant $\mu$ belonging to the $P$ space serves as a reference
state (Fermi vacuum), and the excitation operators $S^{\mu}$ are vacuum
dependent. The nature of the determinants in the model space may be general;
the only requirement is that all determinants belong to the same Hilbert
space. The most useful scheme is probably the HSCC approach with a model space
built of general MCSCF solutions. This will make the HSCC method suitable for
global potential surface calculations. The XIH-HSCC equation in the case of
complete model space $P$ is
\begin{align}
\lbrack S^{\mu},H_{0}+P_{i}\Delta P_{i}]P^{\mu}  &  =\{\beta\Delta S^{\mu
}P_{i}P^{\mu}+V\{\exp(S^{\mu})\}P^{\mu}-\{\exp(S^{\nu})\}P^{\nu}%
H_{_{\mathrm{eff}}}P^{\mu}\}_{\mathrm{conn}}\nonumber\\
P^{\nu}H_{_{\mathrm{eff}}}P^{\mu}  &  =P^{\nu}(H\{\exp(S^{\mu}%
)\})_{\mathrm{conn}}P^{\mu} . \label{eq:heff_hscc}%
\end{align}

The HSCC effective Hamiltonian (\ref{eq:heff_hscc}), unlike the FSCC effective
Hamiltonian, has a non-diagonal structure, coupling different Fock-space sectors
belonging to the same Hilbert space. This leads to better treatment of
nondynamic (static) correlation. The \emph{mixed sector coupled cluster} (MSCC), which
may be regarded as a hybrid approach combining the advantages of FSCC and
HSCC, has been derived~\cite{Landau:04} within the IH2 scheme based on
IHC (\ref{eq:ih2}). The MSCC exponential parameterization of the wave operator
$\Omega$ and the working equation are formally similar to those of FSCC (see
(\ref{eq:exps})--(\ref{eq:FSCC})), but the subsystem embedding condition is
now relaxed, and several sectors of the Fock space belonging to the same Hilbert
space are mixed and diagonalized together. MSCC may thus yield the most balanced
inclusion of dynamic and nondynamic correlation effects. All the
multireference multiroot CC methods described above may be used for the
challenging task of benchmark calculations for heavy quasidegenerate systems
with more than two electrons/holes in the valence open shell. Implementation of the
relativistic XIH method to higher sectors of the Fock space (based on the nonrelativistic methodology with up to six valence electrons/holes developed in the 1990s~\cite{Hughes:95})
is in progress~\cite{Oleynichenko:EXPT:20,ELIAV202479}.  Another
challenging, long-term project is to apply the IH method within the double
FSCC, a covariant MRCC based on QED, presented in~\cite{Eliav:10a}. This
method can be applied to highly charged heavy ions, which exhibit large QED effects.

Summing up, we conclude that the IH method is an efficient and universal tool
applicable to all multi-root multireference methods. It avoids intruder
states while allowing the use of large, complete model
spaces, significantly improving the accuracy of the calculation.

\subsection{Basis sets}
\label{sec:basis}

The spherically symmetric central potential of the atomic nucleus allows for the exact separability of the angular and the radial parts of the atomic electronic structure. While the angular part can be handled analytically in terms of spherical harmonics, the radial part requires numerically solving the Dirac equation (Schr\"odinger equation in the NR case), typically in the context of the mean-field (MC)SCF method, effectively reducing the problem to a single dimension.
To solve the radial Dirac equation, one can choose among multiple viable basis representations of radial one-electron orbitals.
Here, we focus on the two main classes widely used in present-day high-accuracy atomic calculations -- numerical and optimized Gaussian basis sets -- with an emphasis on the latter.

\subsubsection{Numerical basis sets}

Generally, the radial-angular separation of the relativistic wave function can be represented as
\begin{equation}
    \psi_{n\kappa m}(r,\theta,\phi) = \frac{1}{r}
    \begin{pmatrix}
         f_{n\kappa}(r) \Omega_{\kappa m}(\theta,\phi)\\
        ig_{n\kappa}(r) \Omega_{-\kappa m}(\theta,\phi)
    \end{pmatrix},
\end{equation}
where $f_{n\kappa}$ and $g_{n\kappa}$ are the large and small radial functions, respectively, and $\Omega_{\pm\kappa m}$ are spherical spinors built from the coupling of the spherical harmonics $Y_{lm_l}(\theta, \phi)$ and the spin
functions $\chi_{m_s}$. 
The radial part can be represented by the real two-vector
\begin{equation}
    \phi(r) = 
    \begin{pmatrix}
         f_{n\kappa} \\
         g_{n\kappa} 
    \end{pmatrix},
\end{equation}
which we use in the following discussion.

The conceptually simplest numerical representation of the radial function $\phi$ is discretization on a grid~\cite{Salomonson:89}. This approach is used in several popular atomic structure packages~\cite{GRASP2018,Fritzsche:12,FAC,CI+MBPT}. 
This requires solving discretized differential equations by means of numerical finite-difference-based methods.
The technical details of the radial grid construction vary for different programs, however, a fairly common approach is building a logarithmic grid reflecting the need for finer details in the region close to the nucleus.
This way, the radial dependence can be reexpressed in terms of a new coordinate $t=\mathrm{ln}\,r$.
Using different grid spacings or boundary conditions can lead to more efficient calculations. In some cases, the numerical grid is supplemented by an analytic asymptotic part for the wave function. In the CI+MBPT code~\cite{CI+MBPT}, the wave function inside the nucleus is described by Taylor series in terms of $r/r_\text{nuc}$.
Provided the grid is dense enough and spans the entire region where the wave functions are effectively non-zero, the resulting mean-field numerical wave functions are essentially available with arbitrary precision limited only by the technical implementation.

Another popular numerical basis employed in other atomic structure programs~\cite{MCDFGME,AMBiT,DBSR} is based on B-splines. 
In this case, the radial region is subdivided into interavals (significantly sparser compared to the numerical grid) connected by knots $r_i$ and each of these subintervals is represented by a linear combination of piecewise polynomial B-spline functions of a chosen order.
These are constructed recursively up to a desired order $n$ for each interval $i$
\begin{equation}
    B_{i,0}(r) = \begin{cases}
 1 & \text{if } r_i \leq r < r_{i+1}, \\
 0 & \text{otherwise},
\end{cases}
\end{equation}
\begin{equation}
B_{i,n}(r) = \dfrac{r - r_i}{r_{i+n} - r_i} B_{i,n-1}(r) + \dfrac{r_{i+n+1} - r}{r_{i+n+1} - r_{i+1}} B_{i+1,n-1}(r).
\end{equation}
The main advantage of the B-spline representation is its flexibility and the guaranteed smoothness (continuous differentiability) over the connecting knots up to order $n-2$ for a B-spline of order $n$.

Finite-element basis sets are currently used extensively only in the CI-based codes (Section~\ref{subsub:CI}).
While these numerical basis sets are well-suited for a high-accuracy description of the occupied (spectroscopic) orbitals, producing a large set of unoccupied (virtual) orbitals is somewhat impractical. This in turn limits the accuracy of the post-HF electron correlation treatment.

Further recent developments and extensions of different types of finite-element methods applied to relativistic atomic structure (mostly in the context of DFT) are described in~\cite{ertk2024}.
A recent review of nonrelativistic finite-element atomic calculations can be found in~\cite{Lehtola2019}.
Similarly to atoms, diatomics and general linear polyatomic systems can benefit from separability of the analytically treatable angular part and effective reduction of the numerical problem into two dimensions using cylindrical or elliptical coordinates (for diatomics). This was, however, so far only applied in the nonrelativistic domain~\cite{Lehtola2019}.

\subsubsection{Optimized Gaussian basis sets}
Finite element basis sets are well suited for calculations where spherical symmetry can be employed, i.e. atomic calculations. More general codes that are capable of treating both atoms and molecules and of calculations based on reduced symmetry use sets of basis functions instead \cite{NagJen17}. The wave function of an atomic or molecular orbital is then expanded as a linear combination of basis functions. For an orbital $\phi_i$, we have the general expansion
\begin{equation}
    \phi_i = \sum_\mu c_{\mu i} \chi_{\mu},
\end{equation}
where $c_{\mu i}$ are coefficients and $\chi_{\mu}$ basis functions.

One choice of basis functions for use in the equation above are the Slater-type orbitals (STOs), given, for atomic symmetry, by
\begin{equation}
    \chi^S_{\mu} = N_{\mu n} r^{n-1}  e^{-\zeta_{\mu} r} Y_{j,m}(\theta,\phi),
\end{equation}
where, $N_{\mu n}$ is a normalization constant and $Y_{j,m}(\theta,\phi)$ contains the spherical harmonics. However, calculations based on STOs can become prohibitively expensive, in particular for molecules, due to the evaluation of the two-electron integrals. 
An alternative approach is to use Gaussian-type orbitals, GTOs:

\begin{equation}
    \chi_{\mu} = N_{\mu n} r^{n-1}  e^{-\zeta_{\mu} r^2} Y_{j,m}(\theta,\phi).
\end{equation}

The central advantage of GTOs is the reduction in computational costs, in particular for molecular calculations, due to the fact that the product of two Gaussians at different centers is equivalent to a single Gaussian function centered between them \cite{NagJen17}.  However, there is also a disadvantage to using GTOs that to some extent negates the computational advantage, and that is the lower quality description of the atomic orbitals. GTOs fail to reproduce the cusp at the atomic nucleus that is characteristic of atomic orbitals, and do not reproduce well the behavior of the electron distribution at large distances from the nucleus.  Because a GTO provides a poorer representation of the orbitals, a larger basis must be used to achieve accuracy comparable to that obtained from STOs. Although the number of basis functions and, consequently, the number of integral evaluations are increased, the ease by which these integrals can be calculated means that the drawbacks of this type of orbitals are outweighed by the advantages, and thus GTOs are widely used in computational electronic structure approaches, including implementations in the relativistic domain.

In order to increase efficiency further, sets of Gaussian basis functions are optimized. Such optimization can be carried out to reproduce the lowest possible energy for a given type of system, or for the best performance for a  given property. Optimizing a basis set is equivalent to optimizing the set of Gaussian exponents,  $\zeta_{\mu}$. Usually, such optimization is carried out according to the angular momentum $l$ of the orbital, meaning that basis functions for a given angular momentum quantum number $l$ have the same exponents regardless of the quantum number $j$. This approach also implies that basis function exponents for a given $l$ are optimized separately, as opposed to optimizing the exponents of all $l$-spaces simultaneously. This, however, can become inefficient for heavy elements that experience strong spin-orbit splitting and where the two $j$ components of the same orbital will have very different distributions and energies. The consequence of using $l$-based sets in such cases is that more basis functions are required for the same level of accuracy than for $j$-based sets. For example, the p$_{1/2}$ electrons orbit closer to the nucleus compared to their nonrelativistic counterparts (and very similar to the s$_{1/2}$ orbitals), while the p$_{3/2}$ orbital is further removed. Therefore, an optimization that would provide a suitable description for both spinors will require an excessive number of functions, compared to optimizing based on $j$. Thus, basis sets optimized based on $j$ are more suitable for relativistic calculations, where the spin-orbit splitting is large.  However, in practice, most existing programs are based on $l$ optimized basis sets.

The computational cost in nonrelativistic calculations is commonly reduced further by using  the so-called contracted basis sets, where multiple (primitive) Gaussians are combined, with fixed coefficients, into a single contracted function. Using contracted basis sets in relativistic calculations, however, is problematic as it would severely restrict the small component basis that is generated from the contracted large component using the kinetic balance  condition (see Section~\ref{sec:NVPA}) \cite{Visscher:91, Faegri:02}. Thus, usually basis sets used in relativistic calculations are uncontracted. It is, however, possible to generate contracted basis sets for relativistic four-component calculations by starting with an uncontracted large-component basis, and constructing a small-component basis from this basis using kinetic balance. This set is then used in an uncontracted DHF calculation for the atom in question, yielding large- and small-component atomic functions that are kinetically balanced by virtue of the DHF equations. These atomic functions may then be used to select contracted basis functions for large and for small components \cite{Dyall:07a}.

Choosing suitable and sufficiently sized basis sets is crucial for obtaining accurate results in \textit{ab initio} electronic structure calculations. A number of basis sets suitable for all-electron four-component calculations were developed (see Ref. \cite{PanNee14} for a review of different types of relativistic all-electron basis sets). Some examples are the F\ae gri even tempered basis sets \cite{Fae05}, the all-electron basis sets of Koga~\textit{et al.}~\cite{KogTatMat02, KogTosMat03}, that of Parpia and Mohanty~\textit{et al.}~\cite{Parpia1992}  and the basis sets developed by Dyall~\cite{Faegri:02, Dyall:06, Dyall:07b, Dyall:12}.  The latter are are widely used in relativistic calculations, and we will focus on them in the following and their details are outlined below. 
There are many other all-electron relativistically contracted basis sets that were developed in the scalar-and two-component-relativistic molecular context, such as ANO-DK3 basis sets of Hirao~\textit{et al.}~\cite{Tsuchiya2001,Nakajima2002}, SARC basis sets of Neese, Pantazis~\textit{et al.}~\cite{Pantazis2008,Pantazis2009,Pantazis2011,Pantazis2012}, ANO-RCC of Roos, Widmark~\textit{et al.}~\cite{Widmark1990,Roos2008,Roos2005,Roos2005b,Roos2003,Roos2003b} which was later reoptimized and contracted as ANO-R~\cite{Zobel2019,Zobel2021} and the correlation-consistent family of basis sets (aug-)cc-p(wC)V$n$z-X2C/DK/DK3 of Dunning, Peterson~\textit{et al.}~\cite{Feng2017,Hill2017,Lu2016}, basis sets of Jorge~\textit{et al.}~\cite{Jorge2009,Campos2012,Martins2015,Campos2017,deOliveira2018,deOliveira2019}, and basis set of Weigend~\textit{et al.}~\cite{Pollak2017,Franzke2020}. Many of these can be reclaimed for atomic calculations, especially, if reduction of computational cost is required.

The quality of a basis set is denoted by its cardinal number, which indicates the number of functions used to describe each valence orbital. In the nomenclature of the Dyall basis sets, the cardinal number is denoted by the letter z in the naming of the basis sets, e.g., a cardinality of 4 is written as 4z. The optimization of the basis sets ensures that there are extra functions in the core region close to the nucleus, which is important when describing properties involving the nuclear region, such as hyperfine structure. The basis sets can include higher-$l$ polarization functions that increase the flexibility for characterizing bonding in molecular calculations and improve the treatment of electron correlation. These basis sets can also be augmented with diffuse functions (with small exponents) in order to optimally describe the (outer) valence properties, such as certain bonds in a molecule or the electron affinity or polarizability of an atom. In summary, there are three main aspects that can tuned for these basis sets, and the naming scheme that will be used is summarized in Table~\ref{tab:dyallnaming}.
\begin{table}[h!]
    \caption{Naming convention for the Dyall basis sets used by the DIRAC program.}
    \centering
   \begin{tabular}{l l l l l}
   \hline \hline
   Extra diffuse  & Family & Optimized diffuse & Correlation & Cardinality \\
   \hline
   s-aug- & dyall & a & v (valence) & 2z \\
   d-aug- &  &  & cv (core valence) & 3z \\
   t-aug- &  &  & ae (all electron) & 4z \\
    &  &  &  & 5z \\
   \hline \hline
   \end{tabular}
    \label{tab:dyallnaming}
\end{table}
For example, if a basis set of 4z quality is required, with correlation functions for the valence electrons only, and with three layers of diffuse functions, we would call it the \textit{d-aug-dyall.av4z} basis set. Note that here we used the \textit{d-aug-} prefix in combination with the \textit{a} to obtain a total of three diffuse layers. In this case, the first diffuse layer is energy-optimized, while the two additional layers will be generated automatically in an even-tempered fashion, as given by:
\begin{equation}
    \frac{\zeta_{n+1}}{\zeta_{n}} = \frac{\zeta_{n}}{\zeta_{n-1}}.
\end{equation}
Here, $\zeta_{n}$ is the optimized diffuse exponent, and $\zeta_{n+1}$ is the extra diffuse exponent to be added.

\begin{figure}[ht!]
    \centering
    \includegraphics[scale=0.35]{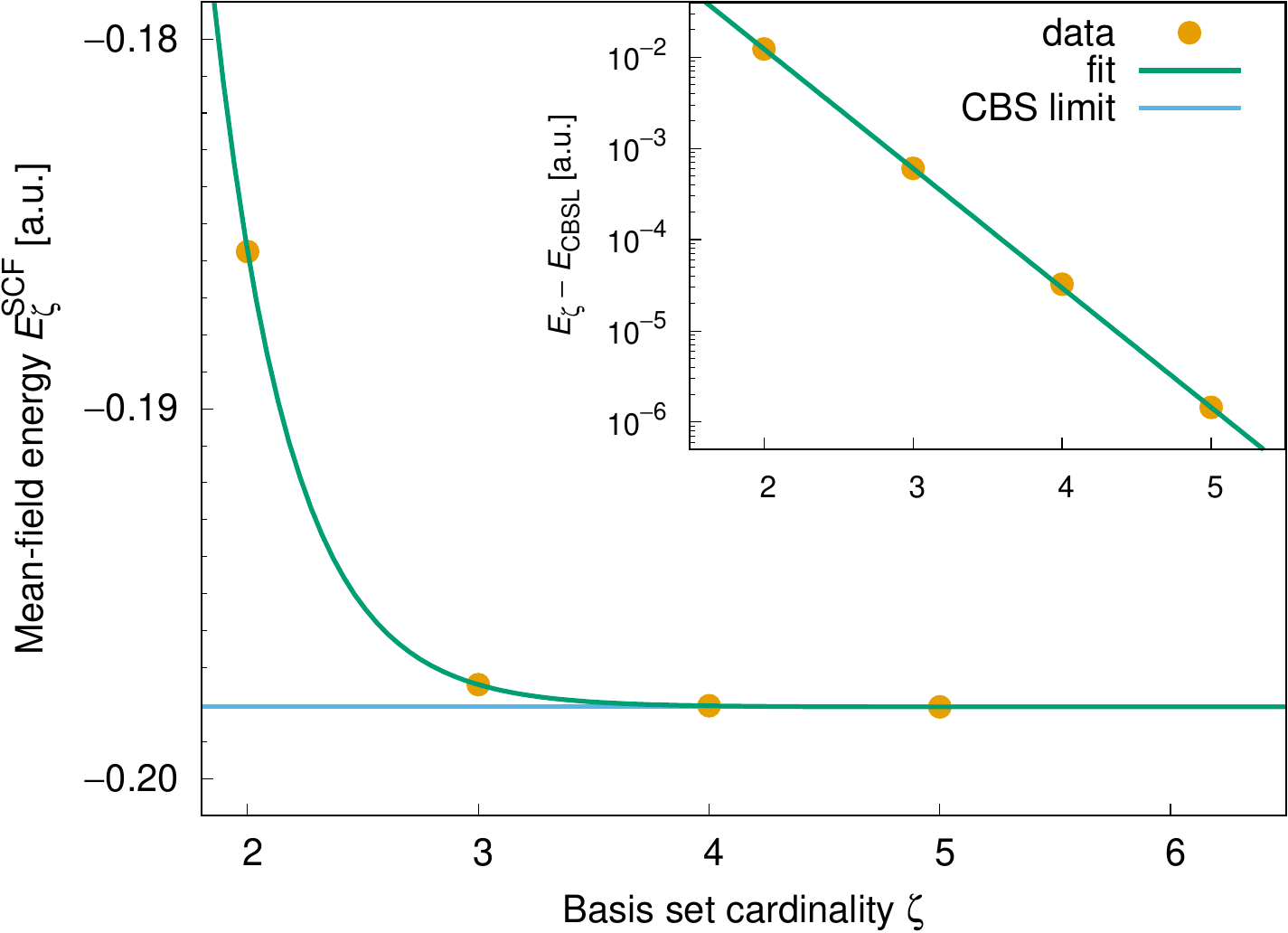} \qquad
    \includegraphics[scale=0.35]{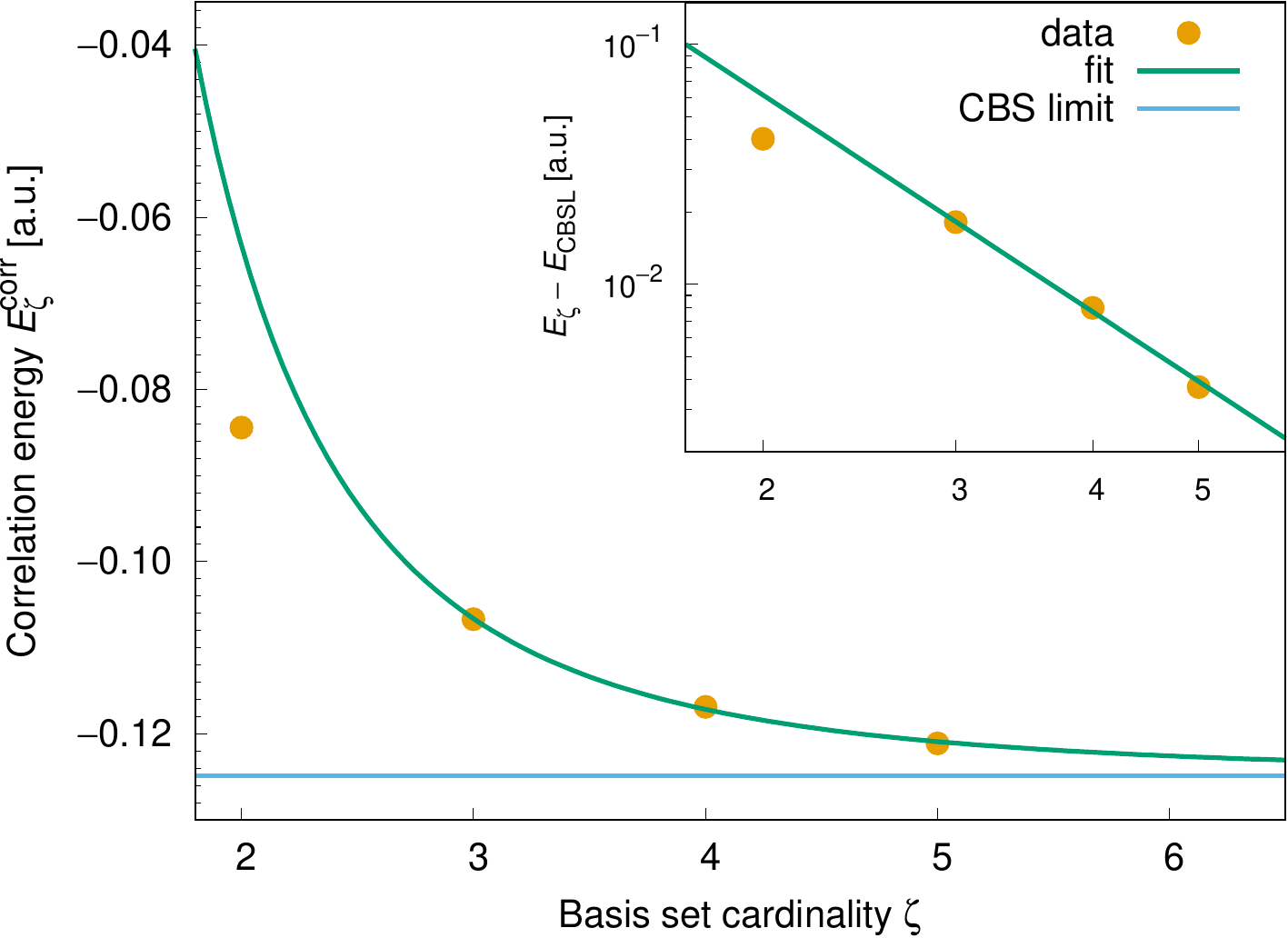}
    \caption{CBS limit extrapolation of the DHF (left) and CCSD(T) correlation energies (right) of Fr$^+$ calculated using the Dyall.av$N$z basis sets and an active space spanning --3 to +10~a.u. Mean-field energies are shifted by 24312~a.u.
    For the respective extrapolation functions, the exponential fit~\eqref{eq:cbs_dunningfeller} and inverse power fit~\eqref{eq:cbs_martin} were chosen. The 2z basis set was omitted from the extrapolation of the correlation energy. The semi-logarithmic (left) and doubly-logarithmic (right) insets highlight the exponential and power-law nature of the extrapolation, respectively.}
    \label{fig:CBSexample}
\end{figure}

The incompleteness of finite size basis sets introduces an inherent error in the calculations, and relativistic correlated calculations that are truly saturated with respect to the size of the basis set are prohibitively expensive. However, it is possible to gradually increase the size of the basis set, where the calculated energy converges as the basis approaches saturation, and we can use this convergence characteristic to extrapolate the result to the complete basis set limit (CBSL). In particular, one can use the correlation-consistent basis sets such as the basis sets originally proposed by Dunning~\cite{dunning1989gaussian}, which are designed for converging post-Hartree–Fock calculations systematically to the complete basis set limit using empirical extrapolation techniques. These basis sets have been extended to the heavy element domain (see, e.g. Refs. \cite{Feng2017,Hill2017,Lu2016}) and the Dyall family basis sets also fall into this category. By performing calculations with 2z, 3z and 4z quality basis sets, and keeping the rest of the computational parameters constant, the results can be extrapolated to the complete basis set limit. 
Several schemes exist for CBSL extrapolation that can be found in Refs.~\cite{NagJen17, Vasilyev:17} and references therein. 
The mean-field SCF and the correlation energies have significantly different rates of convergence to the CBS limit as illustrated in Figure~\ref{fig:CBSexample} and thus are usually extrapolated separately.
The CBSL of the SCF energy is commonly determined using the Dunning--Feller three-parameter extrapolation scheme~\cite{dunning1989gaussian, feller1992application},
\begin{equation}\label{eq:cbs_dunningfeller}
    E_n^\text{SCF} = E_\text{CBSL}^\text{SCF} + B e^{-an},
\end{equation}
where $E_n^\text{SCF}$ are the energies at the $n$z basis set level, and $a$, $B$ and $E_\text{CBSL}^\text{SCF}$ are free parameters, the latter representing  CBSL energy.
A popular scheme for the extrapolation of the correlation energy to the CBSL is the scheme of Martin~\cite{Martin:96}, which relies on an inverse power law
\begin{equation}\label{eq:cbs_martin}
    E_n^\text{corr} = E_\text{CBSL}^\text{corr} + \frac{A}{n^3}.
\end{equation}
It is common practice to entirely exclude 2z energies from the CBSL extrapolation of correlation energies, as these offer relatively poor description. Furthermore, one can choose to extrapolate only from the minimal number of the largest cardinality basis sets, i.e. from two or three points for the two-/three-parameter schemes described here.
An example of CBS limit extrapolation using this scheme is shown in Figure~\ref{fig:CBSexample} using an CCSD(T) calculation of Fr$^+$ at different basis set cardinalities. In case of the Dyall basis set, the highest available cardinality is 5z and we choose to extrapolate using three points 3z -- 5z, omitting the poor-quality 2z basis set result.

Examples of other popular schemes for the extrapolation of correlation energy to the CBS limit are the scheme of Helgaker \cite{HelKloKoc97}, which relies on the $L^{-3}$ error formula, with L the highest angular momentum, and the recent scheme of Lesiuk,  based on an analytic resummation of the missing energy increments using the Riemann zeta function.  

Extrapolation to the CBSL is not limited to energy calculations, as it can be applied to properties as well. Here we distinguish two common types of properties. Many typical properties of interest are obtained from energy derivatives (see Section~\ref{sub:prop} on the finite field approach). It is straightforward to show that the same extrapolation schemes can be applied to the property directly, by differentiation of (\ref{eq:cbs_dunningfeller}) and (\ref{eq:cbs_martin}) with respect to the finite field perturbation strength $\lambda$. Another common set of properties is determined from energy differences, such as electron affinities, ionization potentials, dissociation energies, etc. In this case it is also possible to apply the inverse power law scheme (\ref{eq:cbs_martin}) directly to the extrapolation of the property. The exponential scheme (\ref{eq:cbs_dunningfeller}), however, can only be applied if we impose the restriction that the extrapolation parameters $a$ and $B$ are the same for both energies, which does not need to be true in general. As an alternative to the direct extrapolation of such properties, one can first extrapolate the energies used for the property calculation themselves and then compute their (numerical) derivative or their difference to determine the property at the CBSL.

\newpage
\section{Relativistic coupled cluster in practice}
\label{sec:pract}
The previous section presented the theory behind the relativistic coupled cluster approach. In this section, we will focus on the practicalities of using this approach to calculate atomic structure and properties. We assume that the single-reference coupled cluster method, CCSD(T), is a standard tool in computational chemistry and physics. Hence, energy calculations within this approach do not require further elucidation. Instead, we choose to explain the less commonly used computational schemes. Subsection~\ref{FSCC-calc} presents an example of calculating an atomic spectrum within the FSCC approach. In subsection~\ref{sub:prop} an overview of the procedures that allow calculations of atomic properties within the relativistic CC approach is provided. The various schemes for pushing the accuracy of this method to the meV limit are presented in Subsection~\ref{sub:accur}. Finally, Subsection~\ref{sub:error} introduces the possible strategies used to tackle the challenging problem of assigning error bars on theoretical values within the relativistic coupled cluster approach.

\subsection{Fock-space coupled cluster calculations}\label{FSCC-calc}

The relativistic coupled cluster approach is one of the most powerful tools for calculating spectra of heavy many-electron atoms and molecules. This approach is suitable for calculations on open-shell and multireference systems, thus providing the flexibility to investigate many different atoms, ions, and molecules. It can yield many excitation energies in a single calculation, with accuracy on the order of a few hundred of cm$^{-1}$, and has also demonstrated considerable predictive power. Reviews~\cite{EliBorKal14,Eliav2024} provide an overview of the recent successful applications of this method.  

A usual starting point of the FSCC calculation is a closed-shell reference state. The Dirac--Hartree--Fock equations are solved for this reference state, followed by the coupled cluster equations (equivalent to the single-reference coupled cluster procedure). This corresponds to the sector (0,0) of the FSCC. At the next stage, the electrons are added or removed, one at a time, until the state of interest is reached, corresponding to the sector ($m$,$n$) of the FSCC, where $m$ is the number of valence holes and $n$ the number of valence particles (note that in other works this designation can be reversed). Table~\ref{tab:sectors} contains some examples for calculating atomic spectra in different FSCC sectors. Up to recently, the capability of this method was limited to sectors where $m+n\leq2$ (that is, at most 2 valence electrons or holes, or a single valence electron and a single hole); in 2020 it was extended to allow sectors with $m+n\leq3$ \cite{Oleynichenko:EXPT:20}.

As Table~\ref{tab:sectors} shows, one can treat the same system using different sectors, as given by the example of Ba. Starting with doubly ionized Ba$^{2+}$ and adding two electrons (sector (0,2)) provides us with the spectrum of the neutral atom containing excited states of the type [Xe]$n_1l_1$ $n_2l_2$. Alternatively, if we start with neutral Ba and employ sector (1,1), we will obtain states of the type [Xe]$6s^1$ $nl$. The latter scheme is particularly useful for cases where the desired states can not be reached by using sector (0,2), for example, for spectra of noble gases. It should be noted that when performing the calculation in sectors (0,2), (2,0), or (1,1), we also obtain spectra of states corresponding to the preceding sectors. For example, a sector (0,2) calculation yields not only the spectrum of Ba, but also of Ba$^+$, corresponding to sector (0,1). The hole sectors are not commonly used for neutral systems, as they require a negative ion as the starting point, rendering calculations more challenging. However, they can be used to obtain electron affinities, or to perform calculations on highly charged ions (see, for example, Ref.~\cite{WinCreBek15}), as is presented in Table~\ref{tab:sectors}.

\begin{table}
    \centering
    \caption{Examples of computational schemes within the FSCC approach.}
    \begin{tabular}{llllll}
       \hline \hline
       System   & Configuration & Term & Ref. state & FSCC sector & Calculated energies\\
       \hline 
       Cs & [Xe]$6s$ & $^2$S$_{1/2}$ & Cs$^+$ & (0,1)  & IP and valence spectrum of Cs \\
       Cs & [Xe]$6s$ & $^2$S$_{1/2}$ & Cs$^-$ &(1,0)  & EA and core spectrum of Cs\\
       Ba & [Xe]$6s^2$ & $^1$S$_0$ & Ba$^{2+}$ &(0,2)  & IPs and spectra of Ba and Ba$^+$\\
       Ba & [Xe]$6s^2$ & $^1$S$_0$  & Ba &(1,1)  &IP, EA of Ba and spectra of Ba, Ba$^+$ and Ba$^-$ \\
       Ir$^{17+}$ & [Xe]$4f^{13}5s$ & $^3$F$_4$ &Ir$^{15+}$  &(2,0)  & IPs of Ir$^{15+}$ and Ir$^{16+}$, spectra of Ir$^{16+}$ and Ir$^{17+}$ \\
       \hline  \hline
    \end{tabular}
    \label{tab:sectors}
\end{table}
 
\begin{figure}[h!]
\begin{center}
\includegraphics[scale=0.8]{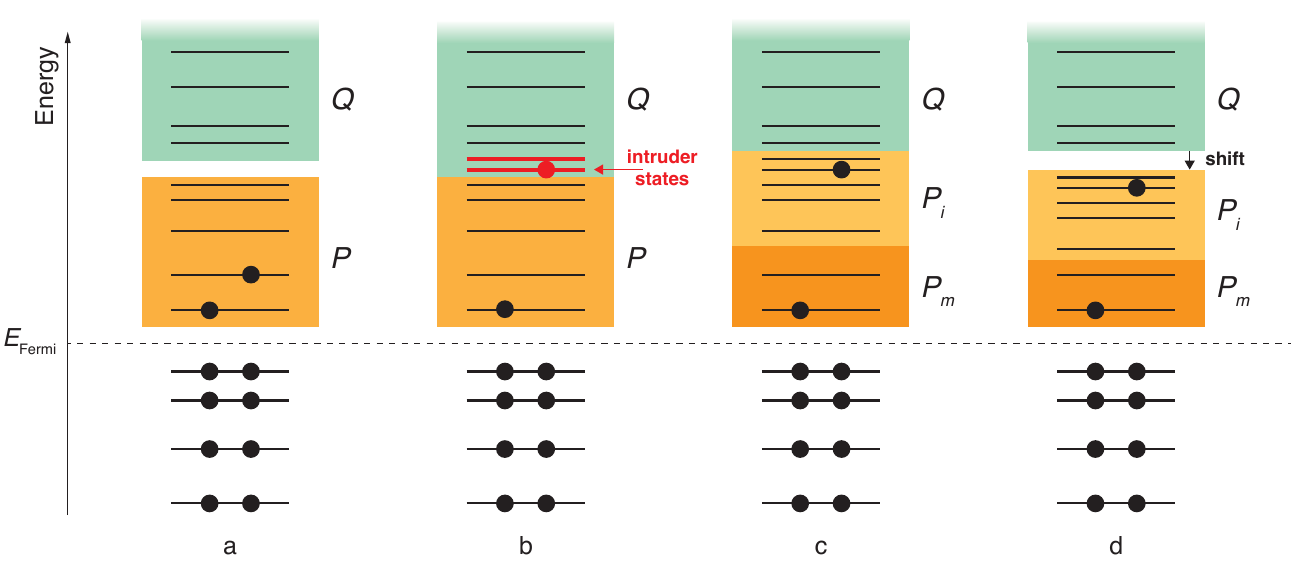}
\caption{Illustration of the intruder state problem and the IH scheme for (0,2) sector of FSCC. The orbitals below the Fermi level are fully occupied, while the two valence electrons are added to the orbitals residing in $P$ space. a) division of the unoccupied orbitals into the $P$ and $Q$ spaces; b) extension of the $P$ space that leads to intruder state problem; c) division of the $P$ space into $P_m$ and $P_i$ subspaces; d) the orbitals in $P_i$ are shifted down to create an energy gap and avoid intruder problems. }
\label{fig:P-space}
\end{center}
\end{figure}

A crucial computational parameter is the choice of the model space, $P$ (Section~\ref{sec:fscc}). One can use a minimal sized $P$. In the case of sector (0,2) calculation on Ba (Table \ref{tab:sectors}), this would be the $6s$ orbital only. In that case, the calculation will yield the ground states of Ba$^{2+}$, Ba$^{+}$, and Ba, and one can extract the first and second ionization potentials. If we are interested in the spectrum of Ba, however, more virtual orbitals should be included in the model space, to allow the calculation of energies that correspond to the excited state configurations. This is illustrated schematically in Figure~\ref{fig:P-space}a. The number of the obtained excited states will correspond to the size of the model space. However, even if we are only interested in the lowest levels, there is a strong advantage in increasing the size of $P$. Including higher-lying virtual orbitals improves the quality of all the calculated states through the diagonalization of the effective Hamiltonian (Section~\ref{sec:fscc}). Thus, to achieve optimal accuracy, the $P$ space should include all functions that are important to the states under study. On the other hand, the convergence of the coupled cluster iterations is enhanced by maximum separation and minimal interaction between $P$ and $Q$, to avoid the problem of intruder states (Sections~\ref{sec:fscc} and \ref{sec:ih}) which are the low-lying $Q$-space states that are close in energy and couple strongly to the higher $P$ states (Figure~\ref{fig:P-space}b), and lead to convergence issues. As the CC method
is an all-or-nothing scheme, if even one $P$-space function has a convergence problem, none of the eigenvalues can be calculated. In order to resolve this issue, one of the variants of the intermediate Hamiltonian approach, described in detail in Section~\ref{sec:fscc} can be used. The selected model space $P$ is divided into two parts, the main model space, $P_m$, which should contain the states of interest, and the intermediate model space, $P_i$, containing all the rest of the states (Figure~\ref{fig:P-space}c). Then, special conditions can be set on the problematic $P\rightarrow Q$ transitions , to avoid intruder problems. Figure ~\ref{fig:P-space}d illustrates a scenario where the states in $P_i$ are shifted  down in energy, creating a significant energy gap between them and the low-lying $Q$ states, corresponding to the XIH variant of the IH approach \cite{Eliav:XIH:05}. 
 To diminish or even fully circumvent the influence of energy shifts of the $P_i$ states, we use a series of calculations with different shifts with \textit{a posteriori} extrapolation to the zero shift value \cite{Eliav:XIH:05}.

\subsection{Properties}\label{sub:prop}
 
Calculating atomic properties using relativistic coupled cluster methods has become indispensable in modern atomic physics and quantum chemistry. These calculations are crucial for understanding fundamental physics, testing QED predictions, and exploring physics beyond the Standard Model. Properties other than energy (e.g., energy derivatives) play an essential role in atomic and molecular structure and spectroscopy research. 

Several almost equivalent formulations exist for property calculations in the single-reference coupled cluster framework (see reviews \cite{bartlett2007coupled,shavitt2009many}). 
The most widely employed such methods are
\begin{enumerate}
    \item expectation value calculations using property operators,
    \item  analytical gradient techniques,
    \item  finite field methods,
\end{enumerate}
Each of these methodologies offers distinct advantages for different types of property calculations in relativistic quantum chemistry. Below, we briefly review these computational approaches and their respective implementations.

The most straightforward approach utilizes the non-relaxed expectation value expression:
\begin{equation}
    \langle P \rangle = \langle\Phi_0|(1 + \Lambda)e^{-S}Pe^S|\Phi_0\rangle,
\end{equation}
where $\Lambda$ represents the de-excitation operator obtained from solving the lambda equations
\begin{equation}
    \langle\Phi_0|\Lambda(H_Ne^S - e^SH_N) = 0,
\end{equation}
with $H_N$ being the normal-ordered Hamiltonian.

The development of analytical gradient techniques for coupled cluster theory represents another significant advancement in property calculations. These methods avoid the numerical instabilities associated with finite difference approaches (see below) and provide a more efficient route to property evaluation. In the case when the energy gradient with respect to a Hamiltonian's parameter $\lambda$ is associated with a particular property, it can be expressed as
\begin{equation}
    \frac{dE}{d\lambda} = \langle\Phi_0|(1 + \Lambda)e^{-S}\frac{dH}{d\lambda}e^S|\Phi_0\rangle + 
    \langle\Phi_0|(1 + \Lambda)e^{-S}H\frac{dS}{d\lambda}e^S|\Phi_0\rangle.
\end{equation}

For relativistic calculations within the four-component Dirac--Coulomb--Breit framework, the gradient formulation must account for the additional complexity of the relativistic Hamiltonian \cite{Shee:16,shee2016relativistic}. In the density matrix formulation, the energy gradient takes the form
\begin{equation}
    \frac{dE}{d\lambda} = \sum_{pq} D_{pq}\frac{dh_{pq}}{d\lambda} + \frac{1}{2}\sum_{pqrs} \Gamma_{pqrs}\frac{d g_{pqrs}}{d\lambda},
\end{equation}
where $D_{pq}$ and $\Gamma_{pqrs}$ are the one- and two-particle density matrices, respectively, and $h_{pq}$ and $g_{pqrs}$ are the one- and two-electron integrals.

The one- and two-particle density matrix elements required for property calculations can be expressed as
\begin{equation}
    D_{pq} = \langle\Phi_0|(1 + \Lambda)e^{-S}\{p^\dagger q\}e^S|\Phi_0\rangle
\end{equation}
and 
\begin{equation}
    \Gamma_{pqrs} = \langle\Phi_0|(1 + \Lambda)e^{-S}\{p^\dagger q^\dagger r\ s\}e^S|\Phi_0\rangle,
\end{equation}
where $\{p^\dagger q\}$ and $\{p^\dagger q^\dagger r\ s\}$ denotes the normal-ordered form of the operators.

For higher-order frequency-dependent properties, such as polarizabilities, the linear response formalism provides an efficient route to calculating
\begin{equation}
    \alpha_{\alpha\beta} = -\langle\langle\mu_\alpha;\mu_\beta\rangle\rangle_\omega,
\end{equation}
where $\mu_\alpha$ and $\mu_\beta$ are components of the dipole operator and $\omega$ is the frequency (for details see reviews~\cite{Watts2008,Sverdrup2023,shee2016relativistic}).

While analytical gradient techniques are preferred when available, the finite-field (FF) approach~\cite{FF:Cohen:1965} remains an important tool, particularly within the FSCC framework, where analytical derivatives are not yet fully developed. The method involves adding  the external field combined from the property operator $P$ and the strength parameter $\lambda$  to the Hamiltonian
\begin{equation}\label{eq:H_tot}
    H(\lambda) = H_0 + \lambda P
\end{equation}
and calculating the energy $E(\lambda)$ by solving the appropriate CC equations for different field strengths $\lambda$. 

As a consequence of the introduction of $\lambda$ in Eq. (\ref{eq:H_tot}), the energy can be expanded in a Taylor series around $\lambda=0$
\begin{equation}
E(\lambda) = E^{(0)} + \lambda E^{(1)} + \frac{1}{2}\lambda^2 E^{(2)} + ... + \mathcal{O}(\lambda^n),
\end{equation}
where $\mathcal{O}(\lambda^n)$ denotes higher-order terms, and $E$ is the total energy of a given electronic state in the presence of the perturbation $P$. The energy $E^{(n)}$ can be associated with the expectation value of a property operator of $n$-th order according to the Hellman--Feynman theorem. The magnitude of $\lambda$ can be chosen such that higher-order terms numerically vanish, and, thus, the first order property $\langle P \rangle$ can be obtained as the  derivative of the energy with respect to $\lambda$
\begin{equation} \label{eq:P_derivative}
\langle P \rangle = E^{(1)} = \left. \frac{d E(\lambda)}{d \lambda} \right| _{\lambda=0}.
\end{equation}

In practice, the most straightforward scheme to calculate $\langle P \rangle$ is to evaluate $E(\lambda)$ at two small values of $\lambda$, which differ only by the sign, and to explore the following simple expression for the numerical differentiation:
\begin{equation} \label{eq:E_derivative}
\left. \frac{d E(\lambda)}{d \lambda} \right| _{\lambda=0} = \frac{E(\lambda)-E(-\lambda)}{2\lambda} + O(\lambda ^3).
\end{equation}
Being, in principle, an all-order method, the FF approach gives rather precise results, at least for the first- and second-order properties (see the applications below and recent reviews~\cite{EliBorKal14,Eliav2024}). However, the method is relatively expensive because several additional calculations are needed for the different sizes of the external fields included in the Hamiltonian. Usually, such inclusion lowers the system's symmetry according to the symmetry of the external field added to the Hamiltonian, making the approach even more expensive. Care is needed in choosing appropriate field strengths because numerical stability must be considered in the differentiation process.
Still, the approach is very popular for property evaluation within the relativistic coupled cluster method in general and with the FSCC approach in particular. Recently, the FF approach has also been extended for transition matrix elements calculations, substantially expanding the applicability of the method~\cite{Zaitsevskii:Optics:18,Oleynichenko:20cpl,Zaitsevskii:TDM:20}. Another significant recent development for the calculation of properties within FSCC, also broadening its usage, is the evaluation of approximated effective property operators and density matrices (both for particular states and transitions between states, see~\cite{SkrOleZai23,oleynichenko2024finite}).   

Implementing property calculations within the relativistic coupled cluster framework requires careful consideration of several practical aspects. 
The accuracy of property calculations depends crucially on the proper treatment of electron correlation effects. For example, the perturbative triples correction significantly improves results for properties sensitive to dynamic correlation, such as electric multipole moments and polarizabilities.  

For properties involving core electrons, such as electric field gradients at nuclei, core-core and core-valence dynamic correlation is essential. This often necessitates using specifically optimized basis sets and carefully considering the correlation space. For the FSCC method, the size of the model space could be used as an additional parameter to increase the precision of property calculations. The relativistic coupled cluster calculations must also account for the nonadditive interplay between relativistic and correlation effects, which becomes particularly important for properties involving the nuclear region or spin-dependent operators.

In recent years, significant advances have been made in relativistic coupled cluster property calculations. The development of exact two-component (X2C) methods has provided an efficient alternative to full four-component calculations for many properties (see reviews~\cite{Liu10,Saue:11,Peng:12}). Within this framework, picture-change effects are appropriately accounted for through the transformation of property operators:
\begin{equation}
    O^{\text{X2C}} = U^\dagger O^{\text{4c}}U
\end{equation}
where $U$ is the decoupling transformation matrix.

Several challenges remain in relativistic coupled cluster property calculations. An important goal is to develop efficient analytical derivative techniques for FSCC and systematically include QED effects in property calculations. Extending to time-dependent properties and novel BSM applications also presents significant opportunities for future development.

Calculating atomic and molecular properties using relativistic coupled cluster methods continues to evolve, driven by theoretical advances and increasing computational capabilities. Combining either analytical gradient or improved FF techniques with systematic addition of the advanced levels of correlation treatment provides a robust framework for high-accuracy calculations. At the same time, newer developments in effective operators, compact efficient natural orbitals basis sets, and QED corrections push the boundaries of achievable precision further.

\subsection{Towards single meV accuracy}\label{sub:accur}

The accuracy of atomic energies calculated within the four-component coupled cluster approach based on a saturated basis set (as illustrated by the black point in Figure~\ref{3axes}) is expected to be on the order of magnitude of 10s of meV, based on extensive comparison of calculated values to experiment, where available.  For excited states treated within the FSCC method, this accuracy strongly depends on the choice of the model space and decreases for high-lying states~\cite{EliBorKal14,Eliav2024}. In the case of atomic properties, such as polarizabilities and hyperfine structure (HFS) parameters, much less comparison with the experiment is available. Still, overall, we can expect theoretical values to fall within about 10\%  of the experimental results~\cite{KanYanBis20,GusRicRei20,HaaEliIli20,DenHaaMoo22}. Such accuracy is often sufficient for estimating the spectra and the relative positions of transition lines. Furthermore, such calculations can indicate the relative magnitude of a certain property in different atomic states or between different species. An example is the identification of atomic levels that are most sensitive to nuclear magnetic dipole or electric quadrupole moments. While the absolute accuracy may not be very high, it is expected to be sufficient for comparison purposes.

However, higher precision is often required, for example, when the position of a given transition line is unknown and theoretical guidance is necessary to search for it. Another example has to do with the interpretation of measurements, such as the extraction of nuclear moments from measured hyperfine structure transitions. These moments will be ultimately determined from a combination of experimental and theoretical parameters. Thus, their quality and reliability will be directly impacted by the accuracy of the theoretical predictions.   

An important challenge is thus to go beyond the state of the art and to push the relativistic coupled cluster approach to the limits of its accuracy. 
In practice, this corresponds to including higher-order corrections that are not addressed within the standard implementation of the method. Since we are dealing with higher-order effects, we address separately the three classes of missing contributions following the axes in Figure~\ref{3axes} -- the higher-order relativistic corrections, the incomplete treatment of electron correlation, and the inherent incompleteness of the employed basis set. Possible strategies for dealing with these corrections are provided below, separately for each class of effects, in Sections~\ref{rel_corr},~\ref{corr_corr} and~\ref{bas_corr}. 

Performing a single calculation including all the contributions necessary to reach the desired high accuracy is not computationally feasible. 
Instead, various composite schemes of incremental corrections are employed.
In these, the higher-order corrections are calculated within a more restricted computational scheme (smaller basis, smaller correlation space, etc.), and added on top of the reference values, obtained in a ``standard'' baseline relativistic coupled cluster calculation (see e.g. Refs.~\cite{FinPet19, LeiKarGuo20,SkrOleZai23}). 
Overall, the optimal (yet realistically achievable, at least in some cases) accuracy to strive for in case of heavy many-electron atoms are single meV in case of energies and single percent for various properties~\cite{Pasteka:17, LeiKarGuo20, Skr21,HaaEliIli20}. 

While the composite treatment of the higher-order effects results from practical necessity, it relies on their effective separability -- an approximation that is increasingly more accurate the further we move in the hierarchy of computational approaches along each of the axes in Figure~\ref{3axes}. 
Gradually, these effects decouple and the resulting contributions become almost exactly additive.
This is a consequence of 
the fast convergence in terms of the size of the individual corrections with the increasing level of theory. While the individual corrections shrink over several orders of magnitude, the residual relative error introduced by the composite scheme remains approximately constant (within an acceptable range). Thus in absolute terms, the error decreases more or less proportionally with the decrease in correction size leading to near-perfect additivity.
Furthermore, the decoupling speeds up even beyond this simple rationalization. For example, it is known that the rate of basis set convergence increases rapidly with the excitation level in CC calculations~\cite{Karton2007,Smith2014,Karton2019,Karton2020}.
Nevertheless, in practical applications, this additivity approximation is not taken blindly, but the associated residual uncertainty is determined as described in Section~\ref{sub:error}.
In the context of molecular quantum chemistry, this additivity approximation 
has been extensively tested against experiment and forms a basis of an entire range of composite approaches developed dating back to Pople's G$n$ schemes~\cite{Pople1989,Curtiss2011}. 
Notable examples of composite schemes reaching high accuracy were developed by the Georgia group~\cite{East1993,Csaszar1998,Schuurman2004}, the Texas group~\cite{DeYonker2006,Wilson2009,Peterson2016}, and the Washington group~\cite{Feller2008,Feller2012,Feller2013,Finney2019}.
For reviews on the topic see Refs.~\cite{Helgaker2008,Peterson2012,Karton2016,Patel2021,Karton2022}.

In most cases, the schemes cited above are focusing on organic molecular compounds and treat the effects of relativity either only as an \textit{a posteriori} correction, or at most at the scalar or ECP level in their baseline.
In contrast, in the applications focused on heavy-element atomic systems described in the following sections, we 
find it necessary to use a 
four-component relativistic baseline.
It should be emphasized that the choice of an appropriate starting baseline 
is essential to ensure the reliability of the composite scheme. 
Ideally, this should be a level of theory that in a single calculation reaches as far as possible in each computational parameter in a balanced fashion.

Therefore visually, in Figure~\ref{3axes}, the shaded cuboid representing the region of the computational space containing all considered effects covered by the baseline calculation (black point) should envelop the largest possible volume, thus carving out the inconvenient domain of appreciable coupling between the cardinal axes.

In the following, we discuss in practical terms the higher-order treatment of relativistic, correlation and basis set contributions.

\subsubsection{Higher-order relativistic effects}\label{rel_corr}
The standard relativistic calculations are based on the four-component Dirac--Coulomb Hamiltonian~\eqref{eq:DCH}. The one-electron part of the DC Hamiltonian is relativistic, while the Coulomb operator can be considered as a nonrelativistic description of the two-electron interaction. In order to correct for the non-instantaneous interactions between the electrons, 
the Breit correction (see Section~\ref{sec:NVPA}) can be added to the two-electron part of $H_\text{DC}$. The effect of this correction on transition energies and ionization potentials is usually small for light elements, on the order of a 100~--~300~cm$^{-1}$ for actinides and transactinides (see, e.g.~\cite{IndSanBou07,ThiSch10}), and can be significantly larger for highly charged ions~\cite{WinCreBek15, WinTorBor16}.

To further improve precision, one should also include the QED corrections. We note that in the context of atomic and molecular physics, QED corrections are understood as effects beyond the Dirac--Coulomb--Breit description. For many-electron systems, these are currently limited to the leading order QED effects, self energy and vacuum polarization.  Rather than employing QED explicitly, which is not computationally tractable for complex systems, effective operators are used. Currently, a number of variants of model Lamb shift operators (MLSO) are commonly used~\cite{flambaum2005radiative,SunSal22,ShaTupYer15,ShaTupYer18}. The different approaches yield comparable accuracy~\cite{SunSal22, KahBerLaa19} and are implemented in a number of computational packages used for coupled cluster calculations (for example, the Tel Aviv TRAFS-3C code~\cite{TRAFS-3C}, the EXP-T program~\cite{Oleynichenko:EXPT:20}, GRASP~\cite{GRASP2018} and the DIRAC package~\cite{Saue:20,DIRAC23,SunSal22}, where it can also be used to treat molecules). Due to their scaling with atomic number (as $Z^2-Z^3$, depending on the orbital~\cite{ThiSch10,SunSal22}), QED effects become significant for heavy and superheavy elements, in particular where the valence electron occupies the $n$s orbital. 
Although vacuum polarization is included up to the fourth order in most implementations, the leading self-energy term is already nonlocal and included by means of approximate radiative potentials. Higher QED corrections are not yet feasible for many-electron systems; furthermore, their size is expected to be small compared to the uncertainty stemming from the incomplete treatment of electron correlation.

\subsubsection{Electron correlation contributions}\label{corr_corr}

Within both the relativistic and the nonrelativistic coupled cluster approach, routine calculations do not usually correlate all the electrons. The correlation space is usually comprised only of a subset of all orbitals (spinors) in the HF reference. A choice is then made to restrict this orbital set either based on the type of orbitals one wants to include or by imposing an energy cutoff criterion (applied separately to the occupied and the virtual orbital sets by means of negative energy and positive energy cutoffs, respectively). Typically, the electrons occupying the valence shell, and the shell below it ($n-1$ shell) (usually corresponding to an active-space orbital-energy cutoff of about --20~a.u.~\cite{Desclaux1973}) are included in the correlation procedure. The rationale behind this choice is that the deeper electrons do not provide a considerable contribution to valence properties, such as ionization or excitation energies or bonding, in the case of molecules. In the nonrelativistic community, it is commonplace to correlate all virtual orbitals, while in the relativistic community, a virtual cutoff is usually imposed as well. The correlation space restriction allows for a reduction of computational costs, without significantly compromising the accuracy of the calculated properties. However, at times, the highest possible accuracy is required. Furthermore, certain properties, such as hyperfine structure parameters, electric field gradients, or various parameters needed for the interpretation of experiments that search for physics beyond the Standard Model, are sensitive to the description of the wave function in proximity to the nucleus. For these properties, freezing the inner electrons can lead to errors of several percent in the calculated values~\cite{HaaEliIli20,HaaDoeBoe21}. Thus, ideally, all electrons should be included in the correlation procedure, which also requires setting a high virtual space cutoff, as it has been shown that high-lying virtual orbitals are important for capturing all the correlation effects related to the core electrons~\cite{TalSasNay18,SkrTit15,SkrMaiMos17}. Thus, while the virtual space cutoff for a standard calculation is usually set at 30~--~50~a.u., when correlating all the electrons all the virtual orbitals up to 1000~--~2000~a.u. is necessary. A common rule of thumb is to keep the positive and negative energy cutoffs (somewhat) symmetrical to achieve a balanced active space description.
For atoms, this computational procedure is usually feasible at realistic computational costs, while for molecules, such calculations can become intractable. To circumvent the high computational costs, an incremental procedure is possible. The reference values are calculated using a saturated basis set. Then, the effect of correlating all the electrons is extracted from a difference between calculations performed within limited and extended correlation spaces, carried out with a smaller basis set (as the effects of correlation space and basis set are virtually independent~\cite{HaaDoeBoe21, KiuPasEli24}). This difference is then used as a correction and added on top of the reference value.
An efficient alternative approach to the reduction of the virtual correlation space is based on natural orbitals (spinors)~\cite{Jensen1988,ChaSurJan22}. Using these, the error introduced by a virtual cutoff is dramatically reduced and a thus much lower cutoff energies can lead to significant savings in computational costs without a notable decrease in accuracy.

For excitation energies calculated within the Fock-space coupled cluster approach, the size of the main model space, $P_m$, and the intermediate space, $P_i$, play a crucial role. The model space can be increased step-wise up to the convergence of the calculated excitation energies of interest.

The most commonly used single-reference coupled cluster approach, CCSD(T), includes single and double excitations, with 
triple excitations treated perturbatively. Up to recently, the relativistic Fock-space coupled cluster approach was limited to single and double excitations, FSCCSD. However, at times, higher excitations become important, and provide a non-negligible contribution to the calculated properties~\cite{Pasteka:17, LeiKarGuo20, SkrOleZai23}. Unfortunately, calculating full triple and higher excitations with an extended basis set and active space is computationally intractable, even for atoms. Thus, the incremental approach is employed in such calculations as well, where the higher excitations (full triples and even quadruple excitations) are calculated using a modest basis set and a limited correlation space (it was shown that higher-order excitations are generally localized in the valence shell region~\cite{Pasteka:17}) and added to the reference results. Examples of programs capable of calculating higher excitations are the MRCC program of Kalay and colleagues~\cite{MRCC,KalSur01,BomStaKal05,KalGau05,KalGau08}, suitable for single-reference coupled cluster calculations, and the recently developed EXP-T~\cite{EXPT:20} program, which allows the inclusion of iterative triple 
excitations in the FSCC procedure. Alternatively, the Psi4 \cite{SmiBurSim20} and CFOUR \cite{MatCheHar20} also allow the inclusion of full triple excitations within the ECP framework.

\subsubsection{Addressing the basis set deficiencies}\label{bas_corr}

While a complete basis set calculation is unrealistic, one can use a set of increasing size (cardinality, z) high-quality basis sets and use one of the available schemes to extrapolate the results to the complete basis set limit, as described in Section~\ref{sec:basis}.

Beyond cardinality, the other two basis set properties important for the quality of the calculations are the presence of core-correlating functions and of diffuse functions. Core-correlating functions are particularly important when all electrons are correlated, and they contain higher angular momentum functions.
Calculations that use such basis sets are computationally expensive, and the incremental approach can aid here as well: the effect of using core-correlating functions can be investigated using lower cardinality basis sets (i.e. 3z instead of 4z), and the effect can be added to correct the reference values. 

Diffuse (low-exponent) basis functions are important for describing the valence region,  bonding in molecules, excitation energies, polarizabilities, electron affinities, and other properties that rely on high-quality descriptions of the region removed from the nucleus. As a rule of thumb, at least one level of augmentation is routinely used in calculations, but often more are needed, and it is prudent to perform the calculations with increasing augmentation level, to verify saturation. It is even possible to extrapolate the results to an infinite augmentation level, as was necessary for the accurate calculation of the electron affinity of the superheavy element Og~\cite{GuoPasEli21} (discussed in Section~\ref{sub:Og}).

Finally, in some cases, nucleus-penetrating tight (high-exponent) functions of s and p angular momenta aiding the description of the region in the vicinity of the nucleus are needed for optimal results. This is the case, for example, for the electronic structure parameters that describe the sensitivity of an atom or a molecule to anapole moments~\cite{BorIliDzu12} or to nuclear Schiff moments~\cite{GauHutYu24}. Then, these functions can also be constructed following the ratio of the highest exponent functions in a given symmetry, similar to diffuse functions. Such cases, however, are less common.\\ 

If the corrections outlined in Sections~\ref{rel_corr} -- \ref{bas_corr} above are employed in a balanced manner, that is, all types of corrections are added to the reference value, extremely high accuracy can be reached for the calculated properties and energies. Sections~\ref{sub:meV},~\ref{sub:Lr} and~\ref{sub:SHE} present some examples of applications of this scheme to heavy atoms.

\subsection{Uncertainty estimates}\label{sub:error}

The question of how to assign uncertainties to theoretical values is non-trivial, and no universal prescription exists. When calculating a property that has not been measured for a certain atom, there are a number of ways that the uncertainty can be estimated: 

\begin{itemize}
    \item One can use the same method to calculate the same property in a similar system where the experiment is available, and use the discrepancy between theory and experiment to estimate the uncertainty of the calculated value. Usually, the proxy system would be the lighter homologue of the atom of interest, as we expect analogous behavior for similar electronic structures. Comparison with experiment in lighter homologs is suitable for testing the quality of treatment of electron correlation but less so for testing approximate methods for treatment of relativity, as relativistic effects are less important in light systems. However, within the four-component relativistic framework we expect the treatment of relativity to be on a sufficiently high level for both light and heavy elements to justify direct comparisons.
    \item An alternative approach is to calculate a different property in the same system where experimental data are available. 
    Here, it is important to select a proxy property that is equivalently sensitive to the quality of the wave function description as the parameter for which we want to assign the uncertainty. 
    For example, to set uncertainty on calculated field shifts, comparison with experiment for magnetic hyperfine structure parameters is a valid strategy, as both properties are sensitive to the quality of the description of the electronic structure of the nuclear region, even if the underlying physical interaction mechanisms differ. In contrast, to assign uncertainty on a calculated electron affinity, the relative uncertainty of the calculated excitation energies is suitable, as both are sensitive to electron correlation in the valence region and to the description of the loosely bound electrons removed from the nucleus. 
    \item Furthermore, we can evaluate the magnitude of the effects missing from the computational description, such as the basis set incompleteness, and the higher-order relativistic and correlation effects. The associated uncertainties can be obtained by performing an extensive systematic computational study based on the incremental improvements in the hierarchy of computational approaches. 
\end{itemize}

Ideally, all three procedures outlined above should be performed for a given calculation to verify the consistency between the uncertainty based on computational considerations alone and based on comparison with the experiment, both for the same property in a similar system and for a different property in the same system.

Below, we present a possible procedure used to assign uncertainties based on computational considerations, which was gradually refined in a series of our previous works, e.g. Refs.~\cite{LeiKarGuo20, GuoPasEli21, HaaDoeBoe21,HaaEliIli20,HaoNavNor20,GusRicRei20,GuoBorEli22}. A similar approach is being employed in recent years by a number of groups working with relativistic coupled cluster methods, differing in how the various contributions to the uncertainty are estimated~\cite{SkrPetTit13,TalSasNay18,TalNayVav19,FinPet19, KaySkrTup21}. There are also analogous schemes for setting uncertainty on calculations carried out using other high-accuracy relativistic approaches, such as CI~\cite{FleNay13,Fle19}, CI+MBPT~\cite{KahBerLaa19}, CI+all order~\cite{PorKozSaf16,SafSafPor18}, or MCDF~\cite{GaiGaiBie08,RadGaiJon14}.  
In a recent perspective article~\cite{Reiher2021}, uncertainty estimation based on systematic improvability and Bayesian statistics is discussed.

To assign a theoretical uncertainty, we again consider different computational parameters separately. Within a robust and transparent method like relativistic coupled cluster, we have a solid understanding of which effects are included in our calculations and which are left out (whether due to lack of implementation or to high computational costs). We can then attempt to estimate the size of the missing effects and use these to evaluate the uncertainty. Since we are dealing with higher-order effects, this procedure can be performed separately for each computational parameter. Overall, one expects higher-order contributions to diminish rapidly in size \cite{Helgaker2000}, which allows us to use the general strategy of setting the upper limit on the size of the missing effects by taking the size of the highest included lower-order contributions or by taking into account also the general order-by-order trend of the included effects allowing for extrapolation to higher orders.

\subsubsection{Neglected relativistic contributions}

A calculation where Breit and leading-order QED effects (SE, VP) are included, either \textit{a priori} (variationally) or as a perturbative correction, is still missing the higher-order QED effects. 
A conservative estimate of their size could be the size of the leading-order QED correction itself. However, this is liable to overestimate the expected uncertainty, as QED effects are expected to decrease rapidly with the order.

Alternatively, one can consider that both the leading-order VP (Uehling) contribution and the SE model potential include the QED contribution of the order $Z \alpha^2$, where $\alpha$ is the fine-structure constant. The next order should thus be $Z^2 \alpha^3$ (in atomic units) from the expansion of the bound-state propagator~\cite{SunSal22}. Thus, the next-order contribution (and the corresponding uncertainty) can be estimated by multiplying the leading-order contributions by the ratio $Z\alpha$, substituting the nuclear charge $Z$ of the investigated atom.

\subsubsection{Neglected electron correlation contributions}\label{corr_error}

The uncertainty due to the treatment of correlation can be divided into two main sources -- the incompleteness of the active space (and the model space, in case of excitation energies) and the truncation of the excitation level.

Ideally, the calculation was performed correlating all the electrons, or alternatively, it was corrected for the effect of freezing part of the core electrons. We are then left with incomplete information about the correlated virtual space. One can then perform smaller basis set calculations correlating all orbitals, or at least orbitals up to a very high active space cutoff (e.g., 10~000~a.u.). 
This is usually performed in several steps, with a gradual increase of the virtual cutoff until reasonable convergence is observed. The remaining effect is assumed to be bound by the size of the included corrections. It is taken as the difference between calculations performed using the two largest virtual space cutoffs.
The contribution of this effect to the overall uncertainty is usually negligible~\cite{HaaDoeBoe21,GuoPasEli21,GuoBorEli22}. 
The uncertainty due to the limited size of the model space is usually taken as the difference between the results obtained using the largest practically possible model space, and the second largest.

 To estimate the uncertainty due to the missing higher excitations beyond CCSDT, we take a conservative 10\%  fraction of the triples contribution. This is justified by the observation that size of the corrections reduces dramatically as one progresses along the excitation rank~\cite{Helgaker2000}, and by the fact that perturbative contributions and the correction with respect to full iterative value usually have the opposite sign (see Figure~\ref{fig:Au-plots} for illustration).

\subsubsection{Basis set incompleteness contributions}\label{bas_error}

The results are usually extrapolated to the complete basis set limit; however, they are not obtained within a truly complete basis set. A conservative way to assign uncertainty on such extrapolated results is to take the difference between the CBSL values, and the values obtained with the largest basis set used (for relativistic basis sets, this is usually a basis of 4z cardinality). At times, to account for the fact that such uncertainty is likely overestimated, this energy difference can be multiplied by a factor of 0.5, for example~\cite{LeiKarGuo20}.

Alternatively, one can use a number of different schemes to perform the CBSL extrapolation (see Sect. \ref{sec:basis} for details), and then take 95\%  confidence interval of the standard deviation between the different schemes as the CBS uncertainty estimate, as detailed in Ref.~\cite{GuoBorEli22}. 
Very recently, an alternative method of estimating the uncertainty of the CBSL extrapolation based on statistics of random walks was proposed~\cite{Lang2025}. The method is free of empirical parameters
and compatible with any extrapolation scheme.

To account for the uncertainty due to the missing core-correlating functions, one can take the difference between an all-electron basis set calculation and a valence basis set calculation; when switching to all-electron basis set is intractable for a saturated and augmented basis set, one can carry out this comparison for a lower cardinality basis. 

Finally, to consider the uncertainty due to an insufficient number of diffuse functions, we can take the difference between the calculation performed with the highest augmentation level (used for the final value) and that performed with one less layer of augmentation. 
For example, if the results were obtained using the doubly augmented basis set, the uncertainty will be taken as the difference between these values and values obtained using a singly augmented basis. 
Alternatively, if the inclusion of multiple diffuse layers is investigated, one can make use of the resulting asymptotically exponential trend to extrapolate to the infinite limit similarly to the CBSL extrapolations based on the basis set cardinality (see Section~\ref{sub:Og} and Ref.~\cite{GuoPasEli21}). The uncertainty can then be taken as the difference between this limit and the highest explicitly included augmentation layer.

\subsubsection{Total uncertainty}\label{tot_error}

Relying on the separability of the higher-order effects described in Section~\ref{sub:accur}, one can assume that the uncertainty contributions stemming from different sources are independent to a large degree. Thus, the total uncertainty is obtained by adding the individual sources of uncertainty using the usual Euclidean norm. Such uncertainties are often dominated by basis set contributions~\cite{LeiKarGuo20,GuoPasEli21,GuoBorEli22,KiuPasEli24}, prompting the development of higher cardinality 5z basis sets, for example~\cite{5z-dyall}. Sections~\ref{sub:meV}, \ref{sub:HFS}, \ref{sub:Lr}, \ref{sub:SHE} provide some examples of such an uncertainty evaluation procedure, together with a breakdown of the various sources of uncertainty.

\newpage
\section{Selected applications}\label{sec:appl}
To illustrate the general concepts of hierarchical improvement of the calculated result together with evaluating the associated uncertainties presented in Section~\ref{sec:pract}, we offer below a selection of case studies where these concepts are practically applied. Section \ref{sub:meV} is based on two publications where the scheme for reaching meV accuracy \cite{Pasteka:17} and for evaluating uncertainties based on the relativistic coupled cluster computational procedure \cite{GusRicRei20} were first introduced. Basic atomic properties, such as the ionization potential and electron affinity presented in these publications, are particularly suitable for straightforward and didactic introduction of these topics. 

Furthermore, we present a diverse selection of applications of the methods at hand, addressing a variety of properties and systems. Some of the presented applications (Sections \ref{sub:At}, \ref{sub:HFS}, \ref{sub:Lr}, and  \ref{sub:IonsE}) were selected to showcase important examples of successful theory-experiment collaborations, where the calculated atomic properties were used to plan, guide, and interpret the measurements. In particular, Section \ref{sub:HFS} presents a number of studies where electronic structure input was used for extracting nuclear information from hyperfine structure measurements. Robust and reliable uncertainties are crucial for the theoretical values used in such a context. Another outstanding example of theory-experiment synergy is presented in Section \ref{sub:Lr}: these are the theoretical investigations of the heaviest actinide Lr, which are used to plan a variety of future experiments on this element. Section \ref{sub:SHE} presents accurate theoretical studies of basic atomic properties of the superheavy elements nihonium ($Z=113$) and oganesson ($Z=118$), where theory precedes experiment. Finally, the investigations of spectra and properties of highly charged ions shown in Section \ref{sub:IonsE} showcase the flexibility and the broad applicability of the relativistic coupled cluster approach.

\subsection{In pursuit of meV accuracy}\label{sub:meV}

Arguably, the two atomic properties most fundamental for atomic physics and chemistry are the ionization potential (IP) and the electron affinity (EA). These are closely related to the ability and proclivity of the given element to bond with others. As such, IP and EA are often among the first properties determined for an atom by means of both theory and experiment. In this section, we present two examples of atomic IP and EA calculations aiming at an accuracy at the meV level to match or rival the experimental precision.

IP is the energy required to release a bound electron, i.e. ionize the system. Although a technically more correct term would be ionization energy, IP is used by convention.
EA is the energy released by the system binding an additional electron.
Simply, IP and EA may be defined in terms of electron detachment and electron attachment energies of the neutral atom, respectively.
Ionization potentials of elements exhibit remarkably regular periodic trends~\cite{NIST_ASD} with a general tendency of gradual increase from the lower left corner of the periodic table represented by Fr with the lowest IP of 4.0727~eV up to the highest value of 24.5874~eV for He in the upper right corner. After each shell closure, a drop in IP can be observed; however, within each block, the trend holds spectacularly well, with only a few exceptions.
Electron affinities, on the other hand, display comparably more pronounced variation across the periodic table~\cite{Andersen1999}. No monotonic trend is observed within most groups; however, a general feature to be noted is the increase of EA along the period as a shell is gradually filled and a sudden drop for the closed-shell atoms (Groups 2, 12, 18), most of which do not form stable negative ions at all, and thus have negative EAs. The group of elements with the highest EAs are the halogens.

Several other chemically relevant atomic properties are defined purely based on IP and EA, stressing the fundamental importance of these properties for bonding an reactivity. 
Among these, perhaps the most widely recognized and applied is the Mulliken scale electronegativity $\chi_\mathrm{M} =\frac{\mathrm{IP}+\mathrm{EA}}{2}$~\cite{Mulliken1934}, which allows for the prediction of charge redistribution in bonding. This definition was later generalized and connected to the chemical potential $\mu=-\chi_\mathrm{M}$~\cite{Parr1978}.
In the context of the theory of hard and soft acids and bases, Pearson defines chemical hardness $\eta =\frac{\mathrm{IP}-\mathrm{EA}}{2}$ and softness $S=\frac{1}{\eta}$~\cite{Pearson1963,Pearson1987,Pearson1988}.
The electrophilicity index $\omega = \frac{\chi^2_\mathrm{M}}{2 \eta}$~\cite{Parr1999} is related to yet another widely applied concept in chemical reactivity, the so-called electrophilic and nucleophilic reactions.

The electronic structure of the involved neutral, cationic, and anionic atoms is qualitatively different, and this also informs the choice of a particular computational strategy. In cations, orbitals are stabilized by the excess charge both spatially and energetically, while the opposite holds for anions. Shifting the orbital energy ladder up or down in energy may result in closing or widening of a relevant energy gap, thus in turn increasing or decreasing importance of electron correlation, depending on the parent neutral species. Spatial stabilization means the diffuse (low-exponent) basis functions are usually less important for cations. Anions, on the other hand, often only loosely bind the excess electron, rendering the use of diffuse functions indispensable. The weaker the binding, the more this is the case. An illustration of this can be found in Section~\ref{sub:SHE}. The excess electron in an anion asymptotically sees a neutral atom. Consequently, the electron correlation plays an important role in the properties of anions, and in particular in electron affinities~\cite{Pegg2004}.

In the following subsections, we offer two examples of calculations of IPs and EAs reaching units of meV accuracy.

\subsubsection{Gold}\label{sub:Au}

Gold has long been in the center stage of relativistic electronic structure theory due to its famously large enhancement of relativistic and QED effects breaking the smooth periodic trends in Group~11~\cite{Pyykko1979,ThiSch10}. 
Formally, Au is a single valence electron system with a deceivingly simple 6s$^1$ ground state configuration.
However, due to the strong relativistic 6s stabilization and the indirect 5d expansion~\cite{Pyykko1988}, the 5d/6s energy gap becomes small with a $^2\text{S}_{1/2}-^2\text{D}_{5/2}$ separation of only 1.14 eV. The resulting diffuse and polarizable 5d shell is responsible for the enhancement of relativistic effects within the Group 11 and 12 elements of the Periodic Table~\cite{Autschbach2002}. 

Experimentally, the IP and EA of gold were known to a high degree of accuracy already since the 1970s with the respective measured values of 9.22553(2)~eV~\cite{Brown1978} and 2.30861(3)~eV~\cite{Hotop1973,Hotop1985}. However, available theoretical values were long at odds with the experiment. 
Earlier state-of-the-art FSCC calculations~\cite{EliKalIsh94} with perturbatively included QED corrections~\cite{ThiSch10} revealed a rather large discrepancy of 0.16 and 0.05~eV for the IP and EA, respectively, compared to the experiment. This spurred a dispute in the relativistic electronic structure community about whether this is due to missing electron correlation in the positive energy spectrum or perhaps due to the neglect of correlating the negative energy states~\cite{Liu2013}. 
In our group's pioneering study~\cite{Pasteka:17}, we employed for the first time the composite calculation scheme accounting for all the relevant contributions in a systematic and balanced fashion as described in the present review. We were able to reach an agreement with the experiment at the meV level with our final values of IP(Au)~=~9.229~eV and EA(Au)~=~2.307~eV, thus resolving the long-standing discrepancy. 
Ultimately, the accurate treatment of electron correlation both in the baseline relativistic CC calculations (in the positive energy spectrum) as well as in the calculation of the Breit and QED contributions was the missing piece, allowing theory and experiment to reunite.
This computational protocol was later further developed, expanded, and applied to other systems (see, e.g., Refs. \cite{LeiKarGuo20, GuoPasEli21,GuoBorEli22,GuoPasNag24, HaaEliIli20}). We provide some such examples in the following sections.

\begin{figure}[h!]
\begin{center}
\includegraphics[width=\columnwidth]{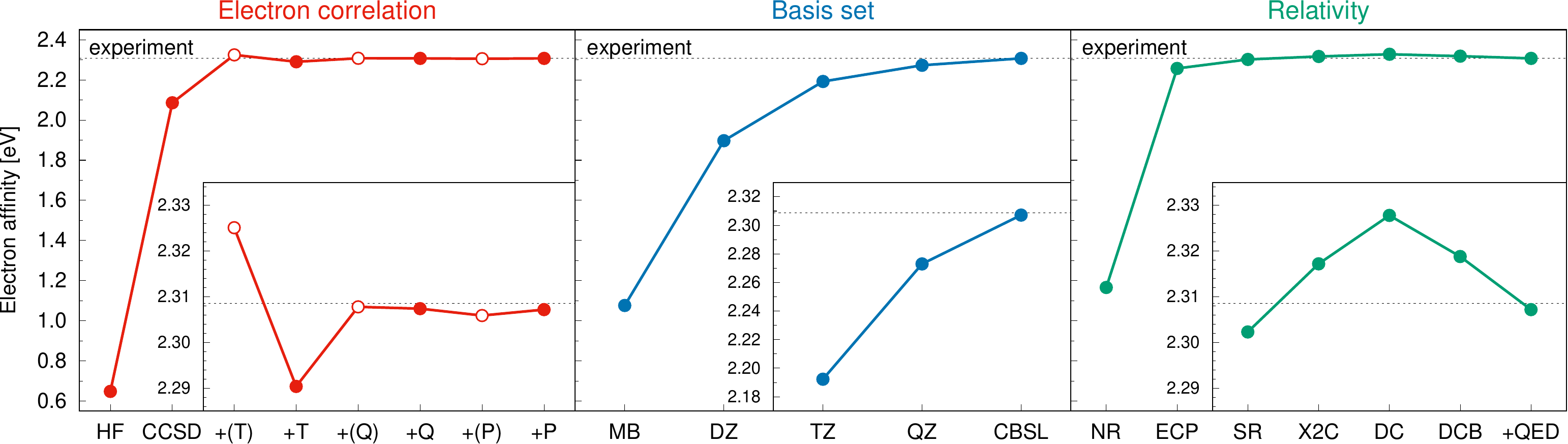}
\caption{Convergence of the electron affinity of gold following the hierarchical improvement of the three main classes of computational parameters -- electron correlation (left), basis set (middle), and relativistic treatment (right) -- as represented by the axes shown in Figure~\ref{3axes}. The insets in the left and right plots show the finer details of the higher-order contributions in a close-up view. 
Open and solid symbols in the left plot represent, respectively, the approximate perturbative and the full iterative treatment of a given excitation level.}
\label{fig:Au-plots}
\end{center}
\end{figure}

A detailed account of the individual contributions can be found in the original work~\cite{Pasteka:17}. Here, we focus on illustrating the concepts introduced in Section~\ref{sec:comp}.
In Figure~\ref{fig:Au-plots}, we showcase the convergence of the EA of gold along each major axis in our conceptual three-dimensional computational parameter space (Figure~\ref{3axes}). 
Trends for IP are qualitatively very similar to those of EA, hence we only focus on the electron affinity. 
In each plot in Figure~\ref{fig:Au-plots}, a single class of computational treatment is explored while the remaining parameters are included to the highest available level. 
In this example, we investigate the full length of each axis starting at the origin point of Figure~\ref{3axes},
in order to fully explore the behavior in the largest possible parameter space. Note, however, that this is somewhat artificial and serves an illustrative purpose only. Practical applications typically start at a much higher and more balanced level of theory, such as DC-CCSD with a 4z basis set (as represented by the black point in Figure~\ref{3axes}). This was also the case in the original work~\cite{Pasteka:17}. 

Most of the data points shown in Figure~\ref{fig:Au-plots} are based on the original data from Ref.~\cite{Pasteka:17}. In order to supplement the additional points to cover the full range for each computational axis, additional calculations were performed in a manner consistent with the original methodology. Nonrelativistic (NR), scalar-relativistic (SR, spin-free), and two-component X2C calculations were performed using the same basis sets as were used in the original work, while for the representative ECP calculation, we used the cc-pV$N$Z-PP basis set~\cite{Peterson2005} combined with the appropriate Stuttgart--Cologne pseudopotential~\cite{Figgen2005}. For the minimal basis calculations, we have constructed a pseudo v1z basis set for Au following the progression of Gaussian exponents in the original Dyall.v2/3/4z basis sets, resulting in a very modest basis comprising 19s, 14p, 8d, and 7f functions.

The starting point in each of the plots in Figure~\ref{fig:Au-plots} is similarly far from the final theoretical result and the experimental reference, highlighting that the relative importance of each parameter class is fairly balanced in this system. While this is not always the case for heavy atoms, this holds fairly generally. For unknown systems, we thus consider it a good practice to give balanced attention to the theoretical description of all three major directions.

A closer inspection of the electron correlation plot reveals that, indeed, the so-called gold standard of quantum chemistry, CCSD(T), delivers a solid theoretical prediction -- at least on the scale covering the entire set of results. This label is awarded to the method based on typical errors in molecular energetics below 1~kcal/mol (about 50 meV). However, in light of the meV accuracy goal, it is more meaningful to look at higher excitation levels beyond CCSD(T). As can be seen in the inset plot, the meV accuracy is first reached at the CCSDT(Q) excitation level, and including higher-order contributions offers very little improvement, with pentuple excitations contributing less than 1~meV. 
Based on our experience and detailed analyses in other systems, this conclusion can be generalized, and one could thus call CCSDT(Q) the platinum standard method.
This designation was also previously awarded to CCSDT(Q) in the context of molecular interaction and reaction energy calculations~\cite{Kodrycka2019,Lesiuk2022}.
Another common trend that can be observed in this example is the zig-zag pattern formed by the alternating perturbative and iterative treatment of a given excitation level (represented by the open and solid markers in the plot)~\cite{Helgaker2000,Simova2013}.
We also note that as we progress to higher excitation levels, the perturbative result provides a gradually worse approximation to the full iterative CC treatment. The highest perturbative correction used in this work, (P), strongly overestimates the full iterative pentuple correction, P, and in fact, $\Delta$(P) and $\Delta$P contribution almost perfectly cancel out.

Moving to the basis set investigation presented in the middle plot of Figure~\ref{fig:Au-plots}, it is clear that the convergence with the basis set cardinality is the slowest, compared to the other two computational parameters. Note that the extrapolated asymptotic CBS limit is shown as a finite point on the horizontal axis in Figure~\ref{fig:Au-plots}. In this work (as well as many subsequent works), we were limited by the available Dyall basis sets reaching only the 4z cardinality level. This resulted in the CBSL extrapolation being the dominant source of uncertainty in most of these studies. Recently, 5z Dyall basis sets were developed 
which will allow us to reduce the associated uncertainties further to achieve a more accurate and balanced description.
While in this example, we only discuss the basis set cardinality, in the following section on At (Section~\ref{sub:At}), other quality aspects of basis sets and the associated trends are illustrated. 

Finally, the relativistic treatment is shown in the right plot of Figure~\ref{fig:Au-plots}. As expected, we can initially observe a very steep improvement in the agreement between theory and experiment, going from nonrelativistic to relativistic calculations. Already, the scalar-relativistic value appears to reproduce the measurement remarkably well (as shown in the inset plot). This is, however, a result of a cancellation of errors. When spin-orbit effects are taken into account, DC theory overshoots the mark, and only the additional Breit and QED contributions finally correct the result. This cancellation is one of the generally recognized trends in relativistic electronic structure studies, and the DC relativistic theory is often being quipped as 101\%  correct~\cite{Labzowsky1999,Pyykk2011,Indelicato:11} due to QED contributions being roughly 1\%  in size as compared to the relativistic effects, but opposite in sign.

\begin{table}[h!]
\centering
\caption{Breit and QED contributions to the IP and EA of Au (meV).}
\setlength{\tabcolsep}{5pt}
\begin{tabular}
[c]{llrrrr}\hline\hline
&& \multicolumn{2}{c}{IP} & \multicolumn{2}{c}{EA}\\
&& DC-HF & $\delta$CCSD & DC-HF & $\delta$CCSD\\\hline
Breit && --12.3 & --2.9 & --5.2 & --4.7 \\ \hline
SE& LGO & --27.2 & & --11.7 & \\
& ENLO & --26.4 & & --11.4 & \\
& MLSO & --26.1 & --7.0 & --11.2 & --3.9\\
VP & U+KS & 5.3 & & 2.3 & \\
& MLSO & 4.9 & --0.2 & 2.1 & --0.1\\
SEVP & MLSO & 0.0 & 2.6 & 0.0 & 1.5\\
total & MLSO & --21.2 & --4.6 & --9.1 & --2.5\\\hline\hline
\end{tabular}
\label{tab:Au-QED}
\end{table}

At this point, it is also valuable to briefly discuss the choice of the model for the Breit and QED contributions. 
The difference between frequency-dependent Breit~\eqref{eq:Breit_omega} and its $\omega \rightarrow 0$ limit~\eqref{eq:Breit1} is typically entirely marginal. Indeed, for the case of IP and EA of gold, it accounts for an additional --0.5~meV and --0.1~meV, respectively, at the mean-field level. Compared to the full Breit contributions shown in Table~\ref{tab:Au-QED}, the frequency dependence is safe to neglect. Thus, the use of the simpler low-frequency approximation is generally preferred and fully justified. 
A more important factor to consider is whether one calculates these contributions at the mean-field or correlated level, as can be seen in Table~\ref{tab:Au-QED}.
In Table~\ref{tab:Au-QED}, a comparison of different QED contribution calculations is shown. In our study, we used the model Lamb shift operator (MLSO) of Shabaev and co-workers
~\cite{ShaTupYer15}. 
This model Hamiltonian uses the Uehling potential and an approximate Wichmann--Kroll term for the vacuum polarization (VP) potential~\cite{BLOMQVIST197295} and local and nonlocal operators for the self-energy (SE), the cross terms (SEVP) and the higher-order QED terms~\cite{Shabaev:13}. 
To test the validity of these results, we also carried out
perturbative QED calculations using the Uehling and K{\"a}ll{\'e}n--Sabry (U+KS)~\cite{BLOMQVIST197295,FulRin76}
terms (as implemented in GRASP~\cite{Parpia:96}) for the VP, and the effective nonlocal SE operator (ENLO) originally introduced by Ginges and Flambaum~\cite{flambaum2005radiative,ThiSch10}. 
Furthermore, we also include the more approximate perturbative SE values obtained by using the local Gaussian-type operator (LGO) of Pyykk{\"o}~\cite{pyykko2003search}.
The different models give very similar results at the mean-field level.
The implementation of the MLSO Hamiltonian into the Tel Aviv atomic computational package~\cite{TRAFS-3C} allowed us to calculate the QED contributions at the correlated DCB-FSCCSD level of theory. 
The overall Lamb shift of the ionization potential is --26~meV, with the CC contribution accounting for about 21\% . In the case of the EA, the overall Lamb shift is only half in size compared to the IP, but the relative CC contribution remains the same. This indicates that electron correlation contributions to QED cannot be neglected when seeking high accuracy.

\begin{figure}[h!]
\begin{center}
\includegraphics[width=\columnwidth]{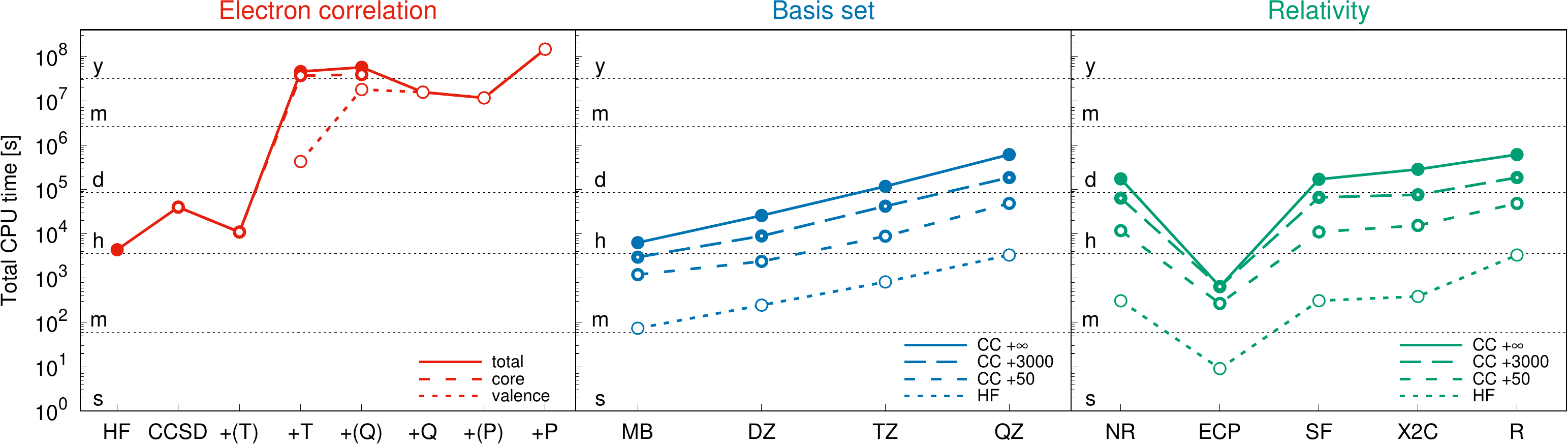}
\caption{(\textit{New figure}) Total CPU times involved in calculations of the electron affinity of gold at various levels of theory following hierarchical improvement of the three main classes of computational parameters -- electron correlation (left), basis set (middle), and relativistic treatment (right) -- as represented by the axes shown in Figure~\ref{3axes}.
The left plot shows the actual production CPU times from the original work~\cite{Pasteka:17} divided into core and valence correlation calculations. The middle and the right plots show scaling of HF and all-electron (except for ECP) CCSD(T) calculations, the latter with different virtual space cutoffs (in a.u.). In the middle plot, the four-component DC Hamiltonian was used in all calculations. In the right plot, the basis was fixed at the QZ level.}
\label{fig:Au-cpu}
\end{center}
\end{figure}

As a complement to Figure~\ref{fig:Au-plots}, we show a concrete breakdown of computational costs involved in calculations of EA of gold at different levels of theory in Figure~\ref{fig:Au-cpu}. The electron correlation plot shows the actual production CPU times from the original work~\cite{Pasteka:17}. Reaching as far as CCSDTQP was only possible by utilizing the composite approach with each successive layer of excitation treatment being gradually tapered in terms of the correlation space and basis set size. For details, we refer the reader to the original work~\cite{Pasteka:17}. As can be seen in the plot, the core correlation was only treated up to the CCSDT(Q) level, further only valence correlation calculations were computationally feasible. 
In the basis set and relativity plots, we offer an example of scaling of HF and all-electron CCSD(T) calculations with increasing basis set size or the level of relativistic treatment using the same basis sets as described above for Figure~\ref{fig:Au-plots}.
The scaling in terms of basis set size is steeper compared to the NR~$\rightarrow$~R scaling.  
Note that the ECP calculations are not strictly all-electron, as the core electrons are replaced by the pseudopotential. This results in a much smaller number of actually correlated electrons as well as a smaller basis set size leading to a significant dip in computational cost as can be seen in the plot.

\subsubsection{Astatine}\label{sub:At}

Astatine, being radioactive and also the rarest (and the heaviest) naturally occurring element on Earth~\cite{Asimov1953}, is relatively sparsely studied, especially compared to its lighter halogen homologs. The minute amounts produced artificially prevent the use of conventional spectroscopic tools. 
Nevertheless, isotope $^{211}$At receives strong attention from the radiopharmaceutical community as the most promising candidate for targeted alpha therapy~\cite{Zalutsky2011,Mulford199S, Teze2017}, due to its favorable half-life of about 7.2~h and its cumulative $\alpha$-particle emission yield of 100\% .
For the development of any such practical application, a solid understanding of the basic chemical properties of astatine is required~\cite{Wilbur2013}. 
Knowledge of the IP, EA, and the related electronegativity, softness, and electrophilicity is essential for the determination of the \textit{in vivo} reaction kinetics as well as the stability of the involved At-containing compounds, especially considering that many of the proposed applications involve an aqueous solution, in which the astatide anion At$^-$ readily forms.

Only as recently as 2013, the IP of astatine was measured using an online laser ionization spectroscopy experiment at CERN-ISOLDE radioactive ion beam facility~\cite{Catherall2017,Rothe2013}, where the experimental results were confirmed by relativistic CCSD(T) calculations. More recently still, an effort was made at the same facility to also measure astatine's EA by means of online laser photodetachment threshold spectroscopy in direct collaboration with our group providing the theoretical prediction~\cite{LeiKarGuo20}.
Unlike in the gold case study (Section~\ref{sub:Au}), here, the theoretical prediction was conducted in parallel to the experimental measurement in an independent, mutually blinded fashion. Removing the target liberated the theory of bias and simultaneously stressed the importance of the \textit{a priori} confidence in the result. 

As in all halogens, the At$^-$ anion is particularly stable due to the simple closed-shell 6p$^6$~$^1$S$_0$ noble-gas-like configuration. The ground state of the neutral At atom is 6p$^5$~$^2$P$_{3/2}$. This is entirely analogous to the light homolog, iodine, which we used as the control system for our study, since its EA was known to high precision (3.059046(4)~eV~\cite{Pel_ez_2009}). The trend of decreasing EAs in the halogen group from Cl to I was expected to continue for At due to the increase in the principle quantum number and the further relativistic destabilization of the p$_{3/2}$-hole.

\begin{table}[h!]
    \centering
    \caption{Electron affinities (eV) of iodine and astatine gradually improving as term-by-term contributions are taken into account within the three major classes of computational parameters. An associated uncertainty is given for each class. The basis set class results are shown calculated at the DC-CCSD level. We use the shorthand notation (34)z to represent CBSL extrapolated results using 3z and 4z basis sets.}
    \begin{tabular}{llllll}
         \hline\hline
         class&  contribution&  EA(I)&  uncert.&  EA(At)& uncert.\\
         \hline
         basis set&  v2z&  2.180&  &  1.535& \\
         &  v3z&  2.707&  &  2.056& \\
         &  v4z&  2.885&  &  2.220& \\
         &  v(34)z&  2.984&  &  2.309& \\
         &  ae(34)z&  2.961&  &  2.296& \\
         &  d-aug-ae(34)z&  2.961&  $\pm$ 0.015&  2.309& $\pm$ 0.015\\
         \hline
         electron&  +$\Delta$(T)&  3.041&  &  2.401& \\
         correlation&  +$\Delta$T&  3.045&  &  2.404& \\
         &  +$\Delta$(Q)&  3.049&  &  2.408& \\
         &  +$\Delta$Q&  3.049&  $\pm$ 0.004&  2.408& $\pm$ 0.004\\
         \hline
         relativity&  +Breit&  3.052&  &  2.411& \\
         &  +QED&  3.055&  $\pm$ 0.003&  2.414& $\pm$ 0.003\\
         \hline
         theory final&  &  \multicolumn{2}{l}{3.055(16)}&  \multicolumn{2}{l}{2.414(16)} \\
         experiment&  &  \multicolumn{2}{l}{3.059046(4)~\cite{Pel_ez_2009}}&  \multicolumn{2}{l}{2.41578(7)~\cite{LeiKarGuo20}} \\
         \hline\hline
    \end{tabular}
    \label{tab:At}
\end{table}

In Table~\ref{tab:At}, a summary of the calculated contributions to EAs of iodine and astatine is collected. Here, we show the gradual evolution of the total EAs as the individual contributions are added term by term to the baseline calculation, bringing the total all the way to the final recommended value.
The three major classes of computational parameters are followed roughly in order of importance -- basis set, electron correlation, and relativistic corrections. For each class, the associated uncertainty is shown as well.
The trends of the calculated contributions are remarkably similar for the two elements. Even the uncertainties, although determined independently, are almost identical. We observe a very fast convergence regarding the excitation order of the CC correlation treatment. Breit and QED corrections are also very similar and rather small. The dominant source of uncertainty is the error associated with the CBSL extrapolation in terms of basis cardinality.  

Accounting for all contributions, we arrived at the final EA values of 3.055(16)~eV and 2.414(16)~eV for I and At, respectively, with uncertainties determined in the manner described in the previous sections . Good agreement with the measured iodine value of 3.059~eV~\cite{Pel_ez_2009} supported our result. 
Once the experiment and theory were able to compare the results for At, a remarkable agreement was revealed between the two. This comparison showed that the procedure to determine uncertainties tends to err on the conservative side, resulting in, perhaps, an overly cautious total uncertainty estimate, especially considering that the actual agreement with the experimental value of 2.41578(7)~eV is much stronger.

Interestingly, the Mulliken scale electronegativity of At determined from the EA and IP to be $\chi_\mathrm{M}= $ 5.87~eV  is significantly lower than that of hydrogen, $\chi _\mathrm{M}=$ 7.18~eV, supporting the calculated bond polarization towards the hydrogen atom in the H--At molecule~\cite{Pilm2014,Saue1996}, resulting in the suggested nomenclature flip from hydrogen halide to halogen hydride. 
Additionally, the value of $\chi_\mathrm{M}$(At) lies between the electronegativities reported for boron (4.29~eV) and carbon (6.27~eV) atoms, which is of high relevance to the use of At in nuclear medicine.

It is instructive to use this case study to investigate the basis set trends in more detail following Figure~\ref{fig:At-basis}. The different sub-families of Dyall basis sets (valence, core-valence, all-electron, augmented) exhibit a different rate of convergence towards the CBS limit. Notably, though, the CBS limit itself is independent of the basis set type, adding confidence to the CBSL extrapolation scheme. Basis sets augmented by diffuse functions perform significantly better in the CBSL convergence compared to the valence basis set. In the inset plot of Figure~\ref{fig:At-basis}, differences with respect to the valence basis sets are shown. These illustrate well the relative importance of adding the core-correlating and diffuse functions and their relative trends. Augmentation corrections are much larger and strongly depend on the cardinality, while the core-correlating functions are almost entirely decoupled from cardinality and showcase near-additive behavior.

\begin{figure}[h!]
\begin{center}
\includegraphics[scale=0.35]{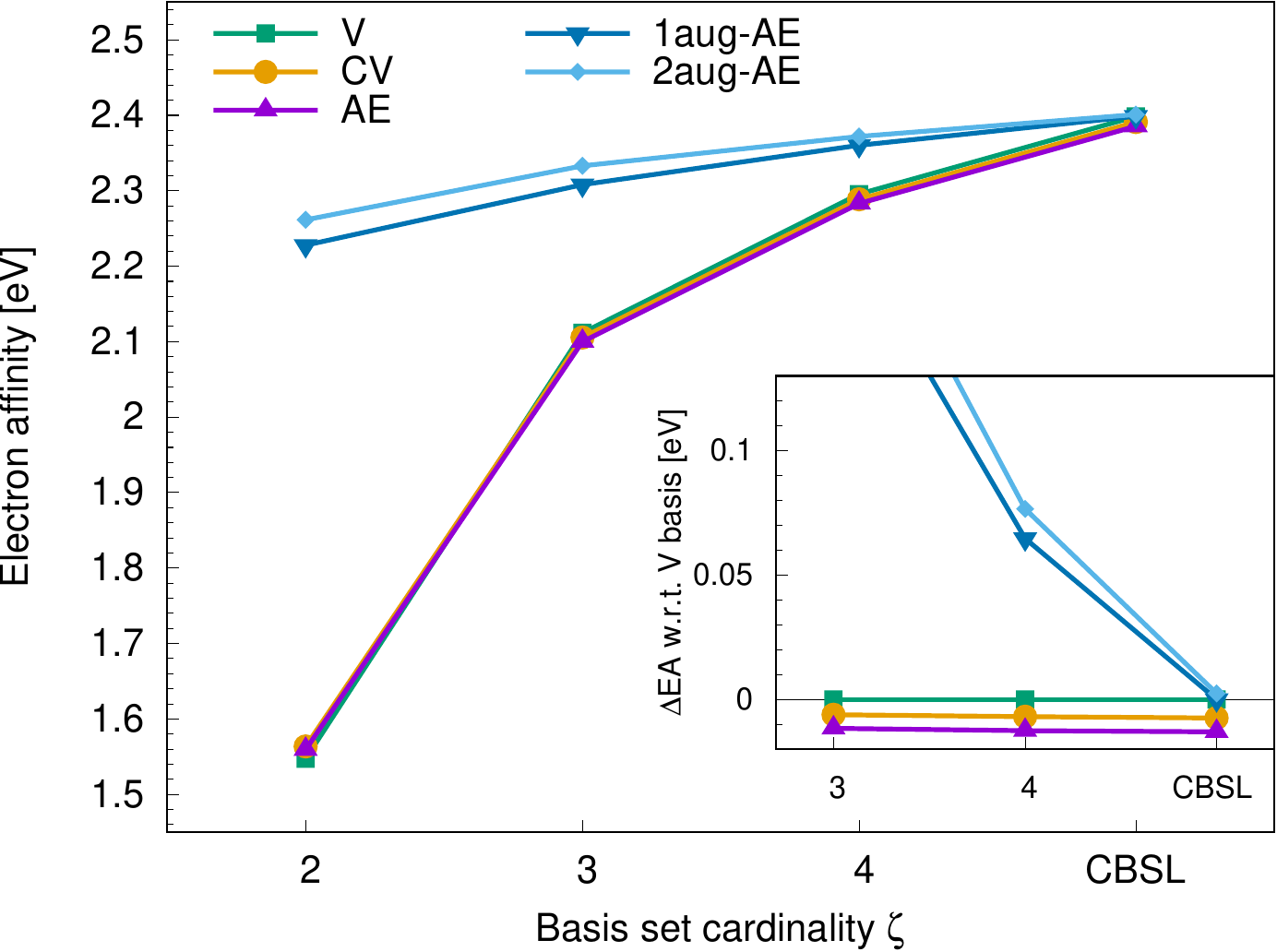}
\caption{Convergence of the electron affinity of At with the increasing basis cardinality. Calculated at the CCSD(T) level of theory using different basis set families. Basis sets are labeled V for valence (Dyall.v$N$z), CV for core-valence (Dyall.cv$N$z), AE for all-electron (Dyall.ae$N$z), prefixes 1aug- and 2aug- correspond to basis set augmented with a single and a double layer of diffuse functions, respectively. Inset shows corrections with respect to the valence basis set baseline.}
\label{fig:At-basis}
\end{center}
\end{figure}

\newpage

\subsection{Nuclear properties from hyperfine structure}\label{sub:HFS}

Atomic spectroscopy experiments can provide access to nuclear properties of exotic elements and isotopes, including the nuclear moments and charge radii.  Such measurements are often accompanied by theoretical investigations necessary for the interpretation of their results.
This section describes how measurements of isotope shifts and hyperfine structure (HFS) of atomic transitions, combined with theoretical input, can be used to obtain information on nuclear properties.
 
\begin{figure}[h!]
     \center
     \includegraphics[scale=0.35]{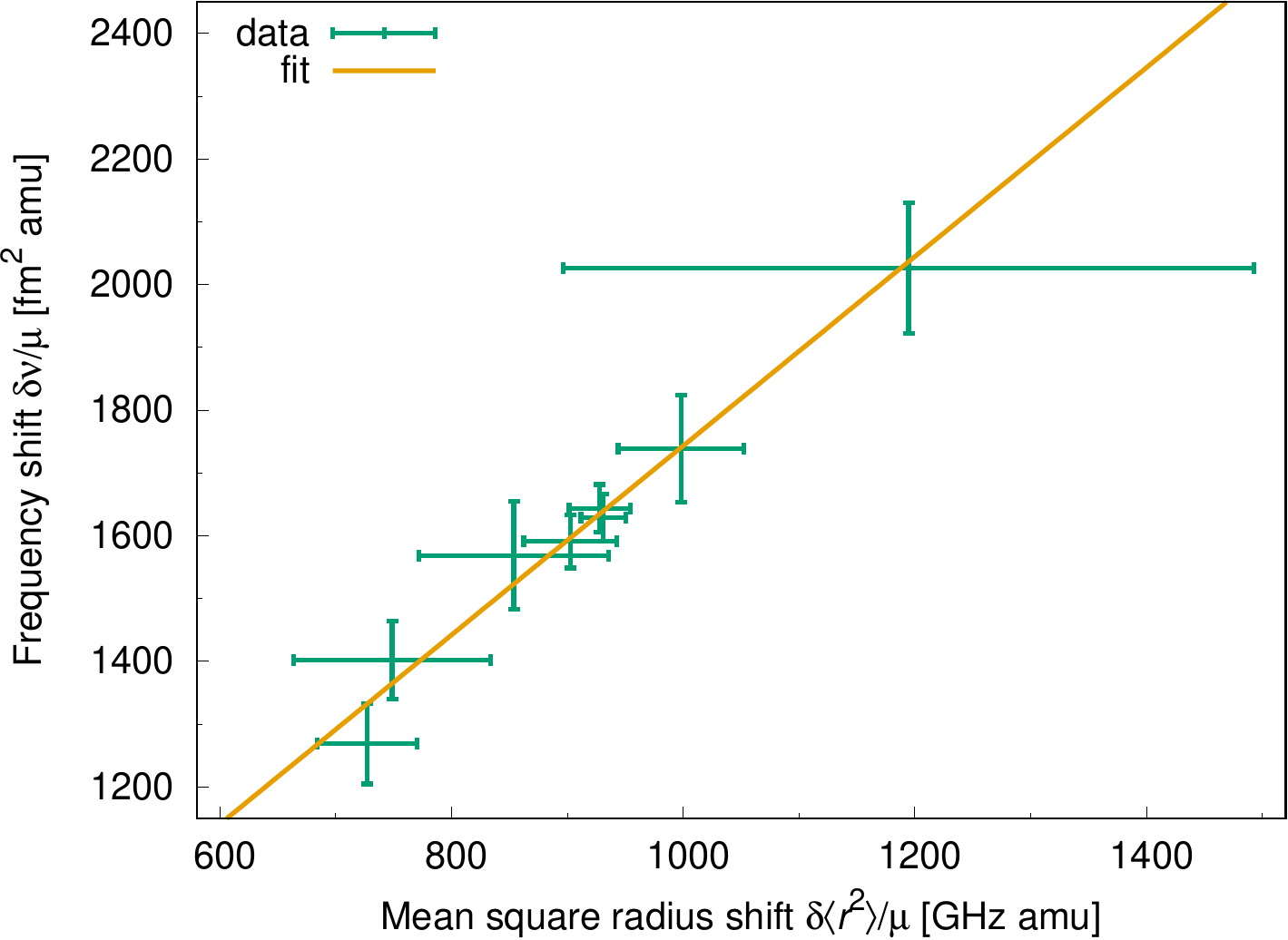}
     \caption{King plot of the $^{1}S _{0} \rightarrow {}^{1}P_1$ transition in tin with $^{120}$Sn as the reference isotope. Based on data from Ref. \cite{GusRicRei20}.}
     \label{fig:kingplot}
 \end{figure}

The isotope shift of an electronic transition between isotopes with mass numbers $A'$ and $A$ can be expressed as
\begin{equation} \label{eq:isotopeshift}
     \delta \nu^{A',A} = F \delta \langle r^2 \rangle^{A',A} + k_\text{MS} \left( \frac{1}{A'} - \frac{1}{A} \right),
\end{equation}
where $\delta \nu^{A',A} = \nu^{A'} - \nu^{A}$ is the frequency shift, $\delta \langle r^2 \rangle^{A',A}$ is the differential mean square charge radius, and $F$ and $k_\text{MS}$ are the field- and mass shift factors. If measurements for a sufficient number of isotopes are available, one can use a King plot~\cite{King:84} to obtain the isotope shift factors, where the linear relation in Eq.~\eqref{eq:isotopeshift} is plotted for the measured frequencies of the same transition $\nu^{A',A}$ for different isotopes to obtain the isotope shift factors. As an example, the King plot of the $^{1}S _{0} \rightarrow {}^{1}P_1$ transition in several tin isotopes from Ref.~\cite{GusRicRei20} is shown in Figure~\ref{fig:kingplot}. Then, the field shift and the mass shift factors are given by the slope and the intercept of the line in Figure~\ref{fig:kingplot}, respectively. These are in turn usually used to extract nuclear radii from new measurements on unstable isotopes.  However, using a King plot is not always possible for rare elements where the number of available isotopes, or the number of measured transitions,  is severely limited. In that case, the field and mass shift factors can be provided by electronic structure theory, permitting the extraction of charge radii.
 
The leading order contributions to the measured HFS splitting $\Delta E_\text{hf}$ are described by~\cite{Hertel:15}
\begin{equation}
    \Delta E_\text{hf} = \frac{A_\text{hf} C}{2} + \frac{B_\text{hf}}{8} \frac{3C(C+1) - 4IJ(I+1)(J+1)}{IJ(2I-1)(2J-1)},
\end{equation}
where $C = F(F+1) - J(J+1) - I(I+1)$ and $F$ takes values $|J-I| \leq F \leq J+I$ in integer steps, for nuclear spin $I$ and total electronic angular momentum $J$. The first term is related to the nuclear magnetic dipole moment $\mu_I$ by 
\begin{equation}\label{eq:hfs_A}
 A_\text{hf} = \frac{\mu_I}{IJ} B_\text{e},
\end{equation} 
where the magnetic field produced by the electrons at the nucleus, $B_\text{e}$ , is a purely electronic factor. It is common to absorb $J$ into the factor and refer to $A_0 = B_\text{e}/J$ instead. Similarly, the second term probes the spectroscopic electric quadrupole moment of the nucleus $Q_\text{s}$ as
\begin{equation}\label{eq:hfs_B}
B_\text{hf} = e q_{zz} Q_\text{s},
\end{equation} 
with the electronic electric field gradient at the nucleus $q_{zz} = \left< \frac{\delta^2 V_\text{e}}{\delta z^2} \right>$. 
In the case where the nuclear moments $\mu_I$ and $Q_\text{s}$ as well as $A_\text{hf}$ and $B_\text{hf}$ are already known for an isotope of the element in question, the nuclear moments $\mu_I'$ and $Q_\text{s}'$ for a different isotope can be determined from the ratios of the HFS parameters of the two isotopes. It should be noted that this approach neglects the hyperfine anomaly that arises due to finite nuclear magnetization \cite{BohWei50}. 
However, when there is no information available on another isotope, such as in the case of heaviest elements, $A_0$ ($B_\text{e}$) and $q_{zz}$ have to be supplied by electronic structure calculations in order to extract $\mu_I$ and $Q_\text{s}$. 
Furthermore, information on the distribution of the nuclear magnetization (i.e. the Bohr--Weisskopf effect~ \cite{KARPESHIN201566}) can be extracted from the measured hyperfine anomaly using  precise atomic~\cite{GinVol18, RobGin21}  (and recently even molecular~\cite{Skri20}) calculations. 

Within the FSCC framework, the finite field method is used for the calculations of these properties (see Section \ref{sub:prop} and Ref. \cite{HaaEliIli20} for details). 
Some recent examples of experiments where relativistic coupled cluster calculations were used to extract nuclear moments and charge radii  measurements are presented below. 

\subsubsection{Isotope shifts and hyperfine structure of Sn}
Tin has the largest number of stable isotopes of all known elements, as it has a nucleus with a closed shell for protons according to the nuclear shell model ($Z = 50$). At the same time, it is the heaviest element with two available isotopes that have both a closed proton and a closed neutron shell: $^{100}$Sn and $^{132}$Sn. These attributes make tin an excellent candidate for studying the evolution of nuclear charge radii and nuclear moments. In 2020, high-precision spectroscopic measurements of a number of transitions  in atomic tin were carried out using the collinear resonance-ionization spectroscopy technique (CRIS) at the ISOLDE facility of CERN~\cite{GusRicRei20}. These measurements were accompanied by FSCC calculations of the field shifts and the hyperfine parameters of the investigated transitions.

The calculations were performed using the X2C Hamiltonian \cite{Ilias:07}. The performance of this approximation was tested for hyperfine parameters by comparing the X2C and four-component DC results obtained within a small basis set and model space. The results differed by 0.5\%  only, justifying the use of this approach in the rest of this work. 

The  recommended values of the HFS constants and field shifts were obtained using the cv4z basis set set of Dyall \cite{Dyall:06}, augmented by four layers of diffuse functions in each symmetry, constructed in an even-tempered fashion (q-aug-cv4z). The importance of using diffuse functions, in particular for the highest levels, can be seen in Table \ref{tab:Sn_uncertainties_aug}. Neutral tin has an open-shell ground-state electron
configuration [Kr]4d$^{10}$5s$^2$5p$^2$. The calculation thus started from the closed-shell Sn$^{2+}$ system ([Kr]4d$^{10}$5s$^2$) and two electrons were added to the orbitals that comprise the model space. A very large model space was used, consisting of the 5p 6s (5d 6p 7s 4f 6d 7p 8s 5f 7d 5g 8p 9s 6f 8d 9p 6g 6h 7g 10p 10s 7f) orbitals, where the orbitals in parentheses are in the intermediate space $P_i$. All the electrons were correlated and virtual orbitals with energies up to 500 a.u. were included in the correlation space.  

\begin{table}[h!]
    \centering
    \caption{Basis augmentation effect on the field shift F$\text{FSCC}$, magnetic field constant $A_0^\text{FSCC}$ and electric field gradient $q_{zz}^\text{FSCC}$ of excited states of Sn. Based on data from Ref. \cite{GusRicRei20}.}
    \begin{tabular}{@{\extracolsep{4pt}}l r r r r r r r r r}   
    \hline\hline
    \multirow{2}{*}{State} & \multicolumn{3}{c}{F$^\text{FSCC}$} & \multicolumn{3}{c}{$A_0^\text{FSCC}$ (MHz)} & \multicolumn{3}{c}{$q_{zz}^\text{FSCC}$ (MHz/b)} \\ 
    \cline{2-4}\cline{5-7}\cline{8-10}
    & t-cv4z & q-cv4z & $\Delta$  & t-cv4z & q-cv4z & $\Delta$ & t-cv4z & q-cv4z & $\Delta$ \\
    \hline
    $5p^2 \; ^3P_1$ & --3492.3 & --3491.5 & --0.02\%  & --253.2 & --249.7 & --1.37\%  & 416.6 & 416.8 & 0.05\%  \\
    $5p^2 \; ^3P_2$ & --3509.8 & --3508.5 & --0.04\%  & 556.3 & 564.3 & 1.43\%  & --692.5 & --693.0 & 0.07\%  \\
    $5p6s \; ^3P_1$ & --1275.7 & --1275.6 & --0.01\%  & 2240.2 & 2255.7 & 0.69\%  & --150.2 & --145.4 & --3.26\%  \\
    $5p6s \; ^3P_2$ & --1311.3 & --1311.8 & 0.04\%  & 742.3 & 748.4 & 0.81\%  & 987.1 & 986.7 & --0.04\%  \\
    $5p6s \; ^1P_1$ & --1404.1 & --1400.3 & --0.27\%  & 120.6 & 130.4 & 7.48\%  & 638.8 & 635.7 & --0.49\%  \\
    $5p7s \; ^1P_1$   & --1914.0 & --1863.2 & --2.72\%  & --43.1 & 554.3 & 107.78\%  & 315.0 & 373.2 & 15.59\%  \\
    \hline\hline
    \end{tabular}
    \label{tab:Sn_uncertainties_aug}
\end{table}
To estimate the theoretical uncertainties an extensive investigation of the effect of different computational parameters on the calculated properties was performed, following the procedure outlined in Section \ref{sub:error}. 
The total conservative uncertainty estimates on the calculated values was about 4\%  for all the states and properties except for the highest lying 5p7s  and 5p6s $^1$P$_1$ states, for which the uncertainties are 8--18\% , depending on the property. 
The uncertainties in this system are dominated by the basis set and correlation effects. 

Table \ref{tab:Sn_F} presents the calculated field shifts of the measured transitions, while Table \ref{tab:Sn_Avaluecompare} shows the final recommended values for the hyperfine-structure constants of the investigated states in Sn, including the corresponding uncertainties and compared to the measured values.
\begin{table}
  \caption{Calculated and experimental field shift factors (MHz/fm$^2$). Based on data from Ref. \cite{GusRicRei20}.}
    \label{tab:Sn_F}
    \centering
    \begin{tabular}{ccc}
    \hline
    \hline
   Transition      & F$^\text{FSCC}$ &F$^\text{Exp}$ \\
   \hline
    $^1 S_0 \rightarrow\ ^1P_1$     & 1552(233) &  1584(209)\\
    $^3 P_2 \rightarrow\ ^3P_2$     &  2217(89)&  2024(184)\\
   $^3 P_1 \rightarrow\ ^3P_2$        & 2200(88) & 2323(395)\\
  $^3 P_1 \rightarrow\ ^3P_1$        &2104(84)  & 2932(1083)\\
   $^3 P_0 \rightarrow\ ^3P_1$        & 2202(88) &2831(546) \\
   \hline
   \hline
    \end{tabular}
\end{table}

The overall agreement between the calculated and experimental values (obtained via the King plot procedure) of the field shifts for each investigated transition is very good, demonstrating the excellent performance of the FSCC approach for this property. 
The measured isotope shifts were used, in combination with FSCC calculation, to extract independent mean-squared nuclear charge radii for $^{112-124}$Sn  using the radius of the reference isotope $^{120}$Sn.

The calculated $A_0$ constants are also found to be in excellent agreement with the measured values, well within the combined uncertainties. The good performance of FSCC for the $A_0$ allows us to expect similar accuracy for the predicted electric field gradients, $q_{zz}$ (Table \ref{tab:Sn_Avaluecompare}). These so far could not be obtained  experimentally since tin has no stable isotopes with $I>\frac{1}{2}$. The $q_{zz}$ predictions provided the sensitivity of the different atomic states to the nuclear quadrupole moments, which serves as a useful foundation for future investigations of short-lived exotic tin isotopes.

\begin{table}[h!]
    \centering
    \caption{Electronic hyperfine structure parameters for excited states of Sn. Experimentally extracted $A_0$ values (in MHz) are compared with FSCC calculations. The $A_0^\text{exp}$ values were determined using magnetic moments from Ref.~\cite{Yordanov2020}. 
    The predictions for the electric field gradients $q_{zz}^\mathrm{FSCC}$ are shown in the last column. Based on the data from Ref. \cite{GusRicRei20}.}
    \label{tab:Sn_Avaluecompare}
    \begin{tabular}{lrrr}
    \hline \hline
    Level            & $A_0^\text{exp}\phantom{.....}$  & $A_0^\mathrm{FSCC}$\phantom{.} & $q_{zz}^\mathrm{FSCC}\phantom{.}$ \\ 
    \hline 
    5p$^2 \; ^3$P$_1$  & --278(45)   & --257(10)\phantom{0} & 419(17) \\
    5p$^2 \; ^3$P$_2$  & 607(4)\phantom{0}      & 598(24)\phantom{0}  & --691(28)\\
    5p6s$ \; ^3$P$_1$  & 2402(11)    & 2352(94)\phantom{0} & --152(6)\phantom{0} \\
    5p6s$ \; ^3$P$_2$  & 777(5)\phantom{0}      & 783(31)\phantom{0}  & 990(40) \\
    5p6s$ \; ^1$P$_1$  & 127(19)    & 145(13)\phantom{0}  & 645(58) \\
    5p7s$ \; ^1$P$_1$  & 638(8)\phantom{0}      & 571(103) & 378(68) \\
    \hline \hline
    \end{tabular}
\end{table}

\subsubsection{Nuclear moments of Ge}
The hyperfine structure of the 4s$^2$4p$^2$~$^3$P$_1\,\rightarrow\,4$s$^2$4p5s~$^3$P$_1$ transition of $^{69,71,73}$Ge isotopes was measured using collinear laser spectroscopy at the ISOLDE facility at CERN \cite{KanYanBis20}, in order to investigate the moments of these nuclei. The measurements were accompanied by FSCC calculations of the HFS constants, $A_0$ and $q_{zz}$. The computational scheme used for tin in the section above was also adopted for these calculations. Here, Ge$^{2+}$, with a [Ar]3d$^{10}$4s$^2$ configuration, was used as the starting point and two electrons were added to reach neutral Ge, corresponding to the (0,2) sector of FSCC; the model space comprised 4p 5s (4d 5p 6s 4f 5d 6p 7s 5f 5g 7p 6d 7d 8p 6g 8s 6f) orbitals.  Besides the two levels directly involved in the transition, also the 4s$^2$4p$^2$~$^3$P$_2$ level was investigated.

To estimate the uncertainties of the calculated values, an investigation of the effect of various computational parameters was performed, treating the main sources of uncertainty (the limited size of the basis set and the missing correlation and higher-order relativistic effects) separately. These sources of error are considered to be independent and the corresponding uncertainties are combined to give a total conservative uncertainty estimate. 

To determine the size of the uncertainties corresponding to the different effects, calculations were performed at different levels of theory. The uncertainty was then determined by taking the difference between the results of the best and the second best calculation for the parameter under investigation, while keeping all the other parameters fixed.

The basis set uncertainty, $\Delta_\text{bas}$, consists of three components: the basis set cardinality, $\Delta_{\text{bas}}^{\text{card}}$, the number of added layers of diffuse functions, $\Delta_{\text{bas}}^{\text{aug}}$, and the number of functions included for the core correlation, $\Delta_{\text{bas}}^{\text{core}}$. 
The uncertainty from the basis set cardinality is determined by taking the difference of the results obtained using the 3z and 4z level basis sets, while keeping all the other parameters fixed. The d-cv3z and d-cv4z results are shown in Table~\ref{tab:Ge-unc-bas-car} together with the uncertainty, $\Delta_{\text{bas}}^{\text{card}}$, derived from their difference. $\Delta_{\text{bas}}^{\text{card}}$ is quite small (0.1--1\% ) for each level and property, indicating that the calculated values are already converged at the 3z level. 
\begin{table}[ht]
    \caption{Calculated energies, $E$, and hyperfine parameters, $A_0$ and $q_{zz}$, obtained using the d-cv3z and d-cv4z basis sets together with the respective absolute differences $\Delta_{\text{bas}}^{\text{card}}$.}
    \label{tab:Ge-unc-bas-car}
    \centering
    \begin{tabular}{@{\extracolsep{4pt}} l rrr rrr rrr}  
    \hline
    \hline
    & \multicolumn{3}{c}{E (cm$^{-1}$)} & \multicolumn{3}{c}{$A_0$ (MHz)} & \multicolumn{3}{c}{$q_{zz}$ (MHz/b)} \\
    \cline{2-4} \cline{5-7} \cline{8-10}
    Level & d-cv3z & d-cv4z & $\Delta_{\text{bas}}^{\text{card}}$ & d-cv3z & d-cv4z & $\Delta_{\text{bas}}^{\text{card}}$ & d-cv3z & d-cv4z & $\Delta_{\text{bas}}^{\text{card}}$ \\
    \hline
    4s$^2$4p$^2~^3$P$_1$& 554.3 & 554.0 & --0.3 & --73.8 & --72.8 & 0.9 & 276.6 & 276.3 & --0.2 \\
    4s$^2$4p$^2~^3$P$_2$& 1398.1 & 1396.4 & --1.7 & 319.8 & 321.5 & 1.7 & --563.4 & --563.0 & 0.5 \\
    4s$^2$4p5s~$^3$P$_1$   & 37364.7 & 37387.8 & 23.1 & 1307.3 & 1312.1 & 4.8 & --202.9 & --202.1 & 0.9 \\
    \hline
    \hline
    \end{tabular}
\end{table}

The augmentation level has a much higher effect on the calculated constants (Table~\ref{tab:Ge-unc-bas-aug}). In this case the uncertainty, $\Delta_{\text{bas}}^{\text{aug}}$, is slightly larger for the HFS constants of the highest $4s^24p5s~^3P_1$ state, resulting in a relative uncertainty of over 5\% for its $q_{zz}$. It appears that this excited state could benefit from a better description of the outer part of the wave function. It should be noted that a smaller model space was used to obtain the results in Table~\ref{tab:Ge-unc-bas-car} compared to those in Table~\ref{tab:Ge-unc-bas-aug}. As a consequence, the d-cv4z results in the two tables are different.

\begin{table}[ht]
    \caption{Calculated energies, $E$, and hyperfine parameters, $A_0$, and $q_{zz}$, obtained using the d-cv4z and t-cv4z basis sets together with the respective absolute differences $Delta_{\text{bas}}^{\text{aug}}$.}
    \label{tab:Ge-unc-bas-aug}
    \centering
    \begin{tabular}{@{\extracolsep{4pt}} l rrr rrr rrr}  
    \hline
    \hline
    & \multicolumn{3}{c}{E (cm$^{-1}$)} & \multicolumn{3}{c}{$A_0$ (MHz)} & \multicolumn{3}{c}{$q_{zz}$ (MHz/b)} \\
    \cline{2-4} \cline{5-7} \cline{8-10}
    Level & d-cv4z & t-cv4z & $\Delta_{\text{bas}}^{\text{aug}}$ & d-cv4z & t-cv4z & $\Delta_{\text{bas}}^{\text{aug}}$ & d-cv4z & t-cv4z & $\Delta_{\text{bas}}^{\text{aug}}$ \\
    \hline
    4s$^2$4p$^2~^3$P$_1$& 554.9 & 555.1 & 0.2 & --75.6 & --73.4 & 2.2 & 276.4 & 276.7 & 0.2 \\
    4s$^2$4p$^2~^3$P$_2$& 1397.5 & 1398.1 & 0.6 & 317.2 & 321.1 & 3.9 & --563.1 & --563.4 & --0.3 \\
    4s$^2$4p5s~$^3$P$_1$   & 37361.7 & 37356.5 & --5.2 & 1297.5 & 1315.0 & 17.4 & --212.2 & --200.8 & 11.4 \\
    \hline
    \hline
    \end{tabular}
\end{table}

To analyze the effect of basis functions for correlation, the results obtained within the d-aug-cv4z and d-aug-ae4z basis sets were compared, where the latter contains additional f and g functions for the correlation of the 1s, 2s and 2p orbitals. The resulting relative uncertainties of a few tenths of a percent,  are very small compared to the other basis set effects.

An important source of uncertainty related to correlation is the size of the model space. An estimate of the subsequent uncertainty, $\Delta_{\text{cor}}^{P}$, is obtained from a comparison of results within the model space used for the final values $P_f$ and a smaller model space, $P_s$, consisting of 4p 5s (4d 5p 6s 4f 5d 6p 7p 6d 7s 5g 5f) orbitals, shown in Table~\ref{tab:Ge-unc-modelspace}. The effect of the model space on the energy of the highest level is the largest compared to the other sources of uncertainty. It does not affect the HFS constants as much, however, which is important since these were used for the extraction of the nuclear moments.

\begin{table}[ht]
    \caption{Calculated energies, $E$, and hyperfine parameters, $A_0$, and $q_{zz}$, obtained within the $P_f$ and $P_s$ model spaces together with the respective absolute differences $\Delta_{\text{cor}}^P$.}
    \label{tab:Ge-unc-modelspace}
    \centering
    \begin{tabular}{@{\extracolsep{4pt}} l rrr rrr rrr}  
    \hline
    \hline
    & \multicolumn{3}{c}{E (cm$^{-1}$)} & \multicolumn{3}{c}{$A_0$ (MHz)} & \multicolumn{3}{c}{$q_{zz}$ (MHz/b)} \\
    \cline{2-4} \cline{5-7} \cline{8-10}
    Level & $P_s$ & $P_f$ & $\Delta_{\text{cor}}^{P}$ & $P_s$ & $P_f$ & $\Delta_{\text{cor}}^{P}$ & $P_s$ & $P_f$ & $\Delta_{\text{cor}}^{P}$ \\
    \hline
    4s$^2$4p$^2~^3$P$_1$ & 555.1 & 554.5 & --0.7 & --73.3 & --75.0 & --1.7 & 276.7 & 276.5 & --0.2 \\
    4s$^2$4p$^2~^3$P$_2$ & 1397.9 & 1397.9 & 0.0 & 321.4 & 318.7 & --2.7 & --563.2 & --563.6 & --0.4 \\
    4s$^2$4p5s~$^3$P$_1$   & 37451.8 & 37330.3 & --121.5 & 1311.7 & 1302.7 & --9.0 & --207.9 & --210.4 & --2.4 \\
    \hline
    \hline
    \end{tabular}
\end{table}

The uncertainty due to the neglect of higher-order relativistic effects for a light element such as germanium are expected to be small, but can sometimes be important for HFS constants nonetheless. The size of the contribution of the Gaunt term was taken as the uncertainty, $\Delta_{\text{rel}}$, and found to be negligible relative to the other contributions.

Finally, the contribution from higher-order excitations was included at a fixed estimate of 3\% , based on earlier estimates of the  size of pertubative triples (CCSD(T) -- CCSD) contributions to HFS constants and other properties~\cite{Borschevsky2015, Pasteka:17, HaaEliIli20, GuoBorEli22}. In later works explicit calculations of the contribution of triple excitations are performed, using the EXP-T program~\cite{EXPT:20}.

\begin{figure}[h!]
    \centering
    \includegraphics[scale=0.35]{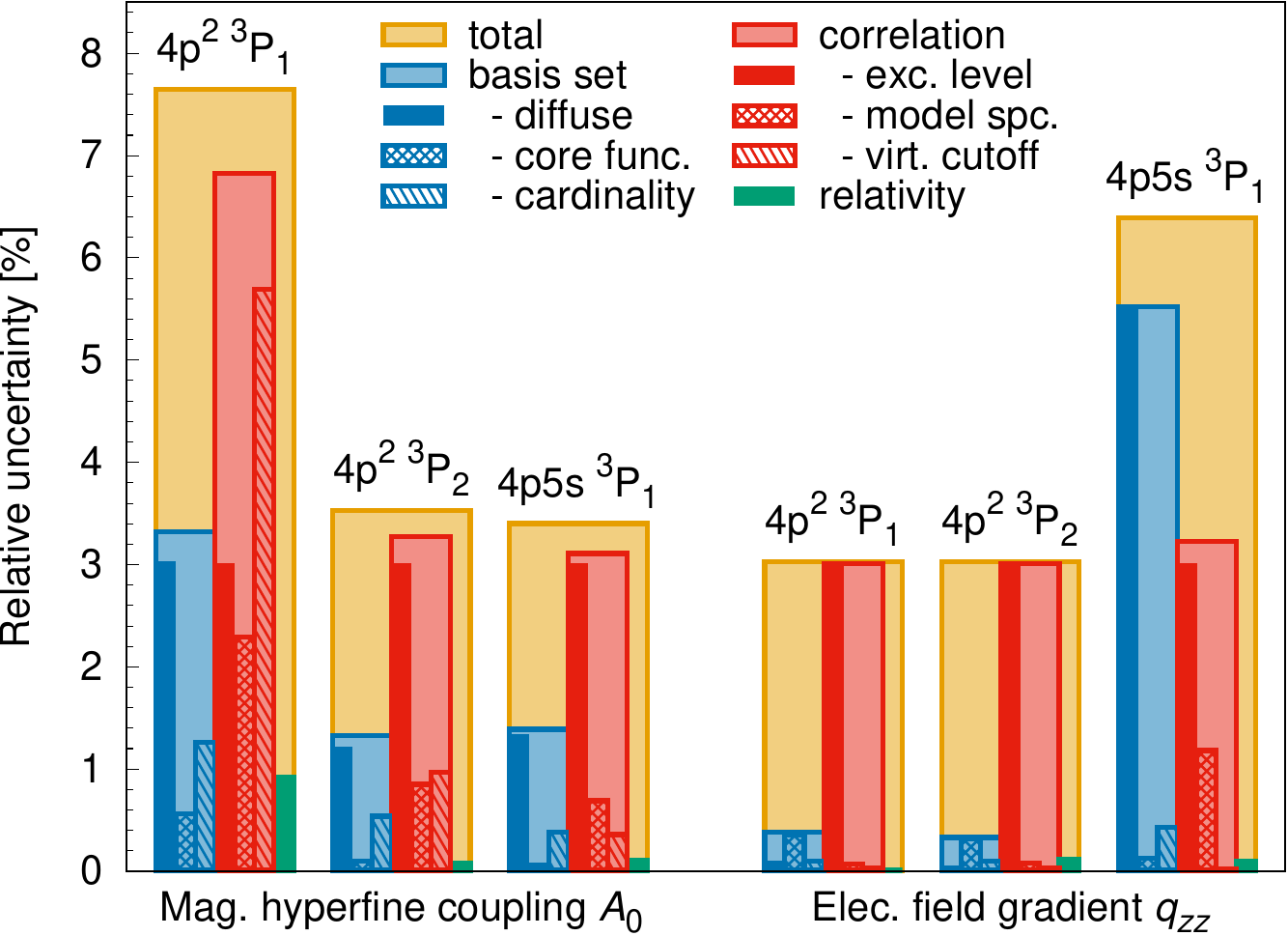}
    \caption{(\textit{Revised figure}) Relative uncertainty contributions for FSCC calculation of $A_0$ (left) and $q_{zz}$ (right) of three states in Ge.}
    \label{fig:Ge_Be_qzz_uncertainties}
\end{figure}

The final uncertainty of the calculated $q_{zz}$ is 3.0\%  for the 4p$^2~^3$P$_1$ and 4p$^2~^3$P$_2$ states and 6.4\%  for 4p5s~$^3$P$_1^\text{o}$, while for $A_0$ the uncertainties are 7.8\% , 3.5\%  and 3.4\%  respectively for these three states. Figure~\ref{fig:Ge_Be_qzz_uncertainties} shows an overview of the different contributions to the total uncertainty for the hyperfine parameters. Both the total uncertainty as well as the size of each contribution depends strongly on the state and property under investigation, demonstrating the importance of a systematic uncertainty estimation procedure, where the contribution from the different sources is estimated separately for each property and electronic state. This investigation also shows that the absolute uncertainties for the HFS constants are of similar size for each level (at least in this case), making the relative uncertainty much larger for states with low HFS sensitivities (small values of $A_0$ and $q_{zz}$).

The final calculated energies and HFS parameters are presented in Table \ref{tab:Ge_final_hfs_Be_qzz} and compared to experimental values, extracted from measurements performed in Ref. \cite{KanYanBis20}. All the presented values are in excellent agreement with experimental results within the theoretical uncertainties. 

\begin{table}[ht]
    \caption{Energies and hyperfine parameters $A_0$ and $q_{zz}$ of excited states in germanium, calculated using the FSCC method, and experimental values. The experimental energies are from Ref. \cite{NIST_ASD},  while experimental $A_0$ and $q_{zz}$ are extracted from measurements in Ref. \cite{KanYanBis20}. Based on data from Ref. \cite{KanYanBis20}.}
    \label{tab:Ge_final_hfs_Be_qzz}
    \centering
    \begin{tabular}{l r r r r r r} 
    \hline
    \hline
    State & E$^{\text{exp}}$ ({\small cm$^{-1}$}) & E$^{\text{FSCC}}$ ({\small cm$^{-1}$}) & $A_0^{\text{exp}}$ ({\small MHz}) & $A_0^{\text{FSCC}}$ ({\small MHz}) & $q_{zz}^{\text{exp}}$ ({\small MHz/b}) & $q_{zz}^{\text{FSCC}}$ ({\small MHz/b}) \\
    \hline
    4s$^2$4p$^2~^3$P$_1$ &   557.13 & 555(17)\phantom{00}   & --79.667(10) & --74(6)\phantom{0}   & 278.4(14) & 277(8)\phantom{0} \\
    4s$^2$4p$^2~^3$P$_2$ &  1409.96 & 1398(42)\phantom{00}  & 330.12(2)\phantom{00}   & 321(11)  & --571(3)\phantom{.00}   & --564(17) \\
    4s$^2$4p5s~$^3$P$_1$   & 37702.31 & 37449(1130) & 1343(6)\phantom{.0000}     & 1314(45) & --204(31)\phantom{.0}  & --208(13) \\
    \hline
    \hline
    \end{tabular}
\end{table}

The measured hyperfine parameters of  $^{71}$Ge and $^{73}$Ge were in good agreement with previously reported values \cite{ChiCoo63,ChiGoo66}. However, a significant discrepancy with the literature value \cite{OluSchShu70} was  found for $^{69}$Ge. The calculated FSCC hyperfine parameters supported the results of the new measurements, leading to a re-evaluation of the magnetic dipole and the electric quadrupole moments of this isotope. The nuclear moments of the three isotopes are presented in Table \ref{tab:Ge_moments}.

\begin{table}[ht]
    \caption{Nuclear moments of different isotopes of Ge from Ref. \cite{KanYanBis20} compared to earlier literature. $\mu^{\text{exp}}$ and $Q_s^{\text{exp}}$ for $^{69,71}$Ge were determined from HFS ratios relative to known values for $^{73}$Ge, while the reported $\mu^{\text{exp}}$ and $Q_s^{\text{exp}}$ for $^{73}$Ge are extracted from the measurements and FSCC calculations of $B_e^{\text{exp}}$ and $q_{zz}^{\text{exp}}$. Based on data from Ref. \cite{KanYanBis20}.}
    \label{tab:Ge_moments}
    \centering
    \begin{tabular}{lrrrr}
    \hline
    \hline
    Isotope & $\mu^{\text{lit}}$ ($\mu_N$) & $\mu^{\text{exp}}$ ($\mu_N$) & $Q_s^{\text{lit}}$ (b) & $Q_s^{\text{exp}}$ (b) \\
    \hline
    $^{69}$Ge & 0.735(7)\phantom{00}~\cite{OluSchShu70} & 0.920(5)\phantom{0} & 0.027(5)~\cite{OluSchShu70} & 0.114(7) \\ 
    $^{71}$Ge & 0.54606(7)~\cite{ChiGoo66} & 0.547(5)\phantom{0} &  &  \\ 
    $^{73}$Ge & --0.87824(5)~\cite{makulski2006gas} & --0.904(21) & --0.196(1)~\cite{Kello1998} & --0.198(4) \\ 
    \hline
    \hline
    \end{tabular}
\end{table}

\subsubsection{Nuclear charge radii of Si}

The study of Si isotopes is of interest for nuclear theory. The form of the nuclear equation of state (EOS)~\cite{Roca-Maza.2018}, which is of importance for properties of neutron stars~\cite{Abbot.2017} as well as superheavy nuclei~\cite{Nazarewicz.2018}, relies on the slope $L$ in the symmetry energy. $L$ is correlated with $|N-Z|$, where $N$ and $Z$ are the neutron and proton numbers, and thus can be constrained by the study of mirror nuclei such as $^{32}$Si and $^{32}$Ar~\cite{Reinhard.2022b,Huang.2023,Bano.2023}. Additionally, isotopes of Si play a role in studies of the (dis)appearance of nuclear magic numbers~\cite{Fridmann.2005, Piekarewicz.2007} and the emergence of exotic states such as bubble nuclei~\cite{Mutschler.2016, Duguet.2017}.

Isotope shift measurements of the $3s^23p^2~^1S_0$ $\rightarrow\ $ $3s^23p4s~^1P^o_1$ transition in the $^{28,29,30,32}$Si isotopes were performed at the BEam COoler and LAser spectroscopy (BECOLA) setup~\cite{Minamisono2013,Rossi2014} at the Facility for Rare Isotope Beams (FRIB) in order to determine the nuclear charge radius of $^{32}$Si \cite{konig_nuclear_2024}. The stable $^{28,29,30}$Si isotopes were used as reference, while the main aim of this work was to extract the charge radius of the radioactive $^{32}$Si. The field and mass shift constants $F$ and $K_{\text{MS}}$ were determined through a King plot procedure, but the limited number of reference isotopes led to a reduced accuracy of the experimental $F$ and $K_{\text{MS}}$. This required theoretical values of the $F$ and $K_{\text{MS}}$ parameters for a reliable extraction of the charge radius of $^{32}$Si.

The field shift of the relevant transition was investigated using the FSCC method, within two different computational schemes.
Scheme A employed the FSCC method with single and double excitations using the DIRAC code~\cite{DIRAC_code:19,Saue:20}. The augmented acv4z basis set ~\cite{Dyall:2016} used, with two additional layers of diffuse functions added in an even-tempered fashion; all electrons and virtual orbitals were included in the correlating space. The model space consisted of the 3p 4s (3d 4p 5s 5p 4f 4d) orbitals, where the orbitals in parentheses were in the intermediate Hamiltonian model space. The uncertainty was estimated using the same approach as described above, but the effect of the triple excitations was included explicitly, by employing the EXP-T code~\cite{EXPT:20,Oleynichenko:EXPT:20} within the v4z~\cite{Dyall:2016} basis set, keeping the 1s electrons frozen and setting the virtual cutoff at 5 a.u.

Scheme B was used to calculate the field shift parameter and the normal and specific mass shift constants, using the EXP-T and DIRAC programs combined with the code presented in Ref.~\cite{Penyazkov:2023}. The calculations were performed within the FSCC method with single, double, and triple excitations (FS-CCSDT) using a manually extended ae3z basis set, which included the addition of 5$s$-, 3$p$-, 2$d$-, and 1$f$-type diffuse functions to the original ae3z~\cite{Dyall:2016} basis set. The results were further corrected for the basis set deficiency by performing FSCC calculations with singe and double excitations, using a larger basis set, namely the manually extended aae4z~\cite{Dyall:2016} basis set. The basis set correction obtained at FSCCSD level was added to the FSCCSDT results. The uncertainty on the obtained values was estimated in a similar procedure as outlined above. The results are presented in Table \ref{tab:Si_F_M}.

\begin{table}
\caption{\label{tab:parameters} The field-shift parameters and the mass shift parameters, $F^\mathrm{\,(el)}$ $K^\mathrm{(MS)}$ obtained via different computational schemes. Based on data from Ref. \cite{konig_nuclear_2024}.}
\label{tab:Si_F_M}
  \centering
\begin{tabular}{lcc}
\hline
\hline
    Scheme   &  $K^\mathrm{(MS)}$ [GHz~u] & $F^\mathrm{(el)}$ (MHz/fm$^2$)  \\ 
\hline    
    Scheme A          &           &       97.0\,(8)              \\ 
    Scheme B   &  $-$373\,(24)         &       93.7(3.7)               \\ 
    Scheme C & $-$367\,(100) & \\
    \hline
\end{tabular}
\end{table}

The two results for the field shift are in excellent agreement with each other, showcasing the fact that the incremental corrections procedure, described in Section \ref{sub:meV} can be constructed in different ways, leading to accurate final numbers. 

Following a different approach (Method C), the mass-shift factors were also calculated within CI+MBPT using the AMBiT program ~\cite{AMBiT}. The obtained result for the total mass shift were in good agreement with the FSCC results (Method B), providing an independent confirmation, as can be seen in Table \ref{tab:Si_F_M}.

The calculated field shift obtained via Scheme A was used to the fit the King plot for the measured transition.  From the combined King fit (based on experimental frequencies and on field shift constrained to the FSCC value), a mass-shift parameter of $K^\mathrm{MS} = -340.8\,(1.4)$\,GHz\,u was extracted, which enabled a reliable determination of nuclear charge radii. The experimental $K^\mathrm{MS}$ was in good agreement with both FSCC and CI+MBPT predictions.

The resulting charge radius of $^{32}$Si ($R(^{32}\textnormal{Si})=3.153\,(12)$\,fm) was used to test nuclear model predictions, yielding an increased understanding of the performance of different nuclear models and of the nuclear EOS, and allowing a new constraint on the EOS L parameter of L $\leq~60$ MeV. Further discussion of the implications of these results in the context of nuclear theories can be found in Ref. \cite{konig_nuclear_2024}.

\subsection{The heaviest actinide: spectra and properties of Lr}\label{sub:Lr}

Lawrencium ($Z=103$) is a synthetic element that was first discovered in 1961 by Ghiorso~\textit{et al.}~\cite{Ghi61}. Interest in Lr, which is the heaviest actinide and the last element before the superheavy series formally begin, has not waned since its discovery, motivated by new insights into the nuclear structure of its isotopes~\cite{Asai:2015}, by the question of whether it, together with lutetium, is a  homolog of scandium and yttrium~\cite{Jensen:1982,Scerri:2009}, and by the strong effects of relativity on its electronic structure, predicted to change its ground state configuration with respect to that of Lu. This configuration change was originally suggested in Ref.~\cite{Desclaux:80} and confirmed in Ref.~\cite{EliKalIsh95} and all the later theoretical works, while experimental confirmation is still awaited. Any experimental investigations on heaviest actinides are inherently challenging due to their low production rates and the short lifetimes of the produced isotopes~\cite{GagJosTur89}. Nonetheless, measurement of the first ionization potential of Lr was achieved in 2015, using a surface ionization process in an atom-at-a-time regime~\cite{SatAsaBor15,SatAsaBor18}. Spectroscopic measurements on both Lr and its singly charged ion, based on different experimental approaches, are planned for the near future. Finally, there are also remote prospects of measurements of the electron affinity of this elusive element. All these studies require accurate predictions of the target properties and various accompanying parameters, motivating many theoretical works within a variety of computational methods. Below, we present some of these investigations, with a particular focus on relativistic coupled cluster studies.

\subsubsection{Ionization potential and electron affinity}

The first accurate calculations of the IP of Lr were carried our in 1995 using the DCB-FSCC approach~\cite{EliKalIsh95}. This study confirmed the earlier MCDF prediction~\cite{Desclaux:80} of the 5f$^{14}$7s$^2$7p$_{1/2}$ ground state configuration of Lr, in contrast to the 4f$^{14}$6s$^2$5d$_{3/2}$ of Lu. This order reversal is due to the relativistic effects that strongly stabilize the 7p$_{1/2}$ orbital in the heavier homologs, bringing it below the 5d$_{3/2}$ one, which is destabilized by relativity~\cite{Pyy78}. The two orbitals are, however, very close in energy, with the lowest excited $^2$D$_{3/2}$ state predicted just about 1500~cm$^{-1}$ above the $^2$P$_{1/2}$ ground state \cite{EliKalIsh95}. 
The IP calculation on Lu gave a result that was higher than the experiment by about 70~meV. These values, and all the later results described in this subsection, can be found in Table~\ref{Lr_IP_EA}.  

This work was followed by two studies based on relativistic pseudopotentials combined with complete active space multiconfiguration self-consistent field approach (and corrected for spin-orbit coupling)~\cite{PhysRevA.58.1103,doi:10.1063/1.1521431}; these have a discrepancy of 0.5~eV for the first IP of Lr (5.28~eV for the earlier and 4.78~eV for the later study), but both predict the second IP to be $\approx$~14.2~eV. The next relativistic FSCC study~\cite{BorEliVil07} used a significantly larger model space than Ref.~\cite{EliKalIsh95}, made possible by augmenting the calculations with the intermediate Hamiltonian~\cite{Landau:IH:99}. However, the predicted IP in this work remained very similar to the 1995 value. The next prediction was made using the CI+all order approach, where the resulting IP was in good agreement with the FSCC results \cite{DzuSafSaf14}.\\

\begin{figure}[h!]
\begin{center}
    \includegraphics[scale=0.35]{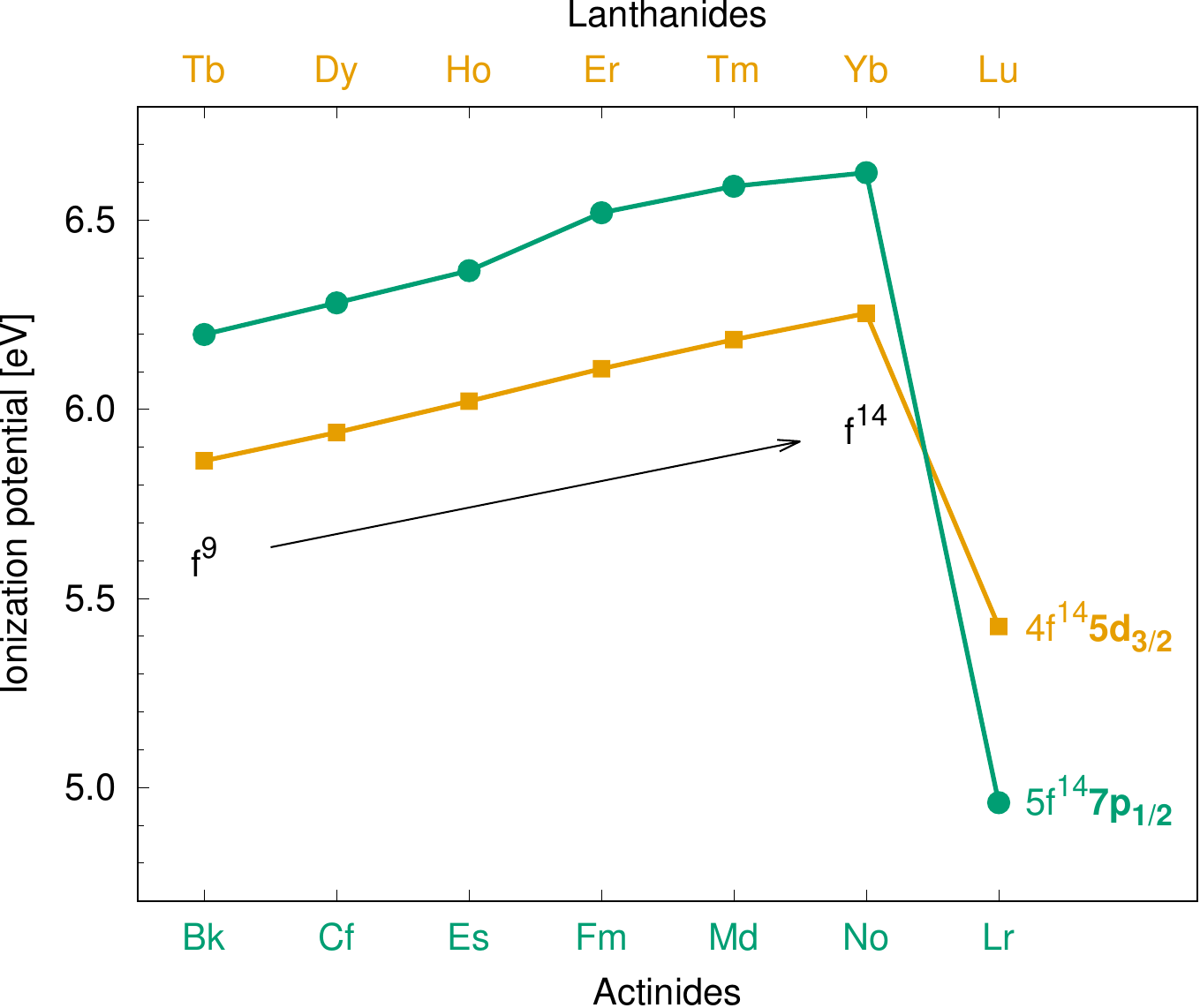}
\caption{Experimental ionization potentials of heavy actinides and lanthanides. Values for Tb -- Lu, Bk, Cf are sourced from~\cite{NIST_ASD}, Es from~\cite{Erdmann1998}, Fm, Md and Lr from~\cite{SatAsaBor18}, and No from~\cite{ChhAckBac18}.}
\label{IPs_Lr}
\end{center}
\end{figure}

In 2014 -- 2015, the IP of Lr was measured in a truly \textit{tour de force} one-atom-at-a-time experiment based on the surface ionization technique~\cite{SatAsaBor15}. The experiment was carried out on the $^{256}$Lr isotope, with a half-life of 27~s and the production rate of a single atom every few seconds. The extraction of the absolute IP of Lr from the measured effective IP (see Ref.~\cite{SatAsaBor15} for details) required the knowledge of the energies of low-lying excited states of the Lr$^+$ ion; dedicated FSCC calculations of these energies were performed. At the same time, relativistic CCSD(T) calculations of the IP of Lu and Lr were carried out in order to provide an independent comparison for the newly measured values (as FSCC energies were used to extract the IP from the measurements, the FSCC IP was not deemed an independent value). The CCSD(T) results were augmented by the Breit and QED corrections. The CCSD(T) IP of Lu (5.418~eV) was found to be in much better agreement with the experiment (5.426~eV) than the FSCC results, highlighting the importance of including the triple excitations for this system. The uncertainty on the predicted IP of Lr was set as twice the difference between the calculated value for Lu and the experiment, i.e. at 15~meV. The calculated and the measured values (see Table~\ref{Lr_IP_EA}) were found to be in excellent agreement with each other, confirming the first determination of the ionization potential of the heaviest actinide. This result was confirmed in a later measurement~\cite{SatAsaBor18}, where also experimental IPs of Fm and Md were presented for the first time, along with the corresponding CCSD(T) values. This work demonstrated that, similar to the lanthanides, the IP values of the heavy actinides up to No increase with filling up the 5f orbital, while that of Lr is the lowest among the actinides, due to the fully filled 5f shell and a weakly bound electron outside the No core (Figure~\ref{IPs_Lr}).

\begin{table*}[!tp]
  \centering
  \caption{A timeline of theoretically and experimentally determined first and second IPs of Lu and Lr and EA of Lr.}
  \begin{threeparttable}[b]    
    \begin{tabular}{@{\extracolsep{4pt}}l lllllll @{}}
    \hline\
    \multirow{2}{*}{Method}&\multirow{2}{*}{Year}&\multicolumn{2}{c}{Lu}& \multicolumn{3}{c}{Lr}&\multirow{2}{*}{Reference} \\
    \cline{3-4}\cline{5-7}
    &&\multicolumn{1}{c}{IP$_{1}$(6s$^2$)}&\multicolumn{1}{c}{IP$_{2}$(6s$^1$)}&\multicolumn{1}{c}{IP$_{1}$(7s$^2$)}&\multicolumn{1}{c}{IP$_{2}$(7s$^1$)}&\multicolumn{1}{c}{EA(7s$^2$7p$^2$)}&\\
   \hline
    \hline
   Experiment&1989&5.4259(13) & & & &&~\cite{MaeMizMat89}\\
DCB+FSCC&1995&5.301&&4.887&&0.307&~\cite{EliKalIsh95}\\
QRPP-CASSCF+APCF&1998&&&5.28&14.21&&~\cite{PhysRevA.58.1103}\\
RPP-CASSCF+APCF	&2003&&&4.78&14.25&&~\cite{doi:10.1063/1.1521431}\\
DCB-FSCC&2007&5.311&&4.894&&0.476&~\cite{BorEliVil07}\\
CI+all order&2014&&&	4.934&&&~\cite{DzuSafSaf14}\\
Experiment&& &&$4.96^{+0.08}_{-0.07}$&&&~\cite{SatAsaBor15}\\
DCB+CCSD(T)+QED&2015&5.418&&4.963(15)&&&~\cite{SatAsaBor15}\\
Experiment&&  &  &$4.96^{+0.05}_{-0.04}$& &&~\cite{SatAsaBor18}\\
FSCC+CI+MBPT&2021&&13.973&&\phantom{$>$}14.500(48)&&~\cite{KahBerLaa19}\\
Experiment&&&  &&$>$13.3(3)\tnote{a}&&~\cite{KwaPorGat21}\\
   CBS-CCSDT+Breit+QED&&5.391(18)&14.026(19)&4.955(9)&\phantom{$>$}14.627(49)&0.446(11)&~\cite{GuoPasNag24}\\
   \hline\hline
  \end{tabular}
  \begin{tablenotes}
    \footnotesize\item [a] Only lower limit is predicted.
  \end{tablenotes}
  \end{threeparttable}
  \label{Lr_IP_EA}
\end{table*}

The second ionization potential of Lr (alongside that of Lu) was calculated using the combination of FSCC and CI+MBPT approaches in Ref.~\cite{KahBerLaa19} and within the relativistic CCSD(T) method, corrected for the Breit and the  QED contributions and for higher-order excitations in Ref.~\cite{GuoPasNag24}. The latter results were also extrapolated to the complete basis set limit and are accompanied by uncertainties, and thus can be considered as recommended values for this property. Recently, an upper limit on the second IP of Lr was established experimentally as 13.3(3)~eV~\cite{KwaPorGat21}, consistent with the theoretical predictions. Furthermore, Ref.~\cite{GuoPasNag24} also provides a prediction of the second ionization potential of Lu, which has so far not been measured.  

Finally, Refs.~\cite{EliKalIsh95,BorEliVil07} present the electron affinity of Lr, calculated with the FSCC approach. It can be seen that, unlike the ionization potential, the electron affinity is increased by about 30\% upon increasing the size of the model space (as was done in the 2007 work~\cite{BorEliVil07}, compared to earlier calculations~\cite{EliKalIsh95}), demonstrating the sensitivity of this property to the description of the electron correlation. The current recommended CCSD(T) value, along with its uncertainty, for the electron affinity is presented in Ref.~\cite{GuoPasNag24}. 

\subsubsection{Transition energies}

Optical spectroscopy of the heaviest elements, such as Lr, can provide us with a wealth of information about these exotic species. Such studies probe the atomic configuration and electronic structure of these atoms and give an insight into the trends in these properties, which are strongly affected by the relativistic effects~\cite{BloLaaRae21, SewBacDre03}. Furthermore, spectroscopic studies can even serve for predictions of chemical behavior and material properties, which is particularly important for the transfermium elements ($Z>100$), where traditional chemical experiments are currently beyond our reach~\cite{Dul17}. Information about the nuclear spin, moments, and radii can also be extracted from the measured optical spectra, complementing the nuclear decay experiments~\cite{RaeAckBac18}. An example of a recent success story is the measurements of atomic levels, the hyperfine structure, and the ionization potential of nobelium~\cite{LaaLauBac16,RaeAckBac18,ChhAckBac18}. Theoretical predictions were important both for the success and for the interpretation of these experiments.
The next element in the Periodic Table is Lr, where prospects for laser spectroscopy are challenged by a tenfold reduced production cross section compared to nobelium~\cite{GagJosTur89}. Spectroscopy of this element is planned at the GSI employing the RADRIS method, which was used for the successful level search in nobelium~\cite{LauChhAck16, LaaLauBac16, WarAndBlo22}. The success of the envisioned measurements will depend on the availability of highly precise theoretical predictions
of the spectral lines, which will be used to develop excitation schemes and to narrow down the search window to be
able to pinpoint the ground-state transitions. Moreover, predictions of lifetimes and branching ratios are needed to quantify experimental parameters such as required detector sensitivities and beam times.

The first relativistic FSCC investigation of the spectrum of Lr was carried out in 1995~\cite{EliKalIsh95}. In this work, along with the ionization potential and electron affinity calculations described above, a number of transition energies of the neutral atom and the anion were calculated, alongside these of its lighter homologue, Lu. The calculations started from singly ionized atoms, and an electron was added in the FSCC sector (0,1) to obtain the lowest levels corresponding to the model space of 6p5d in Lu and 7p6d in Lr. For Lu, spectroscopic data was available, and the average error for the calculated transition energies for this element was on the order of 600~cm$^{-1}$. 
In 2002, an MCDF calculation of the four lowest states of Lr was performed~\cite{ZouFis02}, including the Breit and the Lamb shift contributions. These results were in excellent agreement with the experiment for the corresponding levels in Lu and in very good agreement with the earlier FSCC values for Lr. 

The spectrum of Lr (and Lu) was revisited within the FSCC approach in 2007. The purpose of this work was to suggest favorable transitions for measurements, as first spectroscopy experiments on elements with $Z\geq 100$ were envisioned and attempted at the GSI Helmholtz Centre for Heavy Ion Research (still named simply GSI at that point)~\cite{SewBacDre03,Backe:07}. This time, the calculations were augmented with the intermediate Hamiltonian approach~\cite{Landau:IH:99}, which allowed to obtain many more excited states compared to the earlier work. These calculations also included the Lamb shift, calculated using the approximation scheme by Indelicato~\cite{IndGorDes87}. In this work, the mean average error for Lu (over about 20 transition energies) was 420~cm$^{-1}$. Furthermore, E1 transition strengths were calculated using the RCI approach, along with excitation energies of states with holes in the 7s orbital.
The prime region for observing transitions in the planned GSI experiment was between 20000 and 30000~cm$^{-1}$, and these calculations predicted several strong single electron transitions in this range, e.g. the strongest line at 20100~cm$^{-1}$ corresponding to the to 7p~$\rightarrow$~8s transition and at 28096~cm$^{-1}$, corresponding to 7p~$\rightarrow$~7d transition.

At the same time as the work described above, an MCDF calculation of the spectrum of Lr was also published~\cite{FriDonKoi07}, which focused more on states where an electron is excited out of the s orbital. However, the predictions for the energy of the spectroscopically relevant 7p~$\rightarrow$~8s transition are in good agreement between the two approaches (20131~cm$^{-1}$ for FSCC calculations vs. 20405~cm$^{-1}$ for MCDF, with the estimated uncertainty of the latter prediction of about 1200~cm$^{-1}$). In 2014, spectra of No, Lr, and Rf, along with that of their lighter homologs, were investigated within the combination of the CI method with the linearized single-double CC approach (CI+all order)~\cite{DzuSafSaf14}. The lowest 24 transition energies of Lu were reproduced within an average error of about 300~cm$^{-1}$, and the authors expected similar accuracy for the predicted transition energies of Lr. 

A comprehensive calculation of a large number of excited states of Lr and Lu was carried out by the CI+MBPT approach in 2021, benchmarked by relativistic CCSD(T) calculations for the lowest levels. The latter results were extrapolated to the basis set limit, and an error estimate was performed, following the scheme presented in Ref.~\cite{LeiKarGuo20}. Furthermore, the CI+MBPT procedure also yielded the $g$-factors and the lifetimes of the calculated levels, and the Einstein coefficients for the transitions between them. The average disagreement between the CI+MBPT results and the experimental energies of Lu was just on the order of 150~cm$^{-1}$ for the lowest 32 levels, including doubly excited ones. Similar accuracy can be expected for the predictions for Lr. The accuracy of the CI+MBPT predictions for Lr was also confirmed by their agreement with the CCSD(T) values, where the latter were available. The calculated Einstein coefficients for Lu were in good agreement (within about 40\%) with the available experiment~\cite{NIST_ASD}, and the calculations reproduced correctly the relative strengths for the different transitions.

\begin{figure}[h!]
\begin{center}
    \includegraphics[scale=0.35]{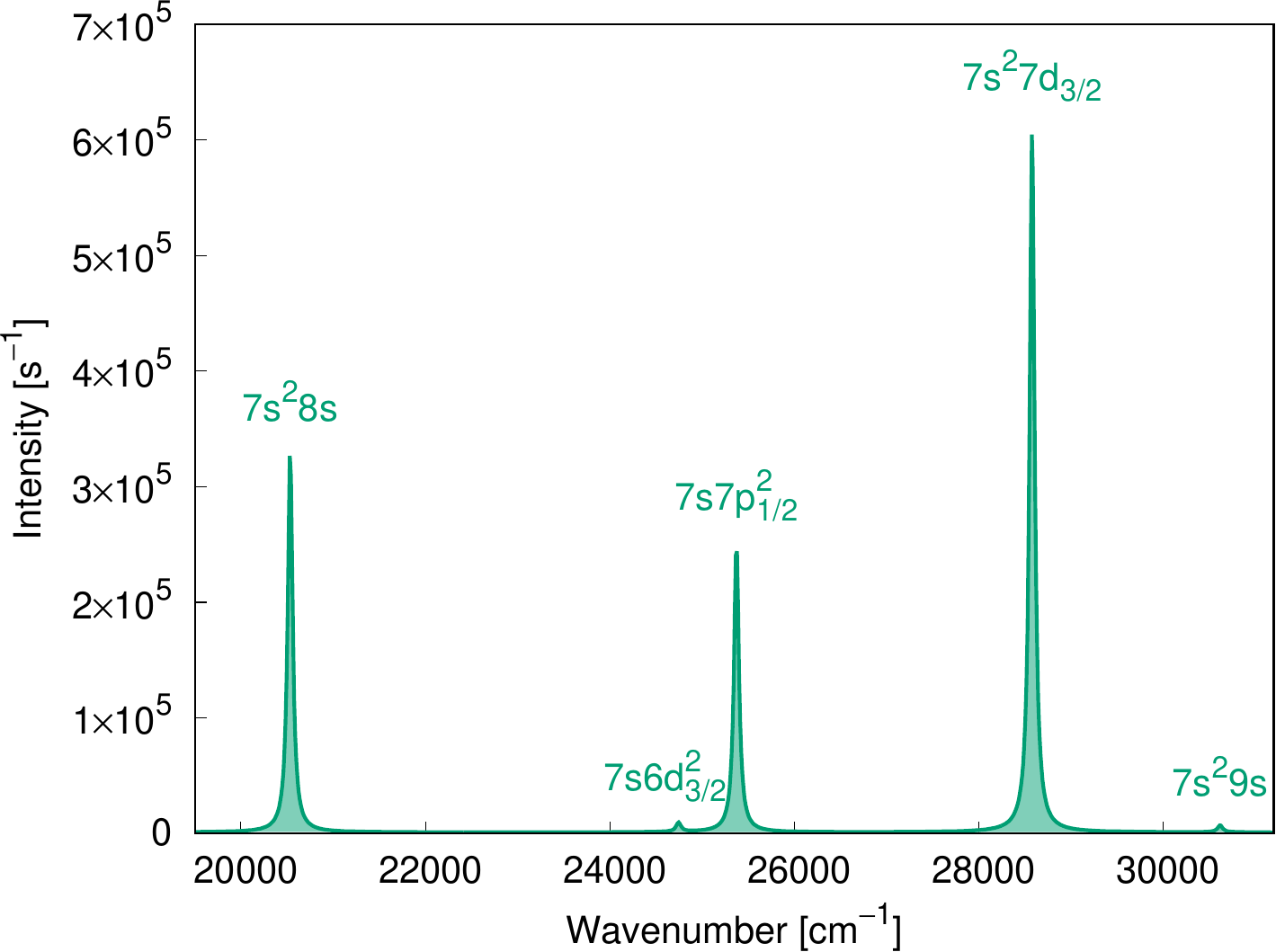}
\caption{Simulated E1 spectrum of Lr in the 20000--30000 cm$^{-1}$ range, based on the calculated energies and transition strengths from Ref.~\cite{KahRaeEli21} using Lorentzian convolution with FWHM = 65~cm$^{-1}$.}
\label{Lr-simulated}
\end{center}
\end{figure}

 \begin{table}
    \centering
    \caption{Calculated energies of the upper states corresponding to transitions of experimental interest. The uncertainty of the CI+all order prediction for the 7s$^2$8s $^2$S$_{1/2}$ state is estimated at 1--3\%. }
    \begin{tabular}{lllllll}
       \hline\hline
       Configuration& Term & FSCC~\cite{BorEliVil07} &MCDF~\cite{FriDonKoi07} & CI+all order~\cite{DzuSafSaf14}&CI+MBPT~\cite{KahRaeEli21}&CCSD(T)~\cite{KahRaeEli21}\\
       \hline 
       7s$^2$7p &$^2$P$_{1/2}$&  \phantom{0000}0&\phantom{0000}0& \phantom{0000}0&\phantom{0000}0&\phantom{0000}0 \\
    7s$^2$8s& $^2$S$_{1/2}$& 20118&20405(1200)&20253&20485&20533(300) \\
    7s7p$^2$& $^4$P$_{1/2}$ &   & & &25380& \\
    7s$^2$7d& $^2$D$_{3/2}$ &  28118&  &&28580& \\
       \hline  \hline
    \end{tabular}
    \label{Lr_transitions_for_exp}
       \end{table}
Based on similar  calculations for Lr, it was possible to identify three transitions from the atomic ground state with suitable transition strengths with Einstein $A$-coefficients above $10^7$\,s$^{-1}$, which is required to ensure an efficient transfer of the population. These transitions target the excited  7s$^2$\,8s~$^2$S$_{1/2}$ level with a transition strength of $3.31\times 10^{7}$\,s$^{-1}$,  the excited  7s$^2$\,7d$~^2$D$_{3/2}$ level with a transition strength of $6.14\times 10^{7}$\,s$^{-1}$ and the excited  7s\,7p$^2$~$^4$P$_{1/2}$ level with a transition strength of $2.51\times 10^{7}$\,s$^{-1}$. The former two transitions were also proposed previously as promising for measurements~\cite{BorEliVil07}. The advantage of the transition to the $^2$D$_{3/2}$ state with $J=3/2$ is its sensitivity to the nuclear spectroscopic quadrupole moment~\cite{BloLaaRae21}. Table~\ref{Lr_transitions_for_exp} summarizes the predictions for the energies of the upper states corresponding to these transitions calculated within the different computational approaches, including the error bars, where provided.
Excellent agreement can be observed for the energy of the 7s$^2$8s~$^2$S$_{1/2}$ state; the fact that the five calculations are based on very different approaches lends a strong confirmation to its predicted position. These state-of-the-art theoretical predictions will guide the frequency scans in future spectroscopic experiments on this element.  The simulated E1 spectrum of Lr in the experimentally relevant frequency range of 20000--30000 cm$^{-1}$ is shown in Figure~\ref{Lr-simulated}.

\subsubsection{Spectrum of Lr$^+$ and its mobility in He gas}

A new development in the field of atomic spectroscopy and ion mobility has been recently proposed under the name of Laser Resonance Chromatography (LRC)~\cite{LaaBucVie20}, which can provide a promising alternative to traditional laser spectroscopy on heaviest elements.

In this method, the ions are subjected to pulsed laser beams for resonant optical pumping into metastable states before their release into a drift tube filled with helium gas. Ions in different electronic states experience different interactions with helium atoms~\cite{LaaBacHab12, vie18} and thus move with different velocities through the drift tube toward the particle detector, enabling state-specific ion separation and resonance detection~\cite{LaaBucVie20}.  The time spectra obtained without resonant excitations characterize initial ground state ions, while the detection of ions at significantly shorter or longer arrival times signals their being in a metastable state. This effect is well established from ion-mobility spectrometry of many transition metals~\cite{KemBow91, IceRueMoi07, ManKem16}. One of the advantages of LRC is the fact that the neutralization step can be omitted (the heavy ions usually emerge from the gas catcher in +1 or +2 charge states~\cite{DroEliBla14}). 
Effective application of the LRC method to superheavy elements requires accurate theoretical predictions of the energies of the relevant levels.  Additionally, calculations of the transport properties involving the interactions between the metal ions and the helium atoms are very useful for optimizing experimental parameters (e.g., temperature or the applied electric field) such that the mobility of the ground state is substantially different from that of the excited state. Recently, two dedicated works were published that used relativistic coupled cluster to address these properties in Lr$^+$~\cite{KahBerLaa19, RamBorBlo23}.

The first comprehensive investigation of the spectra and properties of Lr$^+$ was presented in Ref.~\cite{KahBerLaa19}. The energy calculations were performed using the FSCC approach for both Lr$^+$ and its lighter homologue Lu$^+$. The calculations started by solving the relativistic Hartree--Fock equations and correlating the closed-shell reference states of Lr$^{3+}$ and Lu$^{3+}$. After the first stage of the calculation, two electrons were added, one at a time, to obtain the singly ionized atoms in the sector (0,2) of the FSCC procedure. To achieve optimal accuracy, large
model spaces were used, going up to 13s11p9d8f6g5h for Lu$^+$ and 14s12p10d9f6g5h for Lr$^+$, and the convergence
of transition energies with respect to the model space size was verified. In order to allow the use of such large
model spaces without encountering convergence difficulties in the coupled cluster iterations, the FSCC calculations were augmented by the extrapolated intermediate Hamiltonian approach (XIH)~\cite{Eliav:Alk:05}.

\begin{figure}[ht!]
    \centering
    \includegraphics[scale=0.35]{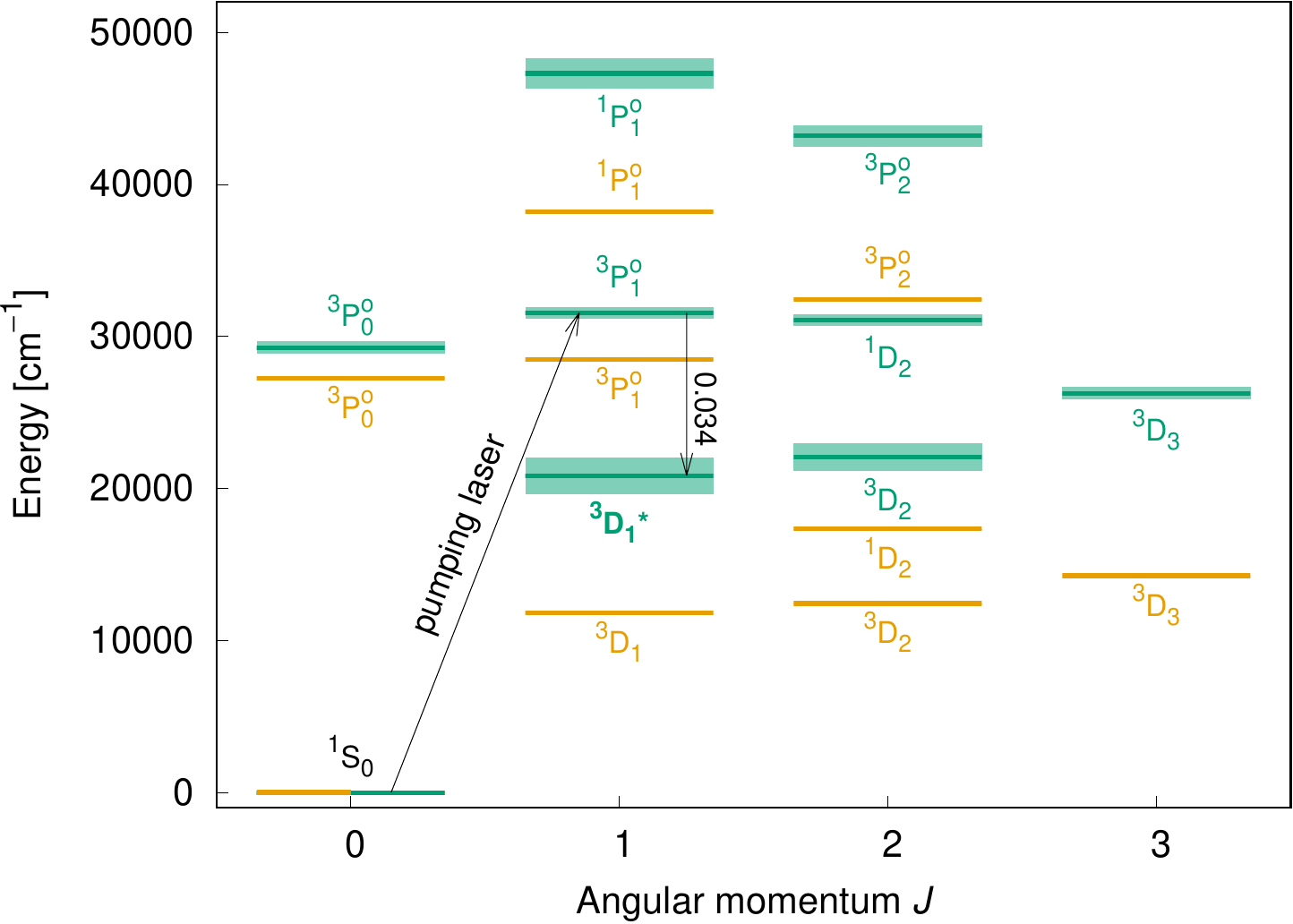}
    \caption{Grotrian diagram of experimental energy levels for Lu$^+$ (orange) and recommended calculated energy levels with the associated uncertainties for Lr$^+$ (green). Black arrows represent transitions relevant for LRC experiments, $^1$S$_0\rightarrow\,^3$P$_1$ and  $^3$P$_1\rightarrow\,^3$D$_1$*, the latter with the corresponding calculated branching ratio. Based on data from Ref. \cite{KahBerLaa19}.}
    \label{fig:Lr+_Lu+}
\end{figure}

CI+MBPT calculations were carried out in the same work, both to test the FSCC predictions and to provide the lifetimes of the levels of interest. The results for Lu (Table~\ref{tab:Lu+}) were in excellent agreement with the experiment for both methods (and the two methods also agreed very well with each other); this led the authors to propose the recommended transition energies for Lr$^+$ as the mean of the calculated FSCC and CI+MBPT results (Table~\ref{tab:Lr+}). The conservative uncertainty estimates on these energies were also provided, given by either the difference between the two calculated energies or the standard deviation of the difference between the CI+MBPT and experimental energy levels for Lu$^+$ (389~cm$^{-1}$), whichever is larger. In all cases, the energies are significantly higher than the
corresponding levels in Lu$^+$ (Figure~\ref{fig:Lr+_Lu+}). This is
due to the relativistic stabilization of the valence 7s shell
in the heavier ion, which makes this system more inert.

\begin{table}
    \centering
    \caption{Excitation energies (cm$^{-1}$) of Lu$^+$ from CI+MBPT, FSCC, and MRCI calculations. All results include the breit and QED corrections. Only levels relevant to the proposed LRC experiment are presented.}    \begin{tabular}{lllllll}
    \hline\hline
 Conf.&State       & FSCC~\cite{KahBerLaa19}& CI+MBPT~\cite{KahBerLaa19} & MRCI~\cite{RamBorBlo22}& Exp.~\cite{MarZalHag78} & \\
 \hline
   6s$^2$ &  $^1$S$_0$     & \phantom{0000}0 & \phantom{0000}0 & \phantom{0000}0 & \phantom{0000}0 \\
  5d6s &$^3$D$_1$       & 12354 &11664  & 12041 &  11796 \\
      &$^3$D$_2$   & 12985 &12380  & 12510 &12432   \\
      &$^3$D$_3$   & 14702 & 14267 & 13814 &  14199 \\
      &$^1$D$_2$   &  17892& 17875 & 16491 &  17332 \\
  6s6p  &$^3$P$_0$     & 27091 & 27303 &28664  & 27264  \\
       &$^3$P$_1$  &  28440& 28520 & 29846 & 28503 \\
      &$^3$P$_2$   &  32294& 32603 &  33863 &32453 \\
     &$^1$P$_1$    &  38464& 37385 & 38433 & 38223 \\
     \hline\hline
    \end{tabular}

    \label{tab:Lu+}
\end{table}

The calculated Einstein coefficients $A$ for Lu$^+$ were in good agreement with experimental values~\cite{Sansonetti:05}, with deviations of about 10--30\%. The relative transition strengths were very well reproduced, and the strongest transitions were identified correctly. Based on the calculated Einstein coefficients, lifetimes of the different levels in Lr were calculated (Table~\ref{tab:Lr+}). Because M1 and E2 transitions are slow, the even-parity states have significantly longer lifetimes than states that can decay via E1 transitions. In particular, the 6d7s~$^3$D$_1$ state can only decay to the ground state via a suppressed M1 transition, leading to a lifetime of $2.2 \times 10^6$ seconds, or about 25 days and making it suitable for LRC experiments. A potential LRC route consists of pumping the ground state $^1$S$_0$ (7s$^2$) to the excited $^3$P$_1$ (7s7p) state, which radiatively decays to the metastable $^3$D$_1$ (6d7s) state with a sizable branching ratio. A later MRCI study of the same levels in both ions was performed~\cite{RamBorBlo22}, with results in good agreement with the FSCC and the CI+MBPT values, see Tables~\ref{tab:Lu+} and~\ref{tab:Lr+} for comparison.

\begin{table}
    \centering
    \caption{Excitation energies (cm$^{-1}$) of Lr$^+$ from CI+MBPT, FSCC, and MRCI calculations. All results include the Breit and QED corrections. The recommended values are obtained as the mean of the FSCC and the CI+MBPT results. Lifetimes (s) derived from CI+MBPT calculations are also included.}
    \begin{tabular}{lllllll}
    \hline\hline
 Conf.&State       & FSCC~\cite{KahBerLaa19}& CI+MBPT~\cite{KahBerLaa19} &MRCI~\cite{RamBorBlo22}& Recommended~\cite{KahBerLaa19}& Lifetimes~\cite{MarZalHag78}  \\
 \hline
7s$^2$ &  $^1$S$_0$     & \phantom{0000}0 & \phantom{0000}0 & \phantom{0000}0 & \phantom{0000}0 & \phantom{0000}-- \\
  6d7s &$^3$D$_1$       & 20265 &21426  & 21563 &20846(1200)  &2.23$\times$10$^6$ \\
      &$^3$D$_2$   & 21623 & 22507 & 22259 & 22065(900) &8.26$\times$10$^{-2}$ \\
      &$^3$D$_3$   & 26210 &26303  & 24630 & 26262(400)&2.97$\times$10$^{-2}$ \\
      &$^1$D$_2$   &31200  &30942  & 28504 & 31071(400)&  1.53$\times$10$^{-3}$ \\
  7s7p  &$^3$P$_0$     & 29487 & 29059 & 31519 &29273(400)  &  2.56$\times$10$^{-7}$\\
       &$^3$P$_1$  & 31610 &31470  & 33710 &31540(400)  &1.45$\times$10$^{-8}$ \\
      &$^3$P$_2$   & 43513 & 42860 & 45451 & 43186(700) & 2.43$\times$10$^{-8}$\\
     &$^1$P$_1$    &47819  & 46771 & 49245 &47259(1000) &1.11$\times$10$^{-9}$ \\
     \hline\hline
    \end{tabular}
    \label{tab:Lr+}
\end{table}

Relativistic coupled cluster and MRCI calculations were also employed for calculations of the interaction potentials of the metal ions in the ground and the metastable $^3$D states with the He gas~\cite{RamBorBlo23}. These calculations were essentially equivalent to calculating the potential energy curves of LuHe$^+$/LrHe$^+$ molecules in different electronic states. Again, the two methods were found to agree. The obtained interaction potentials were used to calculate the ion mobilities corresponding to different experimental conditions, as detailed in Ref.~\cite{RamBorBlo23}. These calculations have shown that at room temperature, the relative drift time differences between ground and metastable states are expected to be about $15$\%  and $13\,$\% for Lu$^+$ and Lr$^+$, respectively, allowing the separation of the electronic states in both species. 
A recent experimental study~\cite{KimJanAry24} confirmed the feasibility of LRC in Lu$^+$. The arrival time distributions of singly charged lutetium revealed two distinct ion mobilities in helium in the ground and metastable states with a relative difference of about 19$\%$, consistent with predictions in Ref.~\cite{RamBorBlo23}. These works open the prospects of application of the LRC technique to Lr$^+$ ions and beyond to the ions of transactinide elements.

\subsection{Superheavy elements}\label{sub:SHE}
Superheavy elements (SHEs) are typically understood to be all transactinide elements ($Z>103$)~\cite{Schdel2006}. 
All SHEs are produced artificially via cold or hot fusion reactions at single-atom quantities~\cite{Moody2013}. As these synthesized nuclei are neutron-poor, they have very short half-lives, further decreasing rapidly with increasing atomic number. This makes direct experimental spectroscopic and chemical investigations of SHEs extremely challenging~\cite{LeNaour2013,Trler2013,Schdel2015,Trler2015}. 
These difficulties motivate theoretical studies of atomic, molecular, and even bulk properties of SHEs, both in support of the planned measurements and, often, as the only route for gaining information about the electronic structure and behavior of these elusive elements. 
Currently, all known SHEs occupy the 7th row of the periodic table, although substantial effort is being expended in the search for the elements 119 and 120 \cite{KhuYakDul20,GanHuaZha22,Bou24, Nel24}, and thus we may expect the 8th period to be added in the foreseeable future.
The last four new elements added to the periodic table in 2016 were nihonium (Nh, $Z = 113$), moscovium (Mc, $Z = 115$), tennesine (Ts, $Z = 117$), and oganesson (Og, $Z = 118$)~\cite{hrstrm2016}. 
In the following two examples, we are extending the methodology previously verified for heavy elements (see e.g. Sections~\ref{sub:Au} and~\ref{sub:At}) to two superheavy elements, Nh and Og.
Relativistic effects play a crucial role in the SHE region of the periodic table, affecting the properties and even, at times, leading to a change in the electronic configuration~\cite{Pershina2019}. For the 7th row elements, both the 7s and the 7p$_{1/2}$ atomic orbitals are strongly contracted and stabilized, and the spin-orbit effects lead to a large splitting between the 7p$_{1/2}$ and the 7p$_{3/2}$ orbitals.

\subsubsection{Nihonium}\label{sub:Nh}

Nihonium (Nh, $Z = 113$), the first element of the heaviest 7p-block of the Periodic Table, was first synthesized in 2004 using the cold fusion reaction of lead with a bismuth target~\cite{Enyo2019}. Its volatility was investigated experimentally through adsorption on gold and quartz surfaces~\cite{Dmitriev2014,Trler2015,Aksenov2017,Yakushev2021}, but no other atomic properties are known.
In our work~\cite{GuoBorEli22}, we aimed to provide a reliable prediction for at least the most basic atomic properties, IP and EA, for Nh. As a control, we used the same methodology to calculate these properties for the lighter homologs In and Tl, for which experimental values are known~\cite{Neijzen1981,Baig1985,Walter2010,Walter2020}.

Table~\ref{tab:Nh} concisely collects the results of our calculations and allows to easily evaluate the importance of the individual contributions. As in the case study of At (Section~\ref{sub:At}), all contributions are sorted into the three major computational parameter classes ordered by their relative importance -- basis set, higher-order electron correlation, and higher-order relativistic corrections. 

The first three lines of Table~\ref{tab:Nh} highlight the much higher relative importance of the basis set cardinality (cv3z vs cv4z) compared to the added outer-core correlating functions (v4z vs cv4z). 
The $\Delta$4z ($E$(4z) -- $E$(3z)) contribution has a growing trend across the IPs of the three elements ranging from 63~meV for In to 149~meV for Nh. On the other hand, for EAs we observe a mostly flat contribution of about 100 meV.
The addition of the outer-core correlating functions is at the order of 10s of meV and shows no regular trend within the group. The further addition of the inner-core correlation functions (ae4z vs cv4z) leads to only a small correction of at most a few meV, with the only notable exception being the increase in IP of Nh by 15~meV.
Adding diffuse functions to the basis set (1-aug-ae4z vs ae4z) has a small effect on the IPs, at the level of a few meV, but a much more pronounced effect on the EAs, up to 65~meV for In. However, a single diffuse layer is sufficient, and all 2-aug results are within 1~meV of the 1-aug values.
Finally, the additional CBSL extrapolation correction based on 3z and 4z values follows the cardinality trend as described above.

The higher-order correlation contributions beyond CCSD(T) are rather small for IPs, but much more significant for the EAs. This is not surprising, as the former deals with the simple p$^0$ and p$^1$ configurations, while the electron correlation is expected to be important for the p$^2$ configuration involved in the EAs. 
In the case of EA of Nh, we can observe the zig-zag pattern between the perturbative and iterative treatments of the given excitation levels. While the perturbative treatment of quadruples is apparently somewhat inadequate, nevertheless, the overall quadruple correction is merely --3~meV signaling convergence in terms of the excitation level.

Finally, the higher-order relativistic corrections (Breit and QED) are quite significant for group 13, and especially for the superheavy Nh, due to the valence p$_{1/2}$ orbital having substantial density close to the nucleus~\cite{Schwerdtfeger2015}.

\begin{table}[h!]
  \centering
    \caption{IPs and EAs (eV) of In, Tl, and Nh are gradually improving as term-by-term contributions are taken into account within the three major classes of computational parameters. 
    The basis set class results are shown calculated at the DC-CCSD(T) level.
    We use the shorthand notation (34)z to represent CBSL extrapolated results using 3z and 4z basis sets.
    Final values include uncertainties and are compared to the experiment~\cite{Neijzen1981,Baig1985,Walter2010,Walter2020}.}
    \begin{tabular}{llllllll}
    \hline\hline
    \multirow{2}{*}{class}&\multirow{2}{*}{contribution}&\multicolumn{3}{l}{IP }&\multicolumn{3}{l}{EA}\\
\cline{3-5} \cline{6-8}
     && In & Tl& Nh & In & Tl& Nh \\
  \hline
  basis set & v4z &5.767&6.055&7.481&0.268&0.234	&0.696\\
  &cv3z &5.695&5.975&7.337&0.152&0.131&0.582\\
  &cv4z &5.758&6.071&7.486&0.241&0.230&0.689 \\ 
  &ae4z &5.756&6.077&7.502&0.239&0.228&0.694 \\
  &1-aug-ae4z &5.759&6.079&7.509&0.304&0.277&0.718\\
  &2-aug-ae4z &5.759&6.079&7.509&0.305&0.277&0.719\\
  &2-aug-ae(34)z&5.804&6.148&7.613&0.315&0.299&0.777\\
  \hline
  electron&+$\Delta$T    & 5.803& 6.145& 7.614& 0.343& 0.298&  0.793\\ 
  correlation&+$\Delta$(Q)   & 5.805& 6.145& 7.608& 0.376& 0.311&  0.777\\
  &+$\Delta$Q   & 5.805& 6.146& 7.610& 0.374& 0.311&  0.790\\
  \hline
  relativity&+Breit & 5.799& 6.131& 7.567& 0.374& 0.309& 0.774\\ 
  &+QED & 5.801& 6.135& 7.569& 0.375& 0.311&0.776\\
  \hline
  theory final&& 5.801(22)& 6.135(32)& 7.569(48)& 0.375(18)& 0.311(12)& 0.776(30)\\
  experiment&&5.786359(1) & 6.108194(2)&   & 0.38392(60)& 0.32005(19)&\\
  \hline\hline
  \end{tabular}
  \label{tab:Nh}
\end{table}

\begin{figure}[h!]
\begin{center}
\includegraphics[scale=0.35]{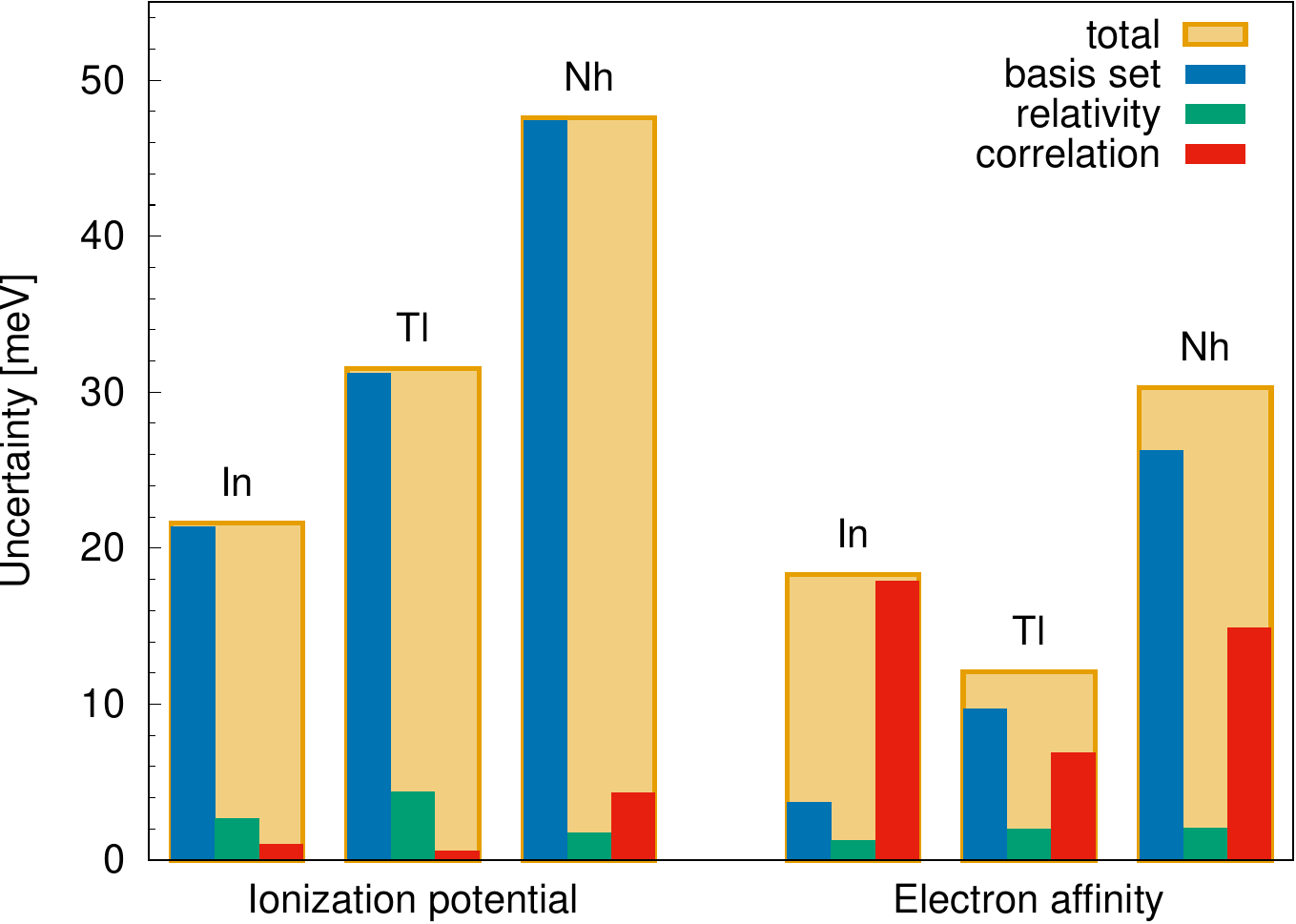}
\caption{Contributions of the different sources of uncertainty for IP and EA of Nh.}
\label{fig:Nh-unc}
\end{center}
\end{figure}

The composition of the determined uncertainties is shown in Figure~\ref{fig:Nh-unc}. The three largest uncertainty contributions are the basis set size, neglected higher-order correlation beyond the CCSDTQ level, and the beyond leading-order QED effects. 
The uncertainty associated with the basis set size grows almost linearly with $Z$ for both the IPs and the EAs.  In most cases, it is by far the dominant source, and only in the case of EA(In) it is smaller than the missing correlation error. The latter source of uncertainty is generally more prominent for EAs than for IPs.

\begin{figure}[h!]
    \centering
    \includegraphics[scale=0.35]{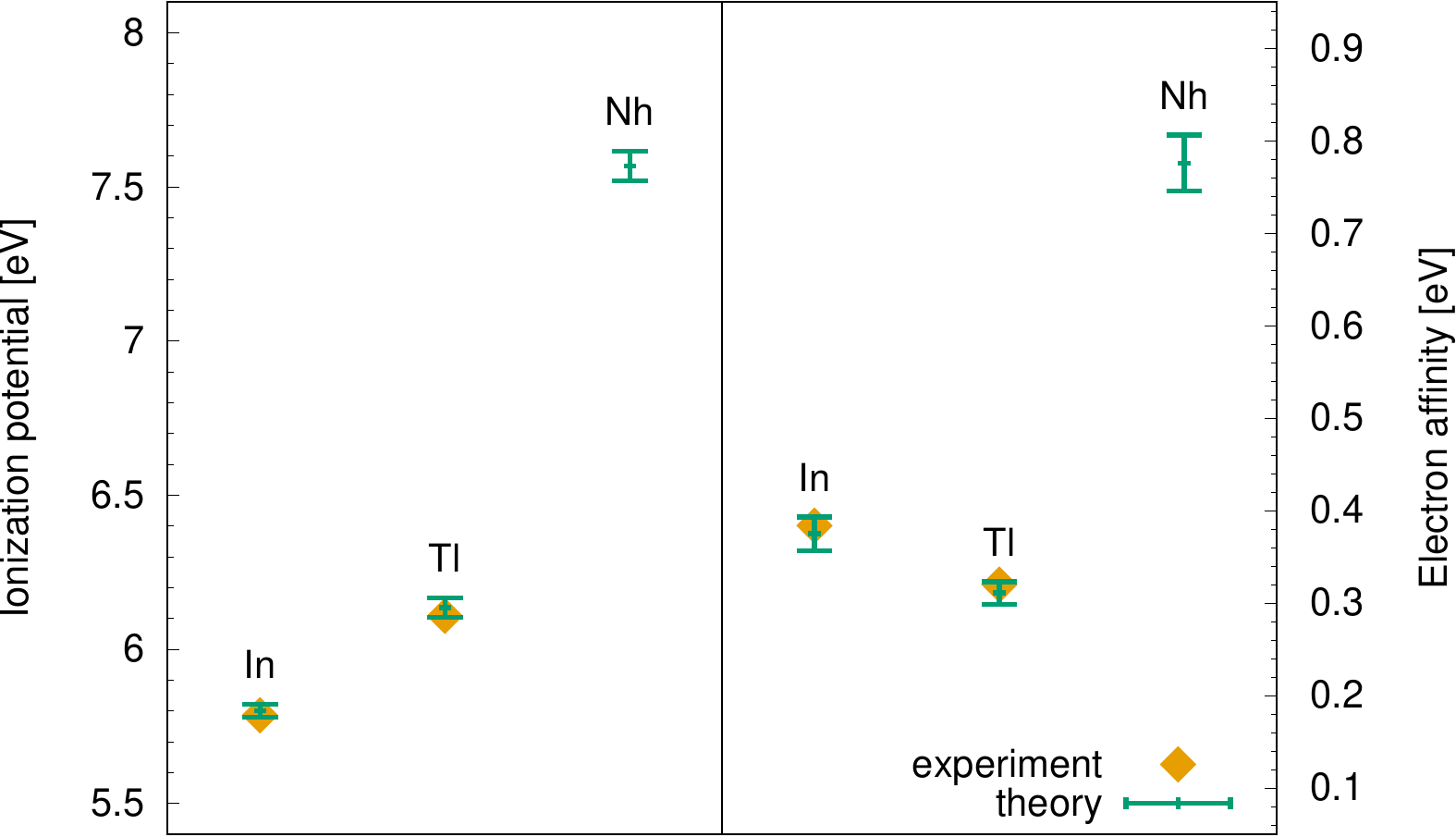}
    \caption{Calculated IPs (left) and EAs (right) with error bars of In, Tl and Nh, compared with experimental values for In and Tl.}
    \label{fig:Nh-IP-EA}
\end{figure}

Our final results compare well with the known experimental measurements for In and Tl. In all cases, measured IPs and EAs of In and Tl fall within the error bars of the theoretical results (Figure~\ref{fig:Nh-IP-EA}), confirming the applicability of the methodology to these atomic systems.
Based on this, we can have the same level of confidence in our predicted values of 7.569(48) eV and 0.776(30) eV, for the IP and EA of Nh, respectively.
Both of these values are significantly higher than those of the lighter homologs, In and Tl, due to the relativistic stabilization of the 7p$_{1/2}$ orbital, producing the typical kink in the periodic trends (Figure~\ref{fig:Nh-IP-EA}). To illustrate, the difference of the 7p$_{1/2}$ and 7p$_{3/2}$ orbital energies reaches 115~mH at the DC-HF level; the corresponding values for In and Tl are only 10~mH and 35~mH, respectively. Consequently, the Mulliken electronegativity $\chi_\mathrm{M} = 4.172$~eV which is significantly higher than the corresponding values $\chi_\mathrm{M}$(In)~=~3.085~eV and $\chi_\mathrm{M}$(Tl)~=~3.214~eV, indicates an increase in reactivity of Nh compared to its lighter homologs.

\subsubsection{Oganesson}\label{sub:Og}

Oganesson (Og, $Z=118$) is the heaviest element unambiguously observed to this date and is the last block in the current periodic table of elements. As such, it serves as the gateway to SHEs with higher atomic numbers that have not yet been discovered~\cite{Giuliani2019}. 
Og formally belongs to the noble gas group 18. However, due to the pronounced relativistic effects, it appears to be neither noble nor a gas. Recent works predict Og to be a solid at room temperature~\cite{Jerabek2019,Smits2020} and to exhibit semiconductor behavior~\cite{Mewes2019}.
Both these properties are uncharacteristic of the rare gases and result from the large spin-orbit splitting of its 7p shell and the stabilization of the vacant 8s orbital~\cite{Pershina2019}) reducing the 7p$_{3/2}-8$s gap.
The polarizability of Og is also predicted to be large (58.0~a.u.)~\cite{Jerabek2018}, which could contribute to stronger dispersion interactions and in turn interesting bulk properties.

Another unusual property of Og is its EA, which was predicted to be positive in earlier studies~\cite{Eliav1996,Goidenko2003,Lackenby2018}, in contrast to the other noble gases. As the predicted values were all rather small (and different from one another), presumably, different higher-order contributions may play a significant role in this property. Thus, we revisit this point using our systematic composite scheme~\cite{GuoPasEli21}.
The usual strategy of using the lighter homologue as a control is complicated by the fact that none of the lighter noble gases bind an excess electron. We thus calculate the first and the second IPs for Og alongside the lighter homologue Rn, for which these are experimentally available.

\begin{figure}[h!]
  \centering
  \includegraphics[scale=0.35]{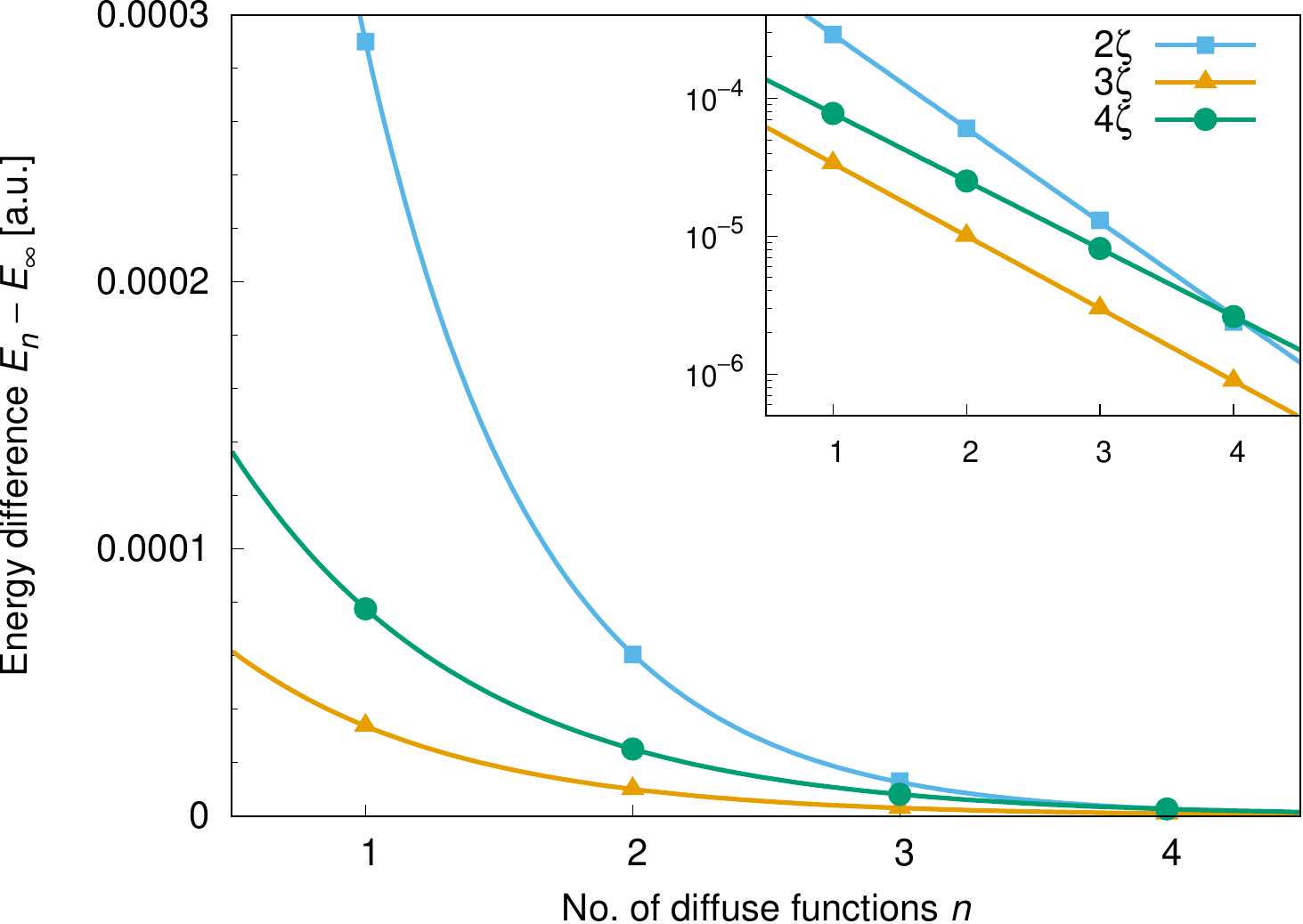}
  \caption{Exponential convergence of the total CCSD(T) atomic energy $E_n$ of Og towards the CBS limit with the increasing level of basis set augmentation $n$ for ($n$-aug)-cv$N$z basis sets. Semi-log inset showcases perfect linearity.}
  \label{fig:Og-dif}
\end{figure}

To obtain quantitatively correct results for the loosely bound Og$^-$ anion, a high-quality description of the region distant from the nucleus, which will host the excess electron, is crucial. 
We have thus investigated the effect of successive augmentation of the basis set (designated (1-aug)- (2-aug)-, and (3-aug)-), generated in an even-tempered fashion based on the first diffuse layer optimized in the original Dyall basis sets~\cite{Dyall:06,Dyall:12}.
We observed near-perfect exponential asymptotics with the increasing number of diffuse functions (Figure~\ref{fig:Og-dif}). This allowed us to extrapolate the total energies to infinite augmentation limit (denoted (($\infty$-aug)-cv$N$z) using a simple exponential function analogous to the Dunning--Feller $e^{-\alpha N}$ scheme~\cite{dunning1989gaussian,feller1992application}.
A similar systematic augmentation expansion was used earlier in the context of the EA of methane~\cite{RamrezSols2014}.
In Figure~\ref{fig:Og-dif}, one can see that the rate of convergence is different for basis sets with different cardinality, and the trend is surprisingly not monotonous.

\begin{figure}[h!]
  \centering
  \includegraphics[scale=0.35]{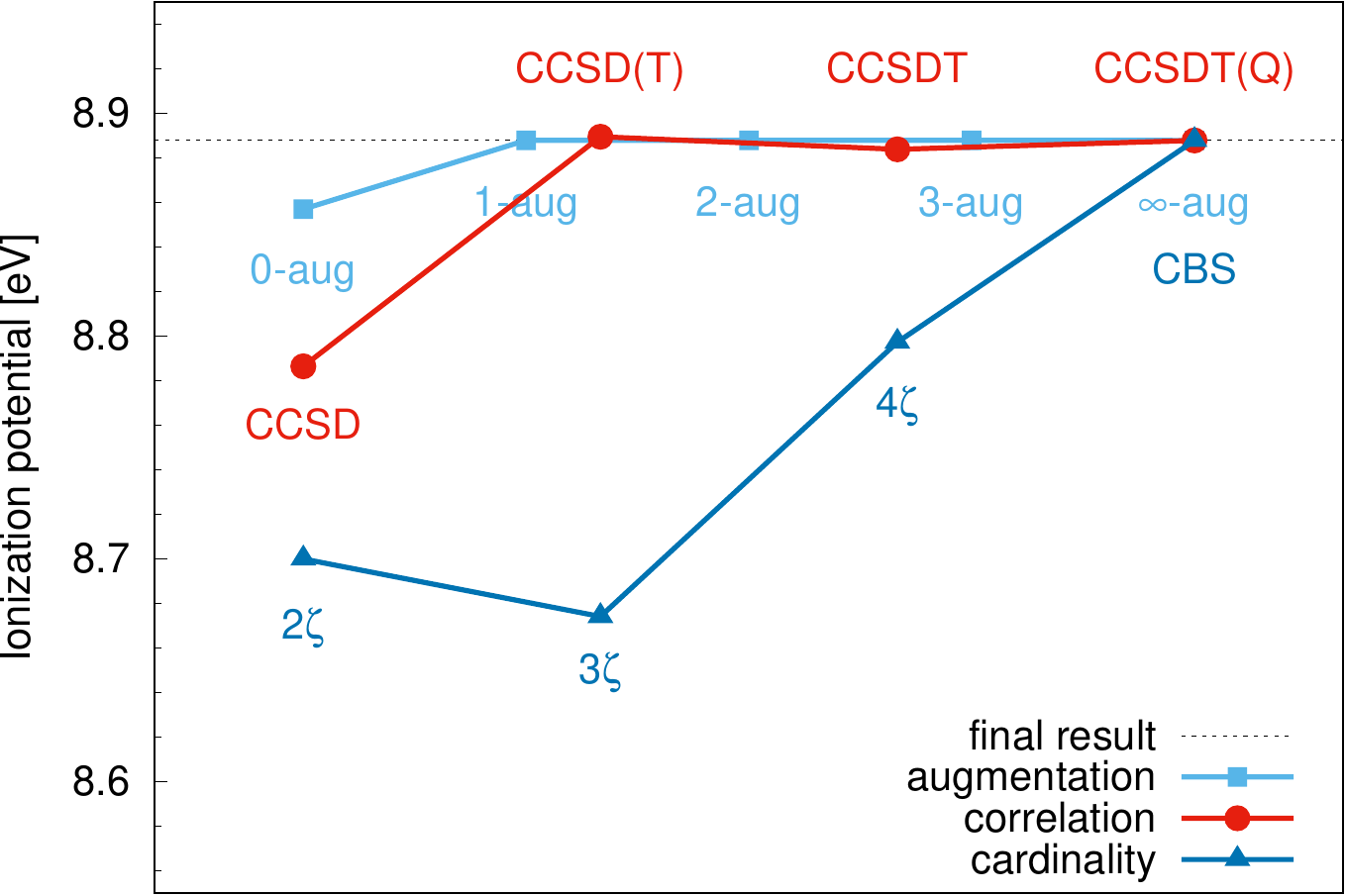} \qquad
  \includegraphics[scale=0.35]{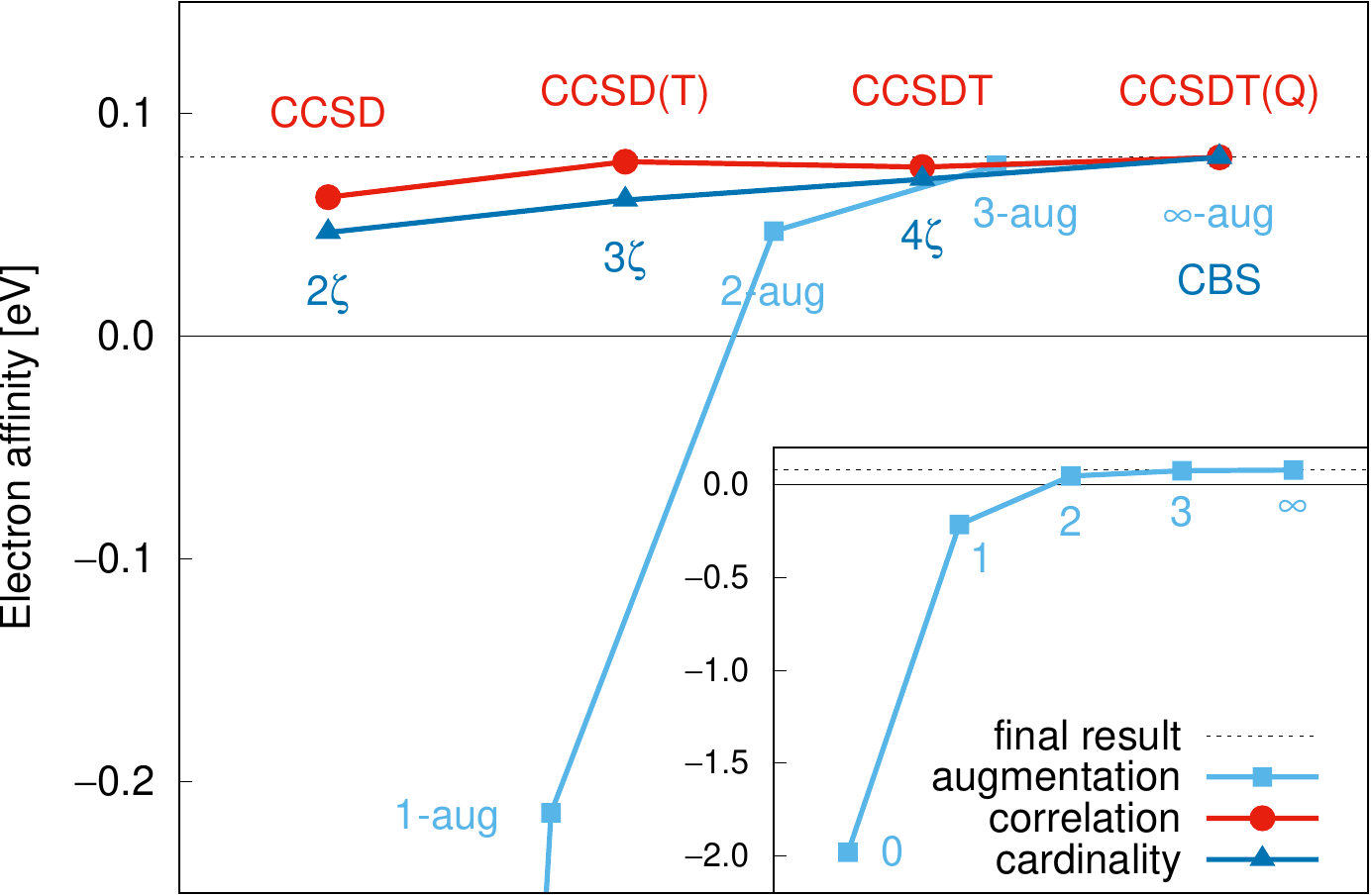}
  \caption{Effect of the basis set cardinality, the level of augmentation, and the treatment of electron correlation on the calculated IP (left) and EA (right) of oganesson. The black dotted lines represent the final result extrapolated to the complete basis set and infinite augmentation limit, corrected for all higher-order contributions. The inset on the right shows the the augmentation trend in full scale.}
  \label{fig:Og-IP-EA}
\end{figure}

We summarize the convergence of the important contributing computational factors  in Figure~\ref{fig:Og-IP-EA}, taking IP and EA of Og as the most exemplar cases exhibiting rather different trends. 
In both plots, each computational parameter is isolated, while the rest are assumed to be converged to the highest available degree of accuracy.
The effect of basis set quality is dramatically different in terms of cardinality and augmentation for the IP and the EA. A single layer of diffuse functions is sufficient for full convergence of the IP value, while it depends significantly more on the basis set cardinality, which converges rather slowly. 
For the EA, the situation is reversed -- even at the 2z level, one obtains a satisfactory result provided the basis set is sufficiently diffuse. Electron affinity, however, only becomes positive after the second diffuse layer is added and almost doubles in value when extrapolated to the infinite augmentation. 
Similar behavior to that of the IP of Og is observed for the IP of Rn  and for both second IPs.

The perturbative triples contribution is more pronounced in the IP compared to the EA. Additional higher-order corrections are small and partially cancel each other out. In radon and for the second IP, the electron correlation converges later at the level of CCSDT \cite{GuoPasEli21}. 
The Breit and QED contributions are of smaller importance here due to the valence p$_{3/2}$ orbital not penetrating the nucleus, and the combined contributions are at the level of a few meV; we thus do not include them in Figure~\ref{fig:Og-IP-EA}.

Based on our investigation, we determined the first and second IP and EA of Og to have values 8.888(44)~eV, 16.195(51)~eV, and 0.080(6)~eV, respectively, thus confirming the previously predicted positive EA for oganesson. Both IPs are lower than the respective values 10.761(57)~eV and 18.990(65)~eV calculated for Rn.
The experimentally determined values for Rn are 10.7485~eV and 21.4(19)~eV, respectively~\cite{NIST_ASD,Finkelnburg1955}. The latter is clearly at odds with our prediction, which raised questions about the result of this 1955 semiempirical study and was later superseded by our calculated value in the NIST data tables~\cite{NIST_ASD}.

\subsection{Highly charged ions}\label{sub:IonsE}

Highly charged ions (HCIs) find a variety of both practical and fundamental applications in the fields of astrophysics, atomic spectroscopy, metrology, accelerator physics, and plasma research. Highly charged ions, in particular those with only a few remaining electrons, can also serve as test systems for accurate bound-state QED calculations~\cite{ShaBonGla18,Indelicato2019}. Optical clocks based on HCIs are a promising alternative to traditional atomic clocks due to their compactness, which makes them significantly less sensitive to external fields and perturbations than neutral systems, leading to decreased systematic uncertainties~\cite{LudBoyYe15,KozSafCre18}. The first realization of an HCI optical atomic clock was recently reported~\cite{KinSpiMic22}. Alongside various metrological applications, HCI clocks are expected to benefit from greatly enhanced sensitivities to new physics. Proposed uses of precision measurement on HCIs and of HCI-based clocks include searches for variation of the fine-structure constant $\alpha$~\cite{BerDzuFla10,BerDzuFla11,BerDzuFla12,KozSafCre18, YuSahSuo23} and tests of Lorentz invariance~\cite{ShaOzeSaf18}. HCIs are also present in various high-temperature plasma environments, such as various stellar objects and around black holes~\cite{Gil01}, and on earth, in plasmas developed, for example, for fusion purposes. An important practical application of HCIs is in nanolithography, especially for application in the semiconductor industry~\cite{BanKosSwi11}.

Accurate calculations of the spectra and properties of highly charged (but many-electron) ions pose unique computational challenges. Due to the high charge, the relativistic effects become crucial, and treatment within the Dirac--Coulomb Hamiltonian is no longer sufficient. Both the Breit and the QED contributions become significant, and their inclusion can affect not just the absolute transition energies but even the ordering of the excited states~\cite{SafSafPor18}. Therefore, these contributions should be included in any calculation that aims to provide reliable predictions of the spectra of these systems. On the other hand, when dealing with many electron systems, correlation effects still remain important, and require a high-accuracy computational approach. Even though relativistic Fock-space coupled cluster is a natural tool of choice for such calculations, most applications of this method to highly charged ions appeared only in the last decade~\cite{NanSah15,WinCreBek15,NanSah16, WinTorBor16} alongside calculations performed using methods such as MCDF~\cite{Indelicato2019}, CI+MBPT~\cite{BerDzuFla10,BerDzuFla11,BerDzuFla12}, or CI+all order~\cite{SafDzuFla14,SafSafPor18} approaches. The subsections below contain examples of the application of this method to the spectra of highly charged tin ions, relevant in the context of nanolithography and semiconductor production, and of heavy, highly charged ions, with a potential to be used as optical standards suitable for search for variation of fundamental constants.

\subsubsection{Tin ion spectra}\label{sub:Tin}

 Highly charged tin ions are essential in bright extreme-ultraviolet (EUV) plasma-light sources for next-generation nanolithography~\cite{VerSheWit22}. However, their complex electronic structure is an open challenge for both theory and experiment, motivating both ambitious measurements and state-of-the-art calculations. Accurate theoretical predictions can be crucial in this context for the identification of the measured lines. 
 
 Tin ions in charge states 7+ to 14+ are used to generate extreme ultraviolet (EUV) light at 13.5~nm wavelength in laser-produced-plasma sources for nanolithographic applications~\cite{BenBanLok08, BanKosSwi11}. These ions have an open [Kr]4d$^n$ shell structure, and the resulting EUV light is generated by thousands of transitions that form the so-called unresolved transition arrays (UTAs) with little dependence on the ion's charge state; the huge number of lines in these UTAs complicates their assignment immensely. Nonetheless, spectroscopic work on these systems was performed using a variety of methods, such as discharge sources, charge-resolved spectroscopy on the EUV regime, charge-exchange spectroscopy, and electron beam ion traps (EBITs) (see Ref.~\cite{WinTorBor16} for the overview of the various measurements). An alternative approach used in the work described here~\cite{WinTorBor16} is to turn to the optical range and to measure the optical, magnetic dipole (M1) transitions between fine structure levels in the ground electronic configuration, thus alleviating some issues of limited resolution and unresolved UTA transitions.

Charge state-resolved optical spectroscopy was performed on Sn$^{11+}$ -- Sn$^{14+}$ using the FLASH-EBIT~\cite{EppCreSim10} at the Max Planck Institute for Nuclear Physics in Heidelberg. Alongside the experiments, high-accuracy Dirac--Coulomb--Breit Fock-space coupled cluster calculations were performed and used to assign the M1 transitions.

The calculations for Sn$^{14+}$, Sn$^{13+}$, and Sn$^{12+}$ started from the closed-shell reference $4s^{2}4p^6$ configuration of Sn$^{14+}$. The limitation of the FSCC codes to atoms with a maximum of two open shell electrons/holes meant that this approach could not be applied to Sn$^{11+}$ with a 4s$^2$4p$^6$4d$^3$ ground state configuration. After the first stage of the calculation, consisting of solving the relativistic Hartree-Fock equations and correlating the closed-shell reference state, different FSCC schemes were used for the different ions. 
In the case of Sn$^{14+}$, a single electron was excited from the 4p to the 4d orbital to reach the $4p^54d^1$ configuration, corresponding to the sector (1,1). For Sn$^{12+}$, two electrons were added to the closed\-shell reference state within the sector (0,2). In this calculation, to achieve optimal accuracy, a large model space was used, comprised of  3\,s, 3\,p, 3\,d, 3\,f, 2\,g, and 1\,h orbitals, and the intermediate Hamiltonian method was employed to facilitate convergence~\cite{Eliav:Alk:05}. The fine structure splitting of Sn$^{13+}$ was also obtained in the framework of this calculation as a result of adding the first electron to the closed-shell reference state. The uncontracted universal basis set~\cite{Malli:93} was used for all the ions, 
consisting of 37\,s, 31\,p, 26\,d, 21\,f, 16\,g, 11\,h, and 6\,h functions; the convergence of the obtained transition energies with respect to the size of the basis set was verified. All the electrons were correlated.
The Lamb energy shift was obtained using the effective potential method, as implemented in the QEDMOD program~\cite{ShaTupYer15}.
All the calculations were performed using the TRAF-3C code~\cite{TRAFS-3C}.

The experimental and FSCC results were further supported by semi-empirical calculations within the framework of the COWAN code~\cite{Cow81,kra19}, which employs isoelectronic scaling methods where data for such procedures is available.  This code relies on calculations using empirically adjusted wave-function scaling parameters and can be used to
identify spectral lines.
The detailed COWAN code computational parameters used for the highly charged tin ions are presented in Ref.~\cite{WinTorBor16}.

\begin{table}
    \centering
    \caption{Experimental energy levels $E_{\mathrm{exp}}$,  ab initio FSCC energy levels $E_{\mathrm{FSCC}}$ (the individual contributions from the Breit interaction $\Delta E_{\mathrm{Breit}}$ and QED $\Delta E_{\mathrm{QED}}$ included in $E_{\mathrm{FSCC}}$ are also shown), as well as semiempirical COWAN code calculations $E_{\mathrm{COWAN}}$ of the fine-structure configurations in Sn$^{12+}$ –- Sn$^{14+}$. All the energies are given in units of cm$^{-1}$. Based on data from Ref.~\cite{WinTorBor16}.}
   \begin{tabular}{llrrrrr }
       \hline\hline
       Ion   & Term & $E_{\mathrm{exp}}$ & $E_{\mathrm{FSCC}}$ & $\Delta E_{\mathrm{Breit}}$ & $\Delta E_{\mathrm{QED}}$ & $E_{\mathrm{COWAN}}$ \\
       \hline
      Sn$^{12+}$  [Kr]4d$^2$   & $^3$F$_2$ & 0  &0   &0   & 0  & 0  \\
        & $^3$F$_3$ & 9786  &9738   &--374   & 29  & 9780  \\
        & $^3$F$_4$ & 18564  &18507   &--655   & 57  & 18563  \\
         & $^3$P$_0$ &    &23642   &--87   & 0 & 22649    \\
         & $^1$D$_2$ &24838    &25285   &--355   & 29 & 24835  \\
         & $^3$P$_1$ & &28750     &--238   & 29 & 27905   \\
          & $^3$P$_1$ &39044 &39636     &--425   & 57& 39042  \\
          & $^3$P$_1$ &39718 &39381     &--983  & 29& 39715   \\
          & $^1$S$_0$ &  &83202     &--381  & 57& 80700    \\
           Sn$^{13+}$  [Kr]4d$^1$   & $^2$D$_{3/2}$ & 0  &0   &0   & 0  & 0  \\
              & $^2$D$_{5/2}$ & 13179  &13144   &--439  & 30  & 12740  \\
              Sn$^{14+}$  [Kr]   & $^1$S$_0$ & 0  &0   &0   & 0  & 0  \\
              Sn$^{14+}$   4p$^5$4d$^1$  & $^3$P$_0$ & &--8692  &--870   &--128      & --8970   \\
              & $^3$P$_1$ &   &0  &--1054   & --128  & 0   \\
               & $^3$P$_2$ & 17311  &17544  &--1262   & --97  &17392  \\
                & $^3$F$_3$ & 19275  &19247  &--1048   & --128  &19115   \\
                & $^3$F$_4$ &    &21027  &--1464   & --97  &20991   \\
                & $^1$D$_2$ &   30278 & 30252 &--1197  & --128  &31354  \\
                & $^3$D$_3$ &   50891 & 51770 &--1510  & --97  &51208  \\
                 & $^3$D$_1$ &    & 76730 &--1535  & 12  &77445  \\
                  & $^3$F$_2$ &  & 91543 & --1996 & 12  &93484  \\
                  & $^3$D$_2$ &   & 109202 &--2348  & 46  &110507  \\
                  & $^1$F$_3$ &   & 117668 &--2445  & 46  &118362  \\
                   & $^1$P$_1$ &   & 209573 &--1853  & 12  &222563  \\
       \hline  \hline
    \end{tabular}
    \label{tab:HCI_tin}
\end{table}

The results of the FSCC calculations are presented in Table~\ref{tab:HCI_tin} and compared with the experimental values and the COWAN results. The importance of the Breit contribution can be seen, in particular in the case of the fine structure levels of the excited 4p$^5$4d$^1$ configuration in Sn$^{14+}$, where it reduces the calculated energies by 1000 -- 2000~cm$^{-1}$. Thus, this contribution should be included in any calculations on HCIs where accuracy is important. The QED contribution is more modest at a few tens to about a hundred cm$^{-1}$. 

The FSCC values are in excellent agreement with the experiment for all the fine structure levels in all charge states where this method could be applied, with a mean average error of just 250~cm$^{-1}$. These \textit{ab initio} results are thus on the same level of accuracy as the semiempirical COWAN code that relies on extensive input data to optimize the adjustable parameters to fit the observed spectra. For most of the transitions where the lines were not experimentally detected, the FSCC and the COWAN predictions are also in good agreement. 
The FSCC calculations also provided the absolute excitation energies for the fine structure levels of the Sn$^{14+}$ 4p$^5$4d$^1$ configuration. Remarkably, these we also found to be in excellent agreement with earlier measurements, such as the calculated energies of the $^3$D$_1$ and  $^1$P$_1$ terms, at 617515~cm$^{-1}$ and 750358~cm$^{-1}$, respectively. These were determined in an earlier EUV experiment to be 616892 and 749429~cm$^{-1}$~\cite{Sug91}. Thus, the FSCC approach maintains its excellent performance and high accuracy also in the high energy range. 

The combination of measurements and calculations performed in Ref.~\cite{WinTorBor16} allowed the re-evaluation and a more accurate determination than was hitherto possible of the fine structure of Sn$^{12+}$ -- Sn$^{14+}$. 
These measurements and identifications provide an input for optical
plasma diagnostic tools. Furthermore, the identifications of the
transitions confirmed the strong predictive power of the FSCC calculations for HCIs, where previous investigations were scarce. 
A similar spectroscopic investigation and analysis were performed on Sn$^{7+}$ to Sn$^{10+}$ ions~\cite{TorWinRya17}; however, due to the limitations of the FSCC approach to two-particle/hole systems, the CI+MBPT approach was used instead. \\

Alongside highly charged tin ions, weakly charged ions also contribute significantly to the light emitted in laser-produced EUV plasmas generated for nanolithography. However, spectroscopic information on the relevant charge states, Sn$^{3+}$ and Sn$^{4+}$, is rather scarce because of the poorly known electronic structure of these ions. Sn$^{3+}$, with its ground electronic configuration  [Kr]4d$^{10}$5s, belongs to the Ag-like isoelectronic sequence. Electronic transitions in this ion were recently investigated by studying its line emission in the wavelength range of 200 -- 800~nm~\cite{SchRyaBor18}. The optical lines belonging to Sn$^{3+}$ were identified among the hundreds of other lines stemming from a laser-produced droplet-based Sn plasma by taking spectra as a function of the laser intensity. The method to single out transitions belonging to an ion in a specific charge state relies on the strongly changing ratio between line intensity and the background emission from the plasma as a function of the laser intensity. The COWAN code was used to assist in identifying the lines, and the consistency of the highly excited levels was checked by quantum-defect scaling.

Relativistic FSCC calculations of the spectrum were performed alongside the experiment. The calculations of the transition energies of Sn$^{3+}$ start from the closed-shell reference  [Kr]4d$^{10}$  configuration of Sn$^{4+}$. After the first stage of the calculation, consisting of solving the relativistic Hartree--Fock equations and correlating the closed-shell reference state, a single electron was added to reach the desired Sn$^{3+}$ state. A large model space was used in this calculation, comprising 10\,s, 8\,p, 6\,d, 6\,f, 4\,g, 3\,h, and 2\,i orbitals in order to obtain a large number of excitation energies and to reach optimal accuracy. The intermediate Hamiltonian method was employed to facilitate convergence~\cite{Eliav:XIH:05}. The rest of the computational details were the same as in the calculations performed for the higher charge states.

The FSCC values are presented in Table~\ref{tab:tin3+} and compared to experiments and earlier relativistic many-body perturbation theory (MBPT) calculations. We can observe that in this case, the Breit contribution is lower compared to Sn$^{12+}$ -- Sn$^{14+}$ due to the smaller charge of the ion, and the Breit and QED contributions are similar in size. The results are in overall excellent agreement with the measurements, with an average error of about 300 cm$^{-1}$, which is on the $10^{-3}$ level of the calculated energies (compared to about 800 cm$^{-1}$ for MBPT), despite addressing very high absolute energies. Generally, the FSCC values overestimate the experiment, while MBPT results underestimate it. The tendency of these methods to have an opposing sign of the error with respect to the experiment was also noted in Ref.~\cite{KahBerLaa19} and used there to set uncertainties on predictions based on the two approaches. Remarkably, the calculated FSCC IP is only 91 cm$^{-1}$ above the experimental value. 

 \begin{table}
    \centering
    \caption{Energy levels of Sn$^{3+}$, relative to its ground state [Kr]4d$^{10}$5s. The statistical uncertainty is presented in parentheses. \textit{Ab initio} FSCC calculations results, $E_{\mathrm{FSCC}}$, are also shown, along with the individual contributions from the Breit interaction $\Delta E_{\mathrm{Breit}}$ and the QED contribution $\Delta E_{\mathrm{QED}}$ (included in $E_{\mathrm{FSCC}}$), and compared to relativistic many-body perturbation theory (MBPT) calculations obtained in Ref.~\cite{SafSavSaf03}. Based on data from Ref.~\cite{SchRyaBor18}.}
    \begin{tabular}{lllllll}
       \hline\hline
       $nl$&$J$   & $E_{\mathrm{exp}}$ &$E_{\mathrm{FSCC}}$ & $\Delta E_{\mathrm{Breit}}$ & $\Delta E_{\mathrm{QED}}$&MBPT \\
       \hline
      5p& 1/2&      &\phantom{0}69741  &\phantom{--0}62   &--171  & \phantom{0}69265\\
        & 3/2&  & \phantom{0}76256 &\phantom{0}--26   &--165&  \phantom{0}75736 \\
         5d& 3/2& 165304(1)    &165646  &--123   &--205  & 164538\\
     & 5/2& 165409 (1) & 166382& --145  &--204& 165283 \\
              4d$^9 5s^2$& 5/2& 169233.6(8)    &  &    &   &  \\
     & 3/2& 177889.0  &  &   & &   \\
      6s& 1/2& 174138.8(4)    &174236 &\phantom{0}--99   &--143  & \\
      6p& 1/2&   197850.6(6)   &198025 & \phantom{0}--74   &--193  & \\
        & 3/2& 200030.1(4) & 200216 &--103   &--193&  \\
        4f& 7/2& 210258.2(6)   &210557  &--158  &--200  & 209418\\
        & 5/2& 219317.9(7) & 210627 &--156   &--200&   209494 \\
         6d& 3/2& 234797.0(1)    &235171 &--134   &--203  &  \\
     & 5/2& 235128.7 (2) & 235497&--144  &--201&  \\
      7s& 1/2& 237617(1)   &237920 &--123   &--175  & \\
     7p& 1/2&   248735.4(2)   &249094 & --110   &\phantom{0}--19  & \\
        & 3/2& 249644.8(1) & 250233 &--124   &\phantom{0}--97&  \\
        5f& 7/2& 251853.0(2)   &252626  &--157  &--201  & 250981\\
        & 5/2& 252162.6(2) & 252666 &--155   &--202&   251025 \\
         5g& 7/2& 258283.3(3)   &258439  &--143  &--201  & 256868\\
        & 9/2& 258283.2(3) & 258439 &--143   &--202&   256828 \\
         7d& 3/2& 267215.5(2)    &267475&--138   &--202  &  \\
     & 5/2&  267394.7(2)& 267647&--143  &--203&  \\
    8s& 1/2& 236544.3(3)   &268895 &--132   &--166  & \\
     6f& 7/2& 275919.8(3)   &276076  &--153  &--201  &  \\
        & 5/2& 276026.2(3) & 276097 &--152   &--201&    \\
         6g& 7/2& 279863.6(2)   &280235  &--143  &--202  & \\
        & 9/2& 279863.6(2) & 280237 &--143   &--201&   \\
        8d& 3/2& 285265(1)   &285497&--140   &--197  &  \\
     & 5/2&  285370(1)& 285597&--143  &--197&  \\
    IP & &  328909.4(3)& 327453&--143  &--201&  \\
       \hline  \hline

    \end{tabular}
    \label{tab:tin3+}
       \end{table}

The overall excellent performance of the FSCC is, however, not entirely uniform and does not extend to a number of anomalous fine-structure splittings in this ion. Table~\ref{tab:tin3+FS} shows the fine structure splittings of the $n$p, $n$d, and $n$f configurations of Sn$^{3+}$. The FSCC results are in excellent agreement with the experiment for the $n$p splittings. However, a discrepancy is observed for the 5d~$^2$D and the $n$f~$^2$F terms. For these splittings, the COWAN calculations perform quite well, as expected for a method based on a fitting procedure to experiment. In contrast, the FSCC \textit{ab initio} calculations fail to reproduce the apparent narrowing of the fine-structure interval of the 5d~$^2$D term and the widening of the 5f, 6f~$^2$F term intervals. For the 5d~$^2$D term, the fine-structure interval is measured at 107~cm$^{-1}$, while the theoretical result is higher by a factor of approximately 7 (Table~\ref{tab:tin3+FS}). 
Furthermore, while the FSCC results reproduce the unusual inverted structure of the 4f~$^2$F term, the magnitudes of the calculated fine-structure splitting of the 5f and 6f states are much smaller than the experimental values. It can be shown that this behavior is due to interaction with the doubly excited 4d$^9$5s$^2$ configuration for the 5d~$^2$D levels and the 4d$^9$5s5p configuration in case of the 5f, 6f~$^2$F levels. 
These configurations belong to sector (1,2), and are thus not included in the FSCC model space. A CI+MBPT calculation that included the doubly excited configuration yielded results that were much closer to experimental values (see Table~\ref{tab:tin3+FS}), resolving the discrepancy.

 \begin{table}
    \centering
      \caption{Fine structure splittings in $n$p, $n$d, and $n$f configurations of Sn$^{3+}$; all results are given in cm$^{-1}$. Based on data from Ref.~\cite{SchRyaBor18}.}
    \begin{tabular}{llllllllll}
       \hline
       &\phantom{0}5p  & \phantom{0}6p &\phantom{0}7p &\phantom{0}5d& \phantom{0}6d & \phantom{0}7d&\phantom{0}4f&\phantom{0}5f &\phantom{0}6f\\
      \hline
     Experiment~\cite{Moore-atomiclevels, SchRyaBor18}& 6508.4 & 2179.5     & \phantom{0}909.1&107.0  &331.1   &179.3  & --60.4& --309.6&--106.4\\
       COWAN& 6417& 2237     & \phantom{0}911&170 &240   &130 & \phantom{--}79& --228&\phantom{0}--73 \\
        FSCC& 6515& 2191    & 1139&736 &326   &172 & --73& \phantom{0}--40&\phantom{0}--21\\ 
        CI+MBPT& &     & &162 &   & & & --620&\phantom{--0}21\\
        MBPT~\cite{SafSavSaf03}&6471 &    &  & 745  & & & --76&\\
     
       \hline  

    \end{tabular}
    \label{tab:tin3+FS}
       \end{table}

The investigations on the tin ions presented above provide a unique opportunity to benchmark the performance of the FSCC approach for spectra of highly and weakly charged ions. The accuracy was shown to be on the same level as when dealing with neutral systems for both M1 and E1 transitions. However, certain limitations of the FSCC approach also showcase the importance of combining it with other complementary methods, such as the CI+MBPT.

\subsubsection{HCI candidates for search for variation of fundamental constants}\label{sub:HCI-VFC}

Several unification theories and standard model extensions predict variation of fundamental constants, such as the fine-structure constant, $\alpha$, or the electron-to-proton mass ratio, $\mu$, in space and in time~\cite{Uza11,Bar05}. One of the promising routes to search for variation of constants is via comparisons, over time, of atomic clocks based on transitions that show a different frequency dependence on the values of fundamental constants (typically, one of the transitions will be insensitive to the relevant constant, while the other should ideally benefit from a high sensitivity). A potential variation would then become observable as a change in the frequency ratio of these clocks.
An example is the measurement of the ratio of aluminum and mercury single-ion optical clock frequencies $\nu_{\text{Al}^+}/\nu_{\text{Hg}^+}$ over the course of about a year, which yielded a constraint on the temporal variation of the fine-structure constant of ${\dot{{\alpha}}}{/}{\alpha}=(-1.6{\pm}2.3){\times}10^{-17}{/}\mathrm{year}$. More recent experiments compared two optical clocks based on two different transitions in the same ion, namely the $^2$S$_{1/2}$~($F = 0$) $\rightarrow$ $^2$D$_{3/2}$~($F = 2$) electric quadrupole (E2) and the $^2$S$_{1/2}$~($F = 0$) $\rightarrow$ $^2$F$_{7/2}$~($F = 3$) electric octupole (E3) transitions in $^{171}$Yb+~\cite{GodNisJon14,LanHunRah21}. The current most stringent limit on the temporal variation of $\alpha$, based on these measurements, is ${\dot{{\alpha}}}{/}{\alpha}=1.0(1.1){\times}10^{-18}{/}\mathrm{year}$.

Electronic transitions in highly charged ions tend to have very high frequencies, in the EUV range or higher, making them impractical for use as clock transitions. However, two types of HCIs can be suitable for optical clock applications. The first are those having convenient magnetic dipole (M1) transitions in the optical range between the fine- or the hyperfine-structure sublevels~\cite{schiller07prl,YuSah13,YudTaiDer14}. The first optical atomic clock based on an M1 transition in Ar$^{13+}$ was recently reported. These transitions, however, exhibit very low sensitivity to the variation of the fine-structure constant, due to both levels involved in the transition being of the same nature and thus having a very similar dependence on $\alpha$. 

The second type of suitable clock transitions was proposed in the pioneering work of Berengut \textit{et al.}~\cite{BerDzuFla10,BerDzuFla12}. These transitions occur in HCIs that benefit from serendipitous orbital crossings between nearly degenerate configurations, causing the frequencies to fall in the optical range. This type of system is an excellent candidate for the search for variation of fundamental constants, in particular for the variation of the fine-structure constant $\alpha$, as the two levels corresponding to the clock transition belong to different configurations, leading to a different dependence on $\alpha$ variation. The relative sensitivity of transition energies to $\alpha$, $K$, can be estimated as~\cite{BerDzuFla10,BerDzuFla12}
\begin{equation}
   K \sim Z^2 (Z_i +1)^2,
    \label{eq:K}
\end{equation}
where $Z_i$ is the charge of the ion. We can thus expect the sensitivity of optical transitions in heavy HCIs to be enhanced by up to two orders of magnitude relative to transitions in neutral atoms or singly charged ions. 

The work of Berengut \textit{et al.}~\cite{BerDzuFla10, BerDzuFla12} offers a guide for identifying the region where HCIs that benefit from crossings of the configurations can be found, based on simple Dirac--Fock calculations of orbital energies. However, to identify the exact element and charge state where these crossings appear, high-accuracy calculations are required. 

Practical implementations of clocks based on HCIs also require knowledge of various systematic effects to determine whether a given transition is suitable for achieving the extremely high precision needed for the detection of possible $\alpha$ variation. It is thus important to have high-accuracy predictions of the various systematic effects prior to experiments. The major systematics affecting the accuracy and stability of a potential  HCI clock are Stark shifts due to lasers, black body radiation shifts, thermal radiation shifts, magnetic field shifts, motion-induced shifts, collisional shifts, and so on~\cite{KozSafCre18,YuSahSuo23}. These effects can be estimated using high-accuracy relativistic calculations, and such estimates can, in turn, help evaluate the feasibility of using a given HCI in an experiment or designing the experimental setup.

Since the original proposal, many promising HCI candidates have been identified based on theoretical investigations. These proposals are summarized in Ref.~\cite{YuSahSuo23}. Many of these calculations were preformed using the CI+MBPT or CI+all order approaches (e.g. Refs.~\cite{BerDzuFla12, SafDzuFla14a, DzuSafSaf15, AllDzuFla24,DzuFla24}). The relativistic coupled cluster approach, in particular its Fock-space variant, is also very well suited for such calculations and was employed for many HCIs with up to two valence electrons or holes~\cite{WinCreBek15,NanSah16,YuSah19,BekBorHar19,JyoChaYu23,YuSah24}. Such theoretical investigations usually provide predictions of the frequencies and the intensities of the optical transition of interest and their sensitivity to variation of the fine-structure constants, along with other properties needed to analyze the systematic effects. 
 
Recently, electron beam ion trap (EBIT) measurements of the spectrum of Pr$^{9+}$  ion were performed, where an orbital crossing transition was observed for the very first time~\cite{BekBorHar19}. Pr$^{9+}$ has two valence electrons and the 5p and 4f orbitals are very close in energy, leading to close 5p$^2$ and 5p4f configurations. Thus, the highly forbidden  5p$^2$~$^3$P$_0$ $\rightarrow$ 5p4f~$^3$G$_3$ transition was identified as the optical transition of interest. The experimentally determined transition energy was 22101.36(5)~cm$^{-1}$, in excellent agreement with the FSCC prediction of 22248~cm$^{-1}$. Interestingly, without HFS, the 5p4f~$^3$G$_3$ state would decay through a hugely suppressed M3 transition with a lifetime on the order of 10 million years. However, the admixture of the 5p4f~$^3$F$_2$ state through the hyperfine coupling induces a much faster E2 transition (lifetime of single years and width on the order of nHz). Such transitions have been probed in the past in state-of-the-art optical clocks~\cite{HunSanLip16} and can be accessed in HCIs using quantum-logic spectroscopy techniques~\cite{SchRosLan05}. 

The FSCC calculations were also used to estimate the size of the expected black-body radiation (BBR) shift. The static polarizability of the clock state was calculated to be 2.4~a.u., showing that it is suppressed due to the contracted size of the valence orbital. 
Furthermore, the ground state polarizability was found to be very similar to that of the clock state, so the differential polarizability of the transition is about 0.05~a.u., making a clock based on this transition extremely resilient to BBR even at room
temperature. The calculations have also shown this transition to be strongly sensitive to both variation of the fine-structure constant and to violation of local Lorentz invariance, making it a promising candidate for an experiment to measure these effects. Based on the theoretical and experimental findings in this work, a detailed experimental scheme was proposed for future precision spectroscopy on this ion.

\newpage
\section{Outlook}\label{sec:out}
The relativistic coupled cluster approach and, in particular, its multireference variant, relativistic Fock-space coupled cluster theory, has emerged as a powerful tool for high-precision atomic (and molecular) simulations, suitable for reaching benchmark accuracy appropriate for modern high-resolution spectroscopy and BSM physics investigations. Recent theoretical developments and applications showcase the  strengths of the CC methodology, particularly its compatibility with relativistic and QED (in case of FSCC) Hamiltonians and its ability to construct consistently improved approximations with predictable accuracy.
Key advantages of CC are the following:
\begin{itemize}
    \item seamless integration with advanced physical theories, 
\end{itemize}
\begin{itemize}
    \item systematic approach to improving accuracy,
\end{itemize}
\begin{itemize}
    \item predictable performance in complex calculations.
\end{itemize}
However, the current  CC formulations face some challenges that limit its applicability and accuracy. Below, we enumerate the challenges with possible strategies for their resolution:

\begin{itemize}
    \item \textit{Valence universality issues in FSCC}: 
   Calculating bound states in higher sectors becomes challenging when lower sectors do not provide satisfactory solutions. For example, describing excitations in closed-shell-like systems with zero electron affinity is difficult. Additionally, the traditional Fock-space approach often suffers from the lack of amplitude relaxation in the lower sectors due to the subsystem embedding condition used in solving coupled cluster equations.
To address these issues, an alternative wave operator ansatz within a simplified state universal framework could be proposed. An example of such an approach is inspired by the work of Banerjee and Simons from 1981~\cite{BanSim81}. It has recently been reformulated into a more efficient and powerful internally contracted multireference coupled cluster (ic-MRCC) method~\cite{Hanrath2000,evangelista2011orbital,hanauer2011pilot}. The ic-MRCC method has been implemented within a nonrelativistic and scalar-relativistic framework in the highly productive and user-friendly open-source GeCCo code, including a universal and efficient coupled cluster automatic code generator~\cite{GeCCo,Hanauer2011,Lipparini2017}.
The ic-MRCC formulation retains many attractive features of the Fock-space approach, except for normal ordering, making it easily implementable using existing FSCC programs. This alternative method effectively resolves the difficulties associated with higher sectors' states calculations and enhances the overall efficiency and accuracy of coupled cluster equation solutions. By adopting this alternative formulation to the relativistic realm, researchers may be able to effectively calculate complicated quasi-degenerate states across the entire Periodic Table, including multivalence open shells of heavy and superheavy elements,  particularly improving the description of excitations in these challenging systems, and thus, ultimately enhancing the benchmark capabilities of coupled cluster theory in quantum chemistry calculations. 
\end{itemize}
\begin{itemize}
    \item \textit{Intruder state problem in FSCC}: Despite recent advancements, a universal solution adequate for FSCCSDT and higher excitations models remains elusive, especially for high sectors of Fock space and geometry-dependent molecular issues. Alternative forms of IH, based on diagonalization of similarity-transformed Hamiltonians, rather than on the Jacobi solution of non-linear coupled cluster equations, developed for the Fock-space approach by Meissner and Musial~\cite{Meissner:98,Meissner:10} could be helpful to overcome the problem of convergence in these cases.
\end{itemize}
\begin{itemize}
    \item \textit{Mixed-sector formulation}: A universal size-extensive method that can handle multiple Fock-space sectors with the same overall number of electrons on equal footing is needed, and its development is in progress.
\end{itemize}
\begin{itemize}
    \item \textit{High-spin electronic states}: An alternative approach using Kramers-unrestricted open-shell vacuum states could potentially simplify calculations for d- and f-element compounds. A similar methodology has been implemented within the nonrelativistic Fock-space coupled cluster approach in the early 1990s~\cite{stanton1992fock} and could be straightforwardly extended to the relativistic realm. 
\end{itemize}
\begin{itemize}
    \item \textit{Analytic property calculations in FSCC}: The evaluation of analytic density matrices for properties is not yet available for intermediate and extrapolated Hamiltonian FSCC. However, recently, essential progress has been made in this direction by the introduction of the finite-order (over excitation operator $S$) method to calculate approximate density matrices and general property effective operators (both for the state average and the transition forms) in the FSCC method~\cite{oleynichenko2024finite}. This development could be regarded as an initial step toward the full implementation of an all-order analytical approach for property calculations. 
\end{itemize}
\begin{itemize}
    \item \textit{High-cost of calculations and availability of compact and flexible basis sets}: There are many efficient techniques used in modern many-body approaches for the reduction of scaling power of the most cost operations, such as diagrams transformation and contraction. One such method, namely tensor train decomposition (TTD)~\cite{oseledets2011tensor}, is particularly promising to use within the relativistic CC method. The expected calculation cost reduction is an order of magnitude. A similar reduction could be expected when using virtual natural orbitals (NO). The effective density matrix approximation derived to a finite order in the power of excitation amplitude for the low excitation approximation (e.g., CCSD) could be used for the derivation of compact virtual natural orbitals for further exploration in higher CC rank calculations (like CCSDT) and higher sectors of the Fock space. The use of NOs could substantially reduce the virtual orbital space and the computational cost as seen in existing implementations for relativistic MP2 \cite{YuaVisGom22} and EOM-CC methods \cite{SurChaNay22,ChaSurJan22} and is now being implemented in FSCC~\cite{oleynichenko2024finite}.
    \end{itemize}
    
\begin{itemize}
\item \textit{Limited availability of high quality analytical basis sets:} The Dyall basis sets, which are considered to be the most suitable for relativistic calculations and were successfully used in the examples discussed in this review, are currently available in the 2z, 3z, and 4z cardinality. The limited size of the basis set is found to be a major source of uncertainty. Larger basis sets are required to reduce uncertainties and to improve the reliability of CBS limit extrapolation schemes. To address this issue, 5z basis sets are in development for the Dyall basis set family~\cite{5z-dyall}. The 5z basis sets for the s- and p-block elements are currently being tested and will be published soon, while basis sets for the other elements, including elements 119 -- 122 are in development.
\end{itemize}

\begin{itemize}
\item \textit{High computational costs:} From a computational perspective, implementing modern quantum and parallel algorithms and exploiting graphics processing units (GPUs) offers the potential for significant speed improvements, particularly for heavy systems and more extensive basis sets.
\end{itemize}

The relativistic coupled cluster approach remains highly suitable for benchmark calculations within the no-virtual-pair approximation (NVPA) and its FSCC variant shows promise for extrapolation beyond NVPA toward integration with QED theory. Recent calculations demonstrate the method's power and reliability in advanced investigations of atomic and molecular systems with complex, quasi-degenerate electronic structures.
In conclusion, while CC theory has some limitations to overcome, it continues to be a valuable and potent approach for high-precision quantum chemical simulations, particularly in relativistic and QED contexts.

\newpage
\section{Summary and conclusions}\label{sec:sum}
This work provides a comprehensive review of high-accuracy relativistic electronic structure calculations tailored for spectroscopic investigations and support of atomic and molecular experiments in the heavy element domain. We explored various computational methodologies, with a specific focus on the four-component Dirac-–Coulomb–-Breit Hamiltonian and relativistic single-reference coupled cluster and Fock-space coupled cluster methods. These approaches effectively account for both relativistic and electron-correlation effects critical for accurately modeling the unique physical behavior of heavy and superheavy elements. The FSCC framework, in particular, demonstrates versatility by providing access to transition energies and hyperfine structure parameters, which are essential for investigating both fundamental physics and potential new physics beyond the Standard Model.

The relativistic computational advancements reviewed here are instrumental in designing and interpreting spectroscopic measurements on heavy elements. 
By combining electron correlation treatments with relativistic effects, this framework provides an effective predictive tool, especially when experimental data is limited or inaccessible, as is common with superheavy elements.

In conclusion, the reviewed methodologies underscore the importance of robust theoretical approaches in supporting experimental physics at the frontier of particle and nuclear physics. Future work may focus on extending these methodologies to incorporate even higher-order correlation, nuclear-structure dependent and quantum electrodynamics (QED) effects, and refining uncertainty quantification for predictions in unexplored elements. By bridging high-level theory with experiments, these methods continue to expand our understanding of atomic structure and fundamental interactions in the heaviest elements of the periodic table.
As experimental precision continues to improve and new applications emerge in areas like quantum metrology and tests of fundamental physics, high accuracy, the possibility to systematically improve the results and rigorous uncertainty evaluation offered by the relativistic CC methods will become increasingly valuable. The ongoing theoretical developments, combined with growing computational capabilities, position the relativistic coupled cluster methods to remain at the forefront of high-accuracy electronic structure calculations for years to come.
Looking ahead, while relativistic CC, and in particular its FSCC variant may not become  routine computational tools due to their inherent complexity and computational demands, they will continue to serve an essential role as a benchmark method for calibrating more approximate approaches and for providing reliable predictions where experimental guidance is limited. The future challenges lie not in establishing the validity of the method, which is now well demonstrated, but in expanding its scope while maintaining its high accuracy and reliability.

\newpage
\section*{Acknowledgements}
 
AB and LFP acknowledge the support from the project number VI.C.212.016 of the talent programme VICI financed by the Dutch Research Council (NWO) and thank the Center for Information Technology at the University of Groningen for support and for providing access to the Peregrine high performance computing cluster and to the Hábrók high performance computing cluster. LFP acknowledges the support from the Slovak Research and Development Agency projects APVV-20-0098 and APVV-20-0127 and from the Scientific Grant Agency of the Slovak Republic (project 1/0254/24). The work of AB was also supported by the project \textit{High-Sector Fock-space coupled cluster method: benchmark accuracy across the periodic table} with project number Vi.Vidi.192.088 of the research programme Vidi financed by the Dutch Research Council (NWO)

\section*{Author's contributions \textit{(optional section)}}
Detailing here the contributions of the authors of the review.

\bibliography{bibliography}

\end{document}